\documentclass[12pt,a4paper]{article} 
\pdfoutput=1
\usepackage{jheppub} 
\usepackage{graphicx} 
\usepackage{amsmath} 
\usepackage{amssymb} 
\usepackage{slashed} 
\usepackage{bigstrut} 
\usepackage{mathtools} 
\usepackage{multirow} 
\usepackage[utf8]{inputenc}
\usepackage{todonotes}
\usepackage{appendix}

\newcommand{\msbar}{\ensuremath{{\overline{\rm{MS}}}}}
\newcommand{\nf}{\ensuremath{n_f}}
\newcommand{\nft}{\ensuremath{\widetilde{n_f}}}
\newcommand{\alphas}[2]{\ensuremath{\alpha_{\rm S}^{#1}{#2}}}
\newcommand{\ud}{\ensuremath{{\rm d}}}

\title{
W$+$charm production with massive $\mathbf c$ quarks in \textsc{PowHel}
}

\author[a]{G.~Bevilacqua,}
\author[b]{M.V.~Garzelli,} 
\author[c]{A.~Kardos,} 
\author[c]{L.~Toth}

\affiliation[a]{ELKH-DE Particle Physics Research Group,\\
H-4010 Debrecen, P.O. Box 105, Hungary}

\affiliation[b]{II.~Institute for Theoretical Physics, Hamburg University
\\
Luruper Chaussee 149, D--22761 Hamburg, Germany}

\affiliation[c]{Department of Experimental Physics, Institute of Physics,\\
Faculty of Science and Technology, University of Debrecen, \\
H-4010 Debrecen, P.O. Box 105, Hungary}

\emailAdd{giuseppe.bevilacqua@science.unideb.hu, maria.vittoria.garzelli@desy.de, kardos.adam@science.unideb.hu, toth.lorant@science.unideb.hu}

\abstract{
The hadroproduction of a $W$ boson in association with a charm quark at the Large Hadron Collider is at the centre of 
current investigations due to its potential to probe the strangeness content of the proton. In this paper we present an implementation of the $W+c$ production process in  the \texttt{PowHel} event generator matched to the \texttt{PYTHIA8} parton shower approach, allowing to obtain predictions for differential cross-sections with NLO QCD accuracy matched to the accuracy of the Shower Monte Carlo event generator. Effects of non-diagonal CKM matrix elements, finite charm quark mass and off-shell $W$ decays including spin correlations are taken into account. We investigate the production of a leptonically decaying $W$ boson in association with either a charmed meson ($W^\pm~+~D^{*\mp}$) or a charmed jet ($W^\pm~+~j_{c}$) and compare our predictions with particle-level measurements by the ATLAS and CMS collaborations at $\sqrt{s} =$ 7 and 13 TeV. The role of the so-called "opposite sign" and "same sign" contributions to the theoretical cross sections is presented and discussed, pointing out the importance of including parton shower effects for a reliable estimate of the latter and a faithful comparison with experimental data.
}
 

\begin{document}
\maketitle 

\section{Introduction}
\label{sec:introduction}

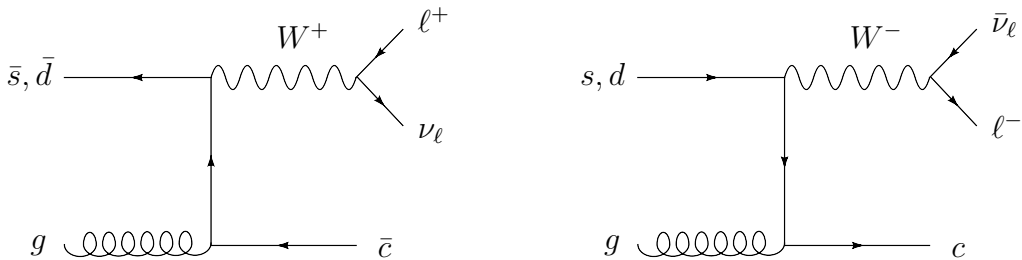
\begin{figure}[h!tb]
\label{fig:LOdiagrams}
\begin{center}
\includegraphics[width=0.3\textwidth]{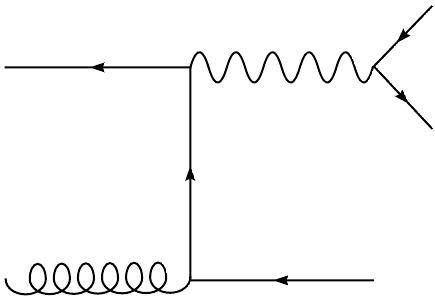}
\put(-48,79){$W^+$}
\put(5,85){$\ell^+$}
\put(5,45){$\nu_\ell$}
\put(-10,0){$\bar{c}$}
\put(-140,3){$g$}
\put(-149,65){$\bar{s},\bar{d}$}
 \hspace{0.19\textwidth}
\includegraphics[width=0.3\textwidth]{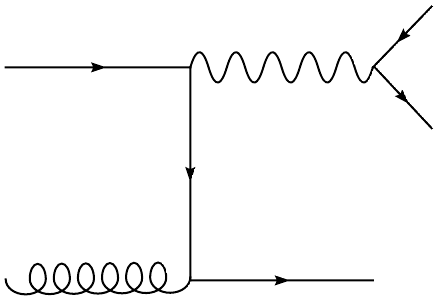}
\put(-48,79){$W^-$}
\put(5,45){$\ell^-$}
\put(5,85){$\bar{\nu}_\ell$}
\put(-10,0){$c$}
\put(-140,3){$g$}
\put(-149,65){$s,d$}
\end{center}
\caption{Feynman diagrams describing $W+c$ production at LO accuracy.}
\end{figure}

Various processes involving the production of heavy flavours at hadron colliders are interesting 
probes of the internal structure of the proton, expressed in terms of parton distribution functions (PDF) 
at leading twist in collinear factorization~\cite{Accardi:2016ndt}. On the one hand, 
single inclusive heavy-meson production has been used to constrain the gluon and sea quark PDFs in 
longitudinal momentum fraction regions not co\-vered by the HERA 
data~\cite{Zenaiev:2015rfa, Bertone:2018dse, Zenaiev:2019ktw}. On the other hand, the production of a 
$W$ boson in association with a charm quark either fragmenting into an isolated $D$-meson reconstructed 
through its decay products or generating a $c$-jet $j_c$ (hereafter we denote both possibilities by $W + c$, 
collectively indicating, with a slight abuse of notation, both the $W^+ + \bar{c}$ and the $W^- + c$ cases), 
can provide important information on strange-quark PDFs and the sea quark flavour composition, 
for which large uncertainties still exist~\cite{Alekhin:2014sya, Alekhin:2017olj, Faura:2020oom}. 
In particular, beside deep inelastic scattering (DIS) HERA data, legacy datasets included in 
PDF fits to constrain the strange sea include DIS dimuon data from massive high-density detectors, 
bubble chamber and nuclear emulsion data on charmed hadron production. The incapability of obtaining good 
fits of all these datasets simultaneously, has led the PDF collaborations to discard or question the validity 
of some of them. Recent precise Large Hadron Collider (LHC) data (in particular Drell-Yan production) are also 
sensitive to strange quark distributions. Their incorporation in PDF analyses has also disclosed potential 
tensions with part of the aforementioned legacy datasets. The present understanding of proton strange sea 
is thus still far from being satisfactory. While waiting for new DIS measurements capable of probing the strange sea 
 at a proposed Large Hadron-Electron Collider~\cite{Abdolmaleki:2019acd} or at a forthcoming 
Forward Physics Facility~\cite{FPF:2021} during the High-Luminosity LHC phase, 
investigating the information we can extract
nowadays
from the $W + c$ process at the LHC  is indeed timely.
In fact this process is sensitive to the (anti-)strange-quark PDF at leading order (LO) accuracy (see Figure \ref{fig:LOdiagrams}). 
The sensitivity to strange PDF sea is somehow weakened by the fact that the Cabibbo-Kobayashi-Maskawa (CKM) 
matrix is non-diagonal and by the inclusion of radiative corrections, which involve the opening of additional 
channels at next-to-leading order (NLO) ($gg$, $q\bar{q}$, $\bar{q}q$, $qq^\prime$, $\bar{q}\bar{q}^\prime$). 
At the LHC, $W+c$ can probe strange-quark PDFs at scales $Q$ of the order of the $W$ mass. 
This scale is larger than the typical $Q \sim 1-10$ GeV reach of the aforementioned legacy 
DIS experiments traditionally used to probe the strange-quark distribution. 
Precise measurements of the cross section ratio $\mathcal{R} = \sigma(W^{+}+\bar{c})/\sigma(W^{-}+c)$ 
can highlight the degree of asymmetry between $s$ and $\bar{s}$ PDFs and help constraining 
the strangeness suppression factor $R_s = (s+\bar{s})/(\bar{u}+\bar{d})$. 
The $\mathcal{R}$ ratio is also sensitive to the $V_{cd}$ parameter of the CKM matrix, 
whose impact on the $W^{-}+c$ channel is substantially larger than on $W^{+}+\bar{c}$ 
due to the fact that in $pp$ collisions the $d$-quark induced parton luminosity is 
larger than the $\bar{d}$-quark induced one. 

On top of that, $W+c$ is a background for intriguing Standard Model (SM) 
processes such as $WH(H \to c\bar{c})$ production, for which experimental analyses have started to be 
performed only recently~\cite{Sirunyan:2019qia, ATLAS:2021zwx}, and for more known processes, 
such as single $t$-quark or $t\bar{t}$ quark production. Beyond the SM, it plays a role in 
searches of Dark Matter final states accompanied by jets \cite{Iwamoto:2017ytj, Aaboud:2017phn, Sirunyan:2017xgm}, 
and in models with an extended Higgs sector, \textit{e.g.} two-Higgs doublet models~\cite{Branco:2011iw} 
with a Higgs boson decaying into a $c\bar{c}$ pair. 

Given all these reasons, $W+c$ production has received considerable attention in the last two decades. 
The first measurements of $W+c$ cross sections, carried out at the Tevatron by the CDF and D0 
experiments \cite{Aaltonen:2007dm,Abazov:2008qz, Aaltonen:2012wn}, collected limited data for $W+j_c$, while more recent measurements have been 
performed at the LHC in Run I and II by the ATLAS (at $7$ TeV) and CMS (both at $7$ and $13$ 
TeV~\footnote{An analysis at $\sqrt{s} = 8$ TeV has also been performed by the CMS collaboration,
for which a paper has been released very 
recently~\cite{CMS:2021oxn}.}) collaborations 
\cite{Chatrchyan:2013uja,Aad:2014xca,Sirunyan:2018hde} for both $W+D$ meson and $W+j_c$ production. 
Results for $W+c$ production in the forward region are also available by the LHCb collaboration \cite{Aaij:2015cha}.

The high phenomenological interest discussed above motivates this work, where we present, 
in the same framework, predictions for both $W + D$-meson and $W~+~j_c$  production in $pp$ 
collisions and compare them to the LHC ATLAS and CMS experimental data 
at the particle level\footnote{In our paper we use particle level for 
hadron level to be in accord with terminology used by the experiments.}. 

Next-to-leading order (NLO) QCD predictions for $W+c$ production in hadronic collisions 
are available since a while~\cite{Giele:1995kr,Campbell:2015qma} and the complementarity of Drell-Yan 
and $W+c$ studies to better constrain NLO PDF fits has already been
pointed out some time 
ago~\cite{Stirling:2012vh}. Meanwhile further progress has been achieved 
with the computation of EW corrections and combined NLO QCD+EW results for $V+{\rm jets}$ final
states \cite{Denner:2009gj,Kallweit:2014xda,Kallweit:2015dum,Biedermann:2017yoi}.

Predictions for $W+c$ hadroproduction at NLO QCD accuracy matched to Parton Shower (PS) 
according to the MC@NLO matching framework~\cite{Frixione:2002ik}
can be obtained using the \texttt{MadGraph5\_aMC@NLO} 
framework~\cite{Alwall:2014hca}, with appropriate modifications of the default version
\footnote{To implement the possibility of massive charm quarks in the \texttt{MadGraph5\_aMC@NLO} 
hard-scattering matrix-elements one has to activate a customized input model file.}. 
Predictions for $W+c$ hadroproduction according to the POWHEG NLO~+~PS matching 
framework~\cite{Nason:2004rx, Frixione:2007vw} with $W$ bosons decaying leptonically are 
presented here for the first time, by making use of the {\texttt{PowHel}} event generator, 
including charm-quark mass in the hard-scattering matrix-elements, treating the  $W$ 
off-shell, as well as including non-diagonal CKM matrix effects.  \texttt{PowHel} produces 
events at the first radiation emission level in the Les Houches Event Format (LHEF). 
Further radiation emissions are simulated by interfaces to standard 
Shower Monte Carlo (SMC) codes.

Very recently, the first, state-of-the-art predictions for $W + j_c$ production at 
NNLO accuracy were presented in Ref.~\cite{Czakon:2020coa}, working in a five flavour number scheme, 
therefore neglecting charm quark finite-mass effects. In an effort to complement this ground-breaking calculation, 
in this work we show predictions for $W + D$-meson production retaining charm quark finite-mass effects in the 
hard-scattering matrix-elements, evaluated in the decoupling renormalization and 
factorization scheme~\cite{Collins:1978wz} with three active flavours.
We also show predictions for $W + j_c$.

A direct comparison of the predictions in the work of Ref.~\cite{Czakon:2020coa}  with the experimental data on differential cross sections at the particle level suffers from some limitations.
This is due, on the one hand, to the lack of parton shower and hadronization 
effects in jet formation (a $c$-jet for the $W + c$ process in a fixed-order NNLO calculation
with five active flavours is formed by one massless charm quark and at most two additional massless partons), 
whereas an experimental jet is obtained by clustering many hadrons, instead of a few partons.
The experimental collaborations, in some analyses, in order to facilitate comparisons with fixed higher-order
predictions unfold their fiducial results from the particle to parton level, 
but this procedure introduces additional systematic uncertainties, depending on the underlying assumptions.
On the other hand, the jet algorithm used in the experimental analyses so far 
(anti-$k_T$~\cite{Cacciari:2008gp}) is not applicable to heavy-flavour jet studies at NNLO
in case the heavy flavour is treated massless in the calculation. When the heavy flavour is treated massless
and a heavy-flavoured pair is produced with high collinearity (due to massless treatment the dead-cone does not
prevent this configuration) the produced singular contribution is not regularized because no subtraction nor 
mass is present to prevent from collinear emissions.
The same issue appears when the heavy-flavoured pair
becomes soft. This implied that the authors of Ref.~\cite{Czakon:2020coa} had to use a different algorithm 
(flavoured-$k_T$ algorithm~\cite{Banfi:2006hf}), which avoids the aforementioned issue by first 
clustering the two quarks together, before a potential recombination with the other final state parton/pseudojet.
On a complementary side in the case of a NLO+PS computation, like the one that we present in this work, 
the accuracy of the hard-scattering matrix-elements is lower (NLO instead of NNLO on inclusive predictions), 
however parton shower, hadronization, multiple parton interaction and beam remnant effects are included, 
allowing for a more realistic modeling of hadronic observables.
We use the anti-$k_T$ jet algorithm like in the already available LHC experimental analyses. 
Let us note that, for the process under consideration, the cross section at Born level does not suffer from infrared singularities when a finite $c$-quark mass is considered. When higher-order QCD corrections are included, the infrared  singularities originating from virtual- and real-radiation contributions cancel for infrared-safe observables and leave the cross section finite. This allows one to cover the whole phase space of $W+c$ production without need of any technical cut.

The main focus of our study is $W+D$-meson production, which allows for looser experimental $p_T$ cuts 
compared to the case of $W+j_c$. For the production of identified hadrons, the complications related to the jet algorithm choice are 
not present. However, uncertainties exist on the hadronization process. Uncertainties regarding the 
heavy-quark treatment in parton shower evolution are present for both the $j_c$ and $D$-meson case. 

We also observe that the experimental collaborations so far have employed $W$~+~jets samples of simulated 
events with hard-scattering matrix-elements evaluated considering five active flavours for their 
published analyses of the $W + c$ data at particle level. On the other hand, they have used predictions 
with massive charm as available in various NLO integrators for comparison with their data unfolded to the parton level.  

After various studies of the compatibility of strange quark distributions traditionally extracted from legacy DIS datasets 
with the LHC $W+c$ data (see e.g. Ref.~\cite{Alekhin:2014sya}), the PDF collaborations have already started to 
include the LHC $W + c$ production data in their fits. In particular, the $W + c$ data analyzed in this paper were very recently included in the new release of the NNPDF PDF fit, 
i.e. NNPDF4.0~\cite{Rojo:2021gdq, Ball:2021leu}, 
after the study of Ref.~\cite{Faura:2020oom} using as a basis the NNPDF3.1 PDF fit~\cite{Ball:2017nwa}, 
which has lead to various variants of the latter, identified by the prefix \texttt{str} (for strange quark). 
Some of these variants include the LHC $W + c$ production data compared with fixed-order NLO predictions 
supplemented by a NNLO $K$-factor although, in the absence of NNLO predictions for the $W + c$ case with massive charm, 
the calculation of a NNLO $K$-factor is indeed challenging. Additionally, such a comparison is possible only 
for experimental data unfolded at the parton level.   
A part of these data (i.e. those in Ref.~\cite{Chatrchyan:2013uja}) has also been included in the very recent 
MSHT20 PDF fit~\cite{Bailey:2020ooq}. The CT18 collaboration, although not including the LHC $W + c$ data in their 
fit~\cite{Hou:2019efy}, has benchmarked their NNLO PDF fits with respect to the ATLAS $W + j_c$ data~\cite{Aad:2014xca}, 
without, however, using them in association with a full NNLO calculation of partonic cross-sections for this production process.
Our work provides new insights about including the LHC $W + c$ data in PDF fits and a new method 
for getting predictions at NLO + SMC accuracy. 

The paper is organized as follows. In Section~\ref{sec:method} we present the computational framework that we used 
for this work and discuss the setup of the calculation, as well as the various options for 
modelling $W$ production and decay available in our $W + c$ implementation. In Section~\ref{sec:pheno} 
we discuss  phenomenological results for differential cross sections adopting the cuts used by the LHC experimental analyses. 
Finally in Section~\ref{sec:conclu} we draw our conclusions.
Details related to the conversions between different factorization and renormalization schemes are presented in Appendix A, 
whereas the role of different contributions to the fiducial cross sections that we compare to experimental data are discussed in Appendix B.

\section{Computational setup}

\subsection{Details of the implementation}
\label{sec:method}
This work is based on the {\texttt{PowHel}} event generator, built on the interface between the {\texttt{POWHEG-BOX-v2}} 
computing framework~\cite{Alioli:2010xd}, which implements the POWHEG NLO+PS matching~\cite{Nason:2004rx,Frixione:2007vw}, 
and {\texttt{HELAC-NLO}}~\cite{Bevilacqua:2011xh}, which provides all matrix elements. 
Infrared divergences are regularized by using the Frixione-Kunszt-Signer subtraction scheme 
\cite{Frixione:1995ms,Frixione:1997np}, as implemented in the {\texttt{POWHEG-BOX-v2}}.
The subtraction of the residual initial-state collinear divergences is achieved in the 
decoupling factorization scheme.  The output of {\texttt{PowHel}} are events at the 
first radiation emission level, i.e. after the emission of the first additional resolved 
or unresolved radiative parton. These are collected in files with Les Houches event format (LHEF). 
Further radiative emissions, as well as the hadronization and hadron decays, 
can be described by providing these files as an input to SMC programs, 
including various versions of \texttt{PYTHIA}~\cite{Sjostrand:2019zhc, Sj_strand_2015, Sjostrand:2006za} and \texttt{HERWIG}~\cite{Bellm:2019zci, Corcella:2000bw, Corcella:2002jc}. 
This way, fully exclusive differential cross-sections for many different observables at the hadron level can be computed.

The {\texttt{PowHel}} event generator was originally developed for phenomenological studies of 
top-antitop hadroproduction in association with a range of different particles. 
In particular, predictions have been presented for several top-quark pair related processes at both the 
Tevatron and the LHC, and used by the experimental collaborations in a variety of SM and BSM analyses. 
LHEF events for all these processes were generated in the ${\overline{\rm{MS}}}$ renormalization and 
factorization scheme with 5 active flavours at all scales above the bottom threshold.
The {\texttt{PowHel}} extension to the decoupling scheme with 4 active flavours at all scales 
was discussed in Ref.~\cite{Bevilacqua:2017cru}, where predictions for $t\bar{t}b\bar{b}$ hadroproduction 
with massive bottom quarks were compared to those with massless bottom previously obtained~\cite{Garzelli:2014aba}.  

In this paper, for the first time, we have extended {\texttt{PowHel}} to the decoupling scheme with 3 active flavours, 
which has required the extension of {\texttt{HELAC-1LOOP}}, part of \texttt{HELAC-NLO}, to include finite-mass effects of 
$c$-quarks in all aspects of the computation. 
Consistently, PDFs in the decoupling scheme with 3 active flavours together with their associated $\alpha_S$ 
evolution have been used as input of the hard-scattering process. 
Several PDF sets extracted according to this scheme are available in the LHAPDF interface \cite{Buckley:2014ana}.
However, only part of them include a number of eigenvectors/set-members, allowing for an estimate of PDF related uncertainties. 
The ABMP16\_3\_NLO PDF fit, that we choose as default in this study, is one of such cases. 

In our analysis we also used examples of PDF sets and corresponding \alphas{}{} in the \msbar{} scheme. 
To be able to use these with hard-scattering matrix-elements with a massive charm quark, in general, 
a compensation term is needed \cite{Cacciari:1998it} 
which appropriately modifies both the coupling and structure functions to decouple the massive degree of 
freedom~\footnote{In the currently available \msbar{} PDF sets and corresponding $\alpha_S$ evolution, 
the number of active flavours below each threshold decreases by just one unity, i.e. 
it switches from 5 light flavours to 4 and 3 light flavours below the bottom and charm thresholds, respectively.}. 
This compensation term is process dependent and the derivation suitable for our purposes 
can be found in Appendix A.

The \texttt{HELAC-NLO} matrix elements have been extensively checked against the public code \texttt{MadGraph5\_aMC@NLO}.
We have also performed comparisons at the integrated level, 
checking our NLO fixed-order predictions against the public code \texttt{MCFM} \cite{Campbell:2019dru}, which includes a $W+c$ implementation by the same authors of the $W+t$ one~\cite{Campbell:2005bb}.
For these cross-checks we used different systems of cuts. In all cases we found perfect agreement.

New updates and features have been also implemented in the \texttt{PowHel} framework. 
These include an interface to the \texttt{POWHEG-BOX-v2} and the implementation of a 
general-purpose phase space generator that can be used for generic multi-leg processes. 
In the all-purpose multi-channel phase space generator we take into account both $s$- and $t$-channel branchings.
For intermediate massive particles with finite width our phase-space generator
maps away the Breit-Wigner resonances.  The phase space is constructed using channels. 
The possible channels are determined by examining the contributing subprocesses undoing 
particle branchings in all possible ways.

The new \texttt{PowHel} version used in this paper got equipped with machinery to 
fully automatically determine all non-zero helicity configurations for each of the subprocesses. 
It only uses a minimal set of subprocesses to obtain all the others by means of crossing.

%-------------------------------------------------
%
\subsection{Input parameters}
\label{sec:input}
%
%-------------------------------------------------

We consider the processes $pp \to \ell^+ \nu_\ell \,\bar{c} + X$ and $pp \to \ell^-  \bar{\nu}_\ell \,c + X$ 
at NLO QCD accuracy, \textit{i.e.} the hard scattering is computed at $\mathcal{O}(\alpha_s^2 \alpha^2)$.
Here and in the following we denote with $\ell$ either an electron, muon or tau, all considered 
massless~\footnote{ For the time being, the  $W^\pm \rightarrow \tau^\pm \overset{(-)}{\nu}_\tau$-decay 
channel is not included in the $W + c$ experimental analyses.}
in the computation of the matrix elements. For this reason the latter are only 
implemented for the electron channel and reused for the other lepton generations.
The possible decay channels can be set through the \texttt{PowHel} \texttt{vdecaymode} reserved word in the 
input card~\footnote{If $n_\ell$ corresponds to the lepton family $\ell = e, \mu, \tau$, 
the ensemble of the allowed decay channels can be set through 
$\mathtt{vdecaymode}=100\cdot n_e + 10\cdot n_\mu + n_\tau$. Using $n_\ell = 1 (0)$ for each channel, 
the channel is turned on (off).}, used for several $W$ boson related processes in \texttt{POWHEG-BOX-v2}. For brevity, 
in the following we will also refer to the above-mentioned processes as to $W^+\bar{c}$ and $W^- c$. 
However it should be clear that we consider realistic final states with hadrons, 
charged leptons and missing transverse momentum. 
Also, unless stated explicitly, no on-shell approximation is applied to model $W$ boson decays.

The CKM matrix is assumed to be non-diagonal. 
More precisely we consider the approximation where mixing occurs only between the first two generations of quarks, 
using $\sin^2\theta_{\mathrm{Cabibbo}} = 0.050934360$ as mixing parameter \cite{PhysRevD.98.030001}. Based on this choice, 
the list of partonic subprocesses that contribute to $W^+ \bar{c}$ and $W^- c$ 
production can be obtained via crossing from the following master amplitudes:
\begin{eqnarray}  \label{Eq:oneloop}
\emptyset \to \ell^+  \nu_\ell \, \bar{c}  s \, g   \hspace{1.5cm}    &    
\emptyset \to \ell^-  \bar{\nu}_\ell \, c  \bar{s} \, g \nonumber \\
\emptyset \to \ell^+  \nu_\ell\, \bar{c}  d \, g    \hspace{1.5cm}    &    
\emptyset \to \ell^-  \bar{\nu}_\ell \, c  \bar{d} \, g
\end{eqnarray}
\begin{eqnarray}  \label{Eq:treelevel}
\emptyset \to \ell^+  \nu_\ell \,  \bar{c} s \, g g  \hspace{1.5cm}    &    
\emptyset \to \ell^-  \bar{\nu}_\ell \, c  \bar{s} \, g g  \nonumber \\
\emptyset \to \ell^+  \nu_\ell \, \bar{c}  d \, g g   \hspace{1.5cm}    &    
\emptyset \to \ell^-  \bar{\nu}_\ell \, c  \bar{d} \, g g  \nonumber  \\
\emptyset \to \ell^+  \nu_\ell \, \bar{c} s \, q \bar{q}  \hspace{1.5cm}    &    
\emptyset \to \ell^-  \bar{\nu}_\ell \, c  \bar{s} \, q \bar{q}  \nonumber  \\
\emptyset \to \ell^+  \nu_\ell \, \bar{c} d \, q \bar{q}   \hspace{1.5cm}    &    
\emptyset \to \ell^-  \bar{\nu}_\ell \, c  \bar{d} \, q \bar{q} \nonumber \\
\emptyset \to \ell^+  \nu_\ell \, c \bar{c} \, \bar{u} d  \hspace{1.5cm}    &    
\emptyset \to \ell^-  \bar{\nu}_\ell \, c \bar{c} \, u \bar{d}  \nonumber \\ 
\emptyset \to \ell^+  \nu_\ell \, c \bar{c} \, \bar{u} s  \hspace{1.5cm}    &    
\emptyset \to \ell^- \bar{\nu}_\ell \, c \bar{c} \, u \bar{s}
\end{eqnarray}
where $q \in \{u,d,s\}$.  The subprocesses in Eq.~(\ref{Eq:oneloop}) 
contribute to the virtual part of the NLO cross section as well as to the integrated subtraction terms, 
while the ones in Eq.(\ref{Eq:treelevel}) contribute to the real part. 
Other input parameters used in our computation are:
\begin{equation}
\begin{array}{lcl}
 G_F=1.16639 \cdot 10^{-5} ~{\rm GeV}^{-2} \,,  &  \quad\quad \quad\quad & m_{t}=172.5 ~{\rm GeV} \,, \\[0.2cm]
m_{b}=4.75  ~{\rm GeV}\,, & & m_{c} = 1.5 ~{\rm GeV} \,, \\[0.2cm]
m_{W}=80.379 ~{\rm GeV} \,, & & \Gamma_{W} = 2.085 ~{\rm GeV} \,, \\[0.2cm]
m_{Z}=91.1876 ~{\rm GeV} \,. \\[0.2cm]
\end{array}
\end{equation}
We work in the $G_\mu$ scheme and the electromagnetic coupling $\alpha$ is derived according to the formula
\begin{equation}
\alpha = \frac{\sqrt{2}}{\pi} \, G_F \, m_W^2 \, \sin^2\theta_W,
\end{equation}
where $\sin^2\theta_W = 1 - m_W^2/m_Z^2$.
The treatment of the resonant $W$ boson is performed in the fixed width scheme.
Leptonic mass effects are neglected in the calculation of the matrix elements. However for compatibility with \texttt{PYTHIA8}, 
for which the default values of all lepton masses are non-zero, we restore leptonic masses at the stage of generation of \texttt{PowHel} events. This is done by rescaling momenta such that energy and momentum conservation as well as masses of resonances are preserved~\footnote{This is done by the built-in \texttt{lhefinitemasses} routine of \texttt{POWHEG-BOX-v2}.}. For the leptonic masses we use the same values encoded in the SMC generator, i.e.  $m_e = 0.511 \cdot 10^{-3}~{\rm GeV}$ and $m_\mu = 0.10566  ~{\rm GeV}$.

In this study we will present results obtained with the following PDF sets:  ABMP16\_3\_NLO~\cite{Alekhin:2018pai}, CT18NLO~\cite{Hou:2019qau} and CT18ZNLO~\cite{Hou:2019qau}, emerged from QCD analyses including different experimental data. Additionally, the analyses performed by the ABMP16 and CT18 collaborations are based on different theory input, heavy-flavour and $\alpha_S$ treatment, different PDF parameterizations and different treatment of uncertainties 
and tolerance criterion. 
We note that the PDF sets adopted in this work assume zero 
strange asymmetry, \textit{i.e.} equal distributions for the $s$ and the $\bar{s}$ sea quarks. 
While it is known that perturbative evolution in QCD generates a $s-\bar{s}$ asymmetry~\cite{Catani:2004nc}, 
the absolute size of the latter is at present unclear and different PDF groups adopt different approaches 
towards it.  
Furthermore we have checked that, at the level of fixed-order NLO calculations in the phase-space region more relevant for current $W + c$ experimental studies, predictions obtained with the NNPDF3.1 NLO PDF set \cite{Ball:2017nwa}, that is currently used in many experimental analyses, lie between those computed with the ABMP16 and CT18 NLO PDFs.
A comprehensive study of PDF uncertainties, based on extensive analysis of all the available PDF sets, is clearly in order. While this study is indeed possible within our framework, it is beyond the scope of the present work and 
therefore left for future analyses.

Each PDF set is provided by the respective PDF collaboration together with a specific value of the strong coupling $\alpha_S$ at the $M_Z$ scale.  The two-loop running of $\alpha_S$, consistent with NLO calculations, is provided by the LHAPDF interface. Let us note that the ABMP16\_3\_NLO PDF set has three active flavours at all scales. The ABMP16 PDF family was derived in a QCD analysis adopting the decoupling scheme for computing the theory predictions used, together with a selection of experimental data, as input to the fit. On the other hand, the CT18 and CT18Z PDF families were derived using a Variable Flavour Number Scheme approach.  The CT18 analysis is regarded as one of the global PDF fits. In case of CT18NLO and CT18ZNLO PDFs, a three-flavour version is not yet available~\footnote{The CT18 collaboration is currently working on such a PDF fit version~\cite{CT18private}.}. In order to use the latter sets in our computations, the matrix elements that {\texttt{HELAC-NLO}} provides in the decoupling scheme have to be properly modified by compensating terms, as already mentioned in section~\ref{sec:method}.  The explicit calculation (see Appendix A) shows that the compensating terms are identically zero when the factorization scale, governing the factorization and PDF evolution, and the renormalization scale, governing $\alpha_S$ evolution, are assumed to be equal.

As for any fixed-order calculation, the hard-scattering process contains a residual dependence on the 
renormalization and factorization scales (denoted by $\mu_R$ and $\mu_F$ respectively). 
We consider $\mu_R = \mu_F = \mu_0 = H_T/2$ to be our default scale choice, where
\begin{equation}
H_T =  \sqrt{p^2_{T,W} + m_W^2} + \sqrt{p^2_{T,c} + m_c^2} \,. 
\label{eq:scale}
\end{equation}
Here $p_{T,W}$ is the transverse momentum of the $W$ boson reconstructed from its decay products. This scale is evaluated using the kinematics of the underlying Born process and is employed for the computation of all (Born, virtual and real) contributions to the NLO cross section with the only exception of the real contribution from $W^\pm c\bar{c}$. For the latter the functional form used for the scale definition reads \smallskip 
\begin{equation}
H_T =  \sqrt{p^2_{T,W} + m_W^2} + \sqrt{p^2_{T,c} + m_c^2}  + \sqrt{p^2_{T,\bar{c}} + m_c^2} \,.
\label{eq:scaleWcc}
\end{equation}
Uncertainties stemming from variation of scales are estimated by rescaling $\mu_R = \xi_R\,\mu_0$ and $\mu_F= \xi_F\,\mu_0$, using the standard seven combinations 
\begin{equation}
\left( \xi_R, \xi_F \right) = \left\{ (0.5, 0.5), (0.5, 1), (1, 0.5), (1, 1), (1, 2), (2, 1), (2, 2) \right\} \,.
\end{equation}
The scale uncertainty band is defined taking the upper and lower bounds of the envelope of predictions obtained with these scale choices.

By default in the \texttt{POWHEG} matching the Sudakov peak is produced by exponentiating the whole real emission contribution. All-order effects should only affect the low $p_T$ region of extra parton emission. However, as it was shown in \cite{Alioli:2008tz}, there can be processes where this method results in an increase of yield in the high $p_T$ region as well.
As a remedy to this problem the \texttt{hdamp} option was introduced \cite{Alioli:2008tz} in order to only exponentiate a
part of the real-emission. With a careful selection of the damping function it was achievable to reproduce the fixed-order curve at high-$p_T$, while simultaneously having a Sudakov shape at low $p_T$.

For the Sudakov peak formation the real-emission contribution is used in the form of ratio of the real-emission and Born
squared matrix elements. There can be processes where the Born squared matrix element vanishes at phase space points.
This results in a numerical instability and was circumvented by the introduction of the \texttt{flg\_bornzerodamp}
parameter which excludes the real-emission contribution from the exponentiation if it is at least five times larger
than the subtractions \cite{Alioli:2008gx}. 
For the process at hand the Born contribution is finite in all phase-phase points, because the charm quark mass is finite, 
although having a small value when compared with some of the
other characteristic scales of the process. The \texttt{bornzerodamp} parameter can also be used to improve the 
efficiency of event generation by excluding those phase-space regions where the extra parton emission
is well resolved, but, because of a peaking contribution, could negatively affect the event generation, 
considering the hit-and-miss technique applied to this purpose.  The Sudakov peak is the consequence of
the all-order summation of Sudakov logarithms. Since Sudakov logarithms have an IR origin, 
the regions where these logarithms are not dominating should not have an effect in the formation of the 
Sudakov peak.

Extensive studies were carried out and it was found that, beside setting the flag \texttt{flg\_bornzerodamp} 
to one, no special damping factor is needed to reach a good efficiency in event generation 
for the process at hand, have the desired Sudakov shape at low-$p_T$ and match the event-level curve to the 
fixed-order one at large $p_T$.

\subsection{$W$ off-shellness and spin correlation effects}

Due to the high efficiency and integrated luminosity reached in the LHC Runs so far, 
enough statistics was collected to apply sophisticated sets of cuts on events potentially contributing to $W+c$ production. 
Due to the decaying $W$ and the exclusiveness of cuts it is not at all trivial to establish a priori up to 
which extent spin correlations are important to make sane predictions. In order to investigate the 
effect of spin correlations or the lack thereof, we consider four different approaches for the modeling of 
$W$ boson decays:

\begin{enumerate}
  \item \emph{Full calculation}: in this case the program uses matrix elements where the leptonic
  final states are present and a finite width is assigned to the $W$ boson. So this option includes 
  full spin correlations in the $W$-boson decay products with finite width effects for the $W$.
  \item \emph{Narrow width approximation {\rm{(NWA)}}}: the same matrix elements are being used as for 
  case 1 but the Breit-Wigner resonance of the $W$ is replaced by a Dirac-delta function
  corresponding to the $\Gamma_W\to 0$ limit:
  \begin{align}
    \left|\frac{i}{p_W^2 - m_W^2 + i m_W\Gamma_W}\right|^2 \to 
    \frac{\pi}{m_W \Gamma_W}\delta(p_W^2 - m_W^2)
    \,.
  \end{align}
  In this approach spin correlations are preserved while finite-width effects of the $W$ boson decay are dropped.
  \item \emph{No spin correlation}: a stable $W$ boson is considered in the matrix elements. 
  The on-shell $W$ boson is decayed in the event generation step, prior to the shower evolution, using isotropic decay.
  This way the SMC receives events with leptonic final state but lacking spin correlations in the
  decay products.
  \item \emph{SMC-decay}: The same matrix elements as for case 3 are adopted. 
  In this approach the $W$ decay is performed directly by the SMC program, 
  which by default does not account for spin correlation effects.
\end{enumerate}

\begin{figure}
  \begin{center}
 \includegraphics[width=0.495\textwidth]{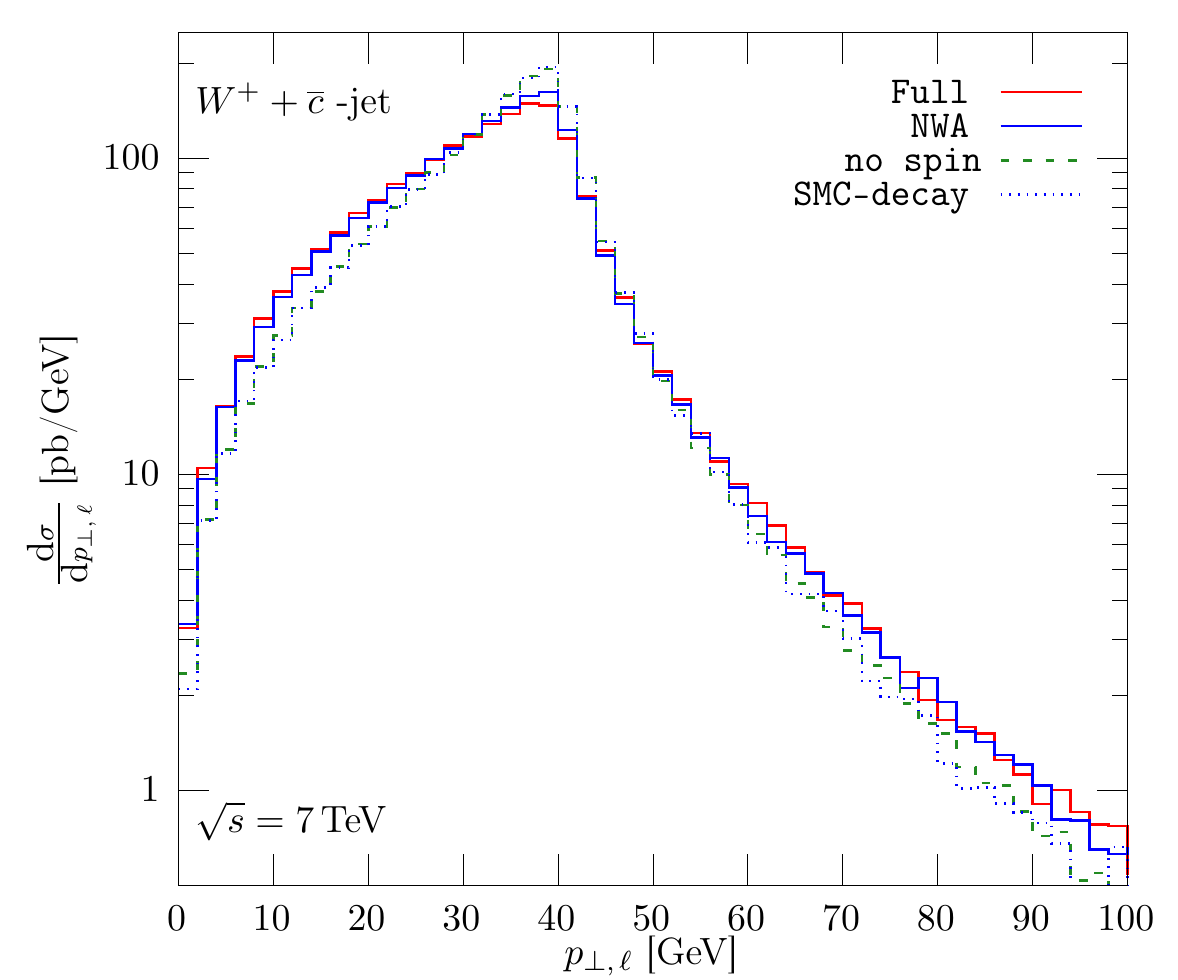}
 \includegraphics[width=0.495\textwidth]{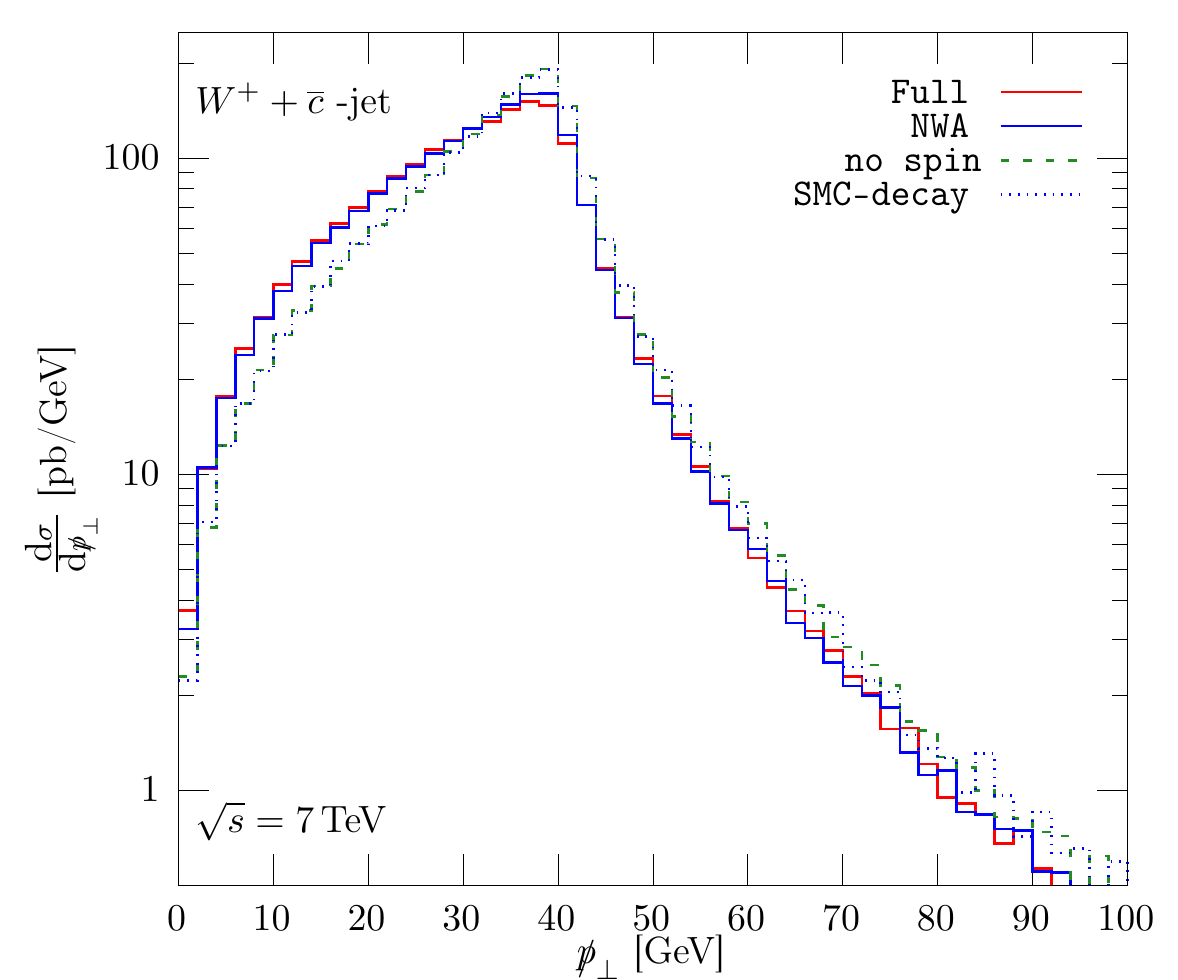}
  \end{center}
  \caption{\label{fig:LHE-spincorr} 
  Distributions for fully inclusive $W^+ \,+\, \bar{c}$ production in the leptonic channel as a function of the tranverse momentum of the charged lepton (left panel) and of the missing $p_T$ (right panel), as obtained from LHEs. Results are based on the ABMP16\_3\_NLO PDF set and on the scale choice is $\mu = \mu_R = \mu_F = H_T/2$. }
\end{figure}

Unless stated explicitly, all predictions shown in Section~\ref{sec:pheno} are based on 
the \emph{Full Calculation} approach.  In Section~\ref{sub:atlas} we will also present a 
comparison of differential NLO+PS cross sections based on the four approaches described above, 
considering the set of cuts adopted by the ATLAS collaboration in the analysis of  $W + c$ at $\sqrt{s} = 7$~TeV. 
For illustrative purposes, we present here a comparison of inclusive predictions for $\ell^{+}\nu_\ell \,+\, \bar{c}$ production at $\sqrt{s} = 7$ TeV ($\ell = e,\mu,\tau$). Since $W$ bosons are always assumed to decay leptonically in our analysis, we will refer to these predictions also with the name "fully inclusive $W + c$ production in the leptonic channel".
These results are obtained from the LHE generated by \texttt{PowHel} and thus refer to the parton level as it emerges after the POWHEG matching. The full available phase space is covered at this stage, namely no kinematical cuts are imposed.
The contribution of $W+c\bar{c}$ final states, 
emerging from the diagrams $u\bar{d}, \bar{d}u \rightarrow c\bar{c} W^+$, which is separately gauge invariant, 
is not included in the plots of Fig.~\ref{fig:LHE-spincorr}.  The contribution of the  
$W+c\bar{c}$ final state to the total inclusive cross section amounts to $\sim 5$\%.
As we will see in the following, the cuts presently applied by the experimental collaborations in $W + c$ analyses greatly suppress it.

Figure \ref{fig:LHE-spincorr} shows distributions of the 
transverse momentum of the charged lepton arising from the $W$ boson decay ($p_{\perp,\ell}$) 
as well as of the total missing momentum ($\slashed{p}_{\perp}$). The two plots report the predictions 
that we obtained using the four above-mentioned approaches, considering the scale setup $\mu = \mu_R = \mu_F = H_T/2$ 
and the ABMP16\_3\_NLO PDF set as input. We observe that predictions labeled with "\texttt{no spin}" 
and "\texttt{SMC-decay}" are very close to each other. This should not come as a surprise since the two 
results adopt identical accuracy for the modeling of the $W$ boson decay. On the other hand spin correlations 
("\texttt{NWA}") have a visible effect on the shape of both distributions, where low $p_T$ regions are 
enhanced for the charged lepton as well as for the missing $p_T$. It is easily understood that such effect 
has an impact on the full calculation chain up to the level of realistic NLO+PS predictions, 
given that the typical setup of the experimental analyses includes
cuts on the lepton $p_T$. 
For comparison, looking at the prediction labeled with "\texttt{Full}", one can observe that effects related 
to the off-shellness of $W$ boson decays play a minor role in this context.

To quantify the numerical impact of $W^{\pm} + c\bar{c}$ contributions, we report in Table \ref{tab:ProdRunsPowHel_incl} the fully inclusive NLO cross sections for $W^{\pm}+c$ and the LO cross sections for $W^{\pm}+c\bar{c}$ 
production for various center-of-mass and PDF setups. One can see that the contribution of $W^{\pm}+c\bar{c}$ to the total cross section is set in the range $2\%-5\%$.

\begin{table}[!th]
\centering
\begin{tabular}{|c|c|c|c|c|}
\hline\hline
%%%%%%%%%%%%%%%%%%%
 Energy & Process& PDF & $m_c$[GeV] &  $\sigma_{\rm NLO}$ [pb]  \bigstrut\\
\hline\hline
  \multirow{12}{*}{$ 13$ TeV } & \multirow{3}{*}{$ W^+ \bar{c} + X$}& {\texttt{ABMP16}} &  $1.5$  &
$4994(7)$ \bigstrut\\
\cline{3-5}
 & & {\texttt{CT18Z}} & $1.4$  &  $5298(6)$ \bigstrut\\
\cline{3-5}
 & & {\texttt{CT18}} & $1.4$  &  $4838(6)$ \bigstrut\\
\cline{2-5}
 & \multirow{3}{*}{$ W^- c + X$} & {\texttt{ABMP16}} &  $1.5$  & $5190(6)$ \bigstrut\\
\cline{3-5}
 & & {\texttt{CT18Z}} & $1.4$  &  $5521(7)$  \bigstrut\\
\cline{3-5}
 & & {\texttt{CT18}} & $1.4$  &  $5053(6)$ \bigstrut\\
\cline{2-5}
  & \multirow{3}{*}{$ W^+ c\,\bar{c}$}& {\texttt{ABMP16}} &  $1.5$  & $156.30(6)$ \bigstrut\\
\cline{3-5}
 & & {\texttt{CT18Z}} & $1.4$  &  $213.36(6)$ \bigstrut\\
\cline{3-5}
 & & {\texttt{CT18}} & $1.4$  &  $209.72(6)$  \bigstrut\\
\cline{2-5}
 & \multirow{3}{*}{$ W^- c\,\bar{c}$}& {\texttt{ABMP16}} &  $1.5$  &   $101.82(3)$  \bigstrut\\
\cline{3-5}
 & & {\texttt{CT18Z}} & $1.4$  &  $138.97(6)$  \bigstrut\\
\cline{3-5}
 & & {\texttt{CT18}} & $1.4$  &  $136.64(6)$ \bigstrut\\
\cline{1-5}
  \multirow{4}{*}{$ 7$ TeV } & $ W^+ \bar{c}+ X$ & {\texttt{ABMP16}} &  $1.5$  &  $2009(2)$ \bigstrut\\
\cline{2-5}
 & \multirow{1}{*}{$ W^- c + X $} & {\texttt{ABMP16}} &  $1.5$  & $2113(2)$ \bigstrut\\
\cline{2-5}
  & \multirow{1}{*}{$ W^+ c\,\bar{c}$}& {\texttt{ABMP16}} &  $1.5$  & $86.70(3)$ \bigstrut\\
\cline{2-5}
 & \multirow{1}{*}{$ W^- c\,\bar{c}$}& {\texttt{ABMP16}} &  $1.5$  &  $51.89(2)$  \bigstrut\\
%%%%%%%%%%%%%%%%%%%
\hline\hline
\end{tabular}
\caption{
NLO QCD cross sections for fully inclusive $W^- \,c + X$ and $W^+ \, \bar{c} + X$ production in the leptonic channel at $\sqrt{s} =$ $13$ and $7$ TeV for different PDF sets and charm quark on-shell mass values adopted in the rest of this work. The contributions from $W^\pm c\,\bar{c}$ final states (contributing at the same order and multiplicity as real radiation, though finite) to the full NLO cross section are also reported explicitly. See text for more detail. }
\label{tab:ProdRunsPowHel_incl}
\end{table}

\section{Phenomenological results}  
\label{sec:pheno}
As mentioned in the Introduction, public results for $W+c$ production at the LHC
have been released by both the ATLAS and CMS collaborations.
We start the analysis by comparing our theory predictions for $D$-meson production 
in association with a $W$ boson decaying leptonically with the data of the CMS analysis at 
$\sqrt{s}~=~13$~TeV~\cite{Sirunyan:2018hde} and of the ATLAS one at $\sqrt{s} =$~7~TeV~\cite{Aad:2014xca}. 
In Ref.~\cite{Aad:2014xca} ATLAS has also published data on $c$-jets. We also report a comparison with the latter.
Because of the massive charm the Born cross section is IR-finite, hence the
total inclusive cross section is well-defined. As already mentioned, the total inclusive NLO QCD cross sections we obtained in
this paper (for both energies and all used PDFs) are listed in Tab. \ref{tab:ProdRunsPowHel_incl}. It should be noted
that 
the listed cross sections for the $W^\pm\,c\,\bar{c}$ process are at leading order only due to the fact that 
only tree-level diagrams for this process might contribute as real corrections to associated $W+c$ production at NLO.

\subsection{Comparison with CMS data: $W^\pm + D^{*}(2010)^{\mp}$ production at 13 TeV} \label{sub:cms}

The CMS analysis of $W^\pm + D^{*}(2010)^{\mp}$ events published in Ref.~\cite{Sirunyan:2018hde} 
provides differential distributions of the absolute value of the pseudorapidity for the charged 
lepton associated with the $W$ boson leptonic decay.  In particular, a $W$ boson is identified 
when it decays into a muon ($p_{T,\mu} >$ 26 GeV and $|\eta_\mu| < 2.4$) accompanied by missing energy. 
Predictions are presented for the individual $\mu^+$ and $\mu^-$ pseudorapidity distributions, 
as well as for their sum.

The potential contribution of the $W^\pm c\bar{c}$ process is greatly suppressed with 
respect to that of the $W^+ \bar{c}$ and $W^- c$ processes through analysis cuts,
which make $W^\pm c\bar{c}$ contributing in equal proportion on average to 
both the signal and the background. When computing the cross sections, the contribution of the events including a 
$D^*(2010)$-meson ($j_c$) with charge having the same sign (SS) as the $\ell$ charge (background) is subtracted 
from that of the events where the $D^*(2010)$-meson ($j_c$) and $\ell$ charges have opposite sign (OS) (signal).
If two $D^*(2010)$-mesons with opposite charges in a same event are both within the experimental cuts, 
the event equally contributes to both the signal and the background, so its net contribution to the (OS - SS) cross-section is null. 
All figures reported in the following refer to (OS - SS) cross-sections. On the other hand, 
the separate roles of the OS and SS contributions to the fiducial cross-sections are quantified and discussed in detail in Appendix B.

As mentioned in the Introduction, on the one hand, the CMS collaboration compared the data at the particle level 
with simulations of both $W^+ \rightarrow \mu^+ \nu_\mu$  and $W^-~\rightarrow~\mu^-~\bar{\nu}_\mu$ production in 
association with light jets made with \texttt{MadGraph5\_aMC@NLO} interfaced to \texttt{PYTHIA8}. 
On the other hand, they also unfold their experimental distributions at the parton level, and, in this case, 
they compare the corresponding results to the fixed-order NLO QCD implementation available in \texttt{MCFM}, 
involving instead massive charm quarks. In their analysis central
$D^{*}(2010)^\mp$-mesons ($|\eta_{D^{*}(2010)^\mp}| < 2.4$), with a relatively low transverse momentum 
$p_{T,D^*(2010)^\mp}$ ($p_{T,D^*(2010)^\mp} >$ 5 GeV), are measured.  
This fact motivates the investigation of charm-quark mass effects in the calculation, 
given that the charm-quark mass value ($\sim \mathcal{O}(1 - 2)$ GeV) has the same order of 
magnitude of the minimum $p_{T,D^*(2010)^\mp}$ value explored by the experimental collaboration.
The $p_{T,D^*(2010)^\pm}$ distributions are peaked at low $p_{T,D^*(2010)^\pm}$ values, 
implying that the typical differential cross sections for the observables considered 
in the experimental analysis, which integrate over the whole measured range of $p_{T, D^*(2010)^\pm}$, 
are dominated by the contribution of events with $D^*(2010)$-mesons with $p_{T,D^*(2010)^\pm}$ 
values close to the kinematical cut. 

\begin{figure}
  \begin{center}
 \includegraphics[width=0.495\textwidth]{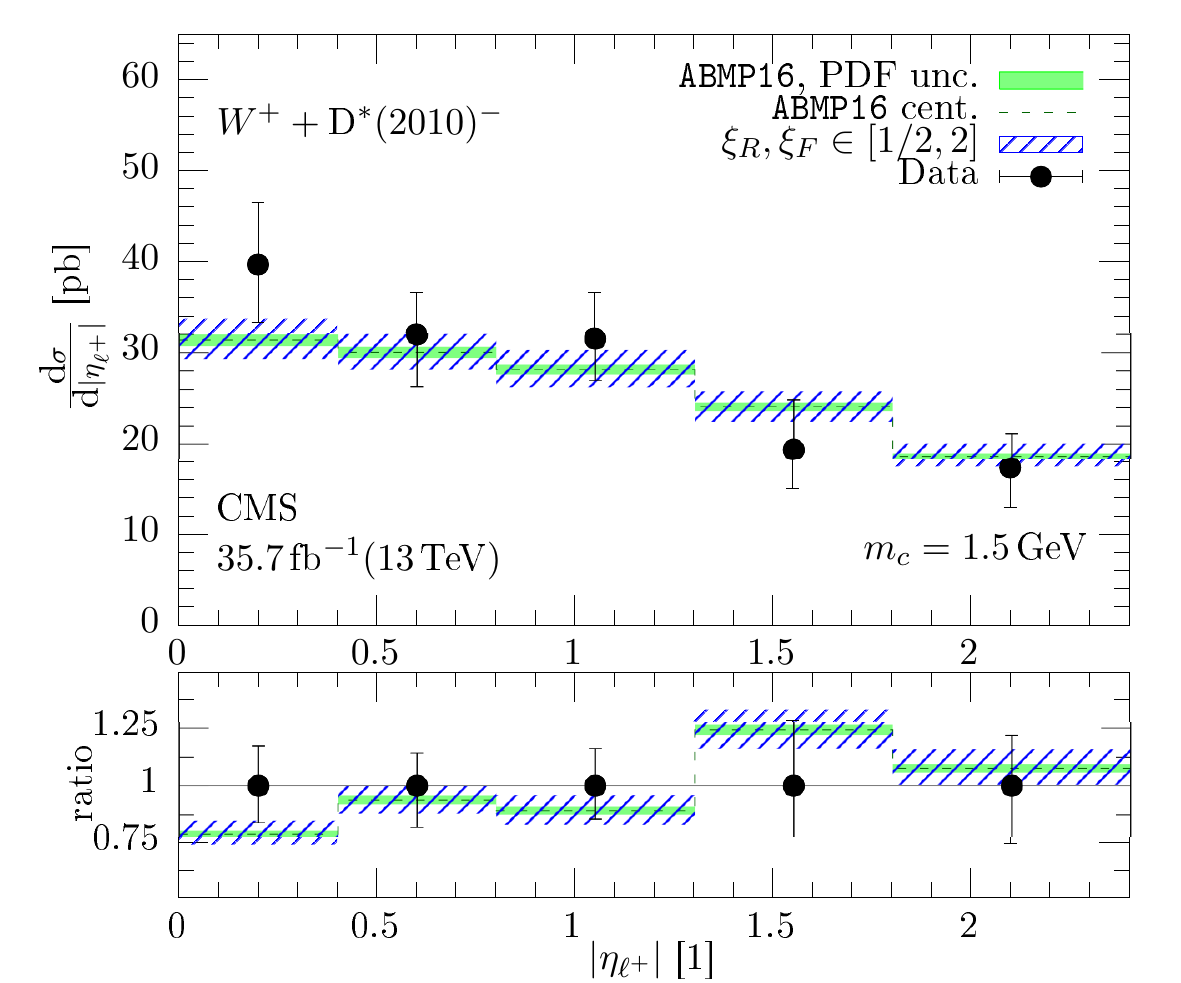}
 \includegraphics[width=0.495\textwidth]{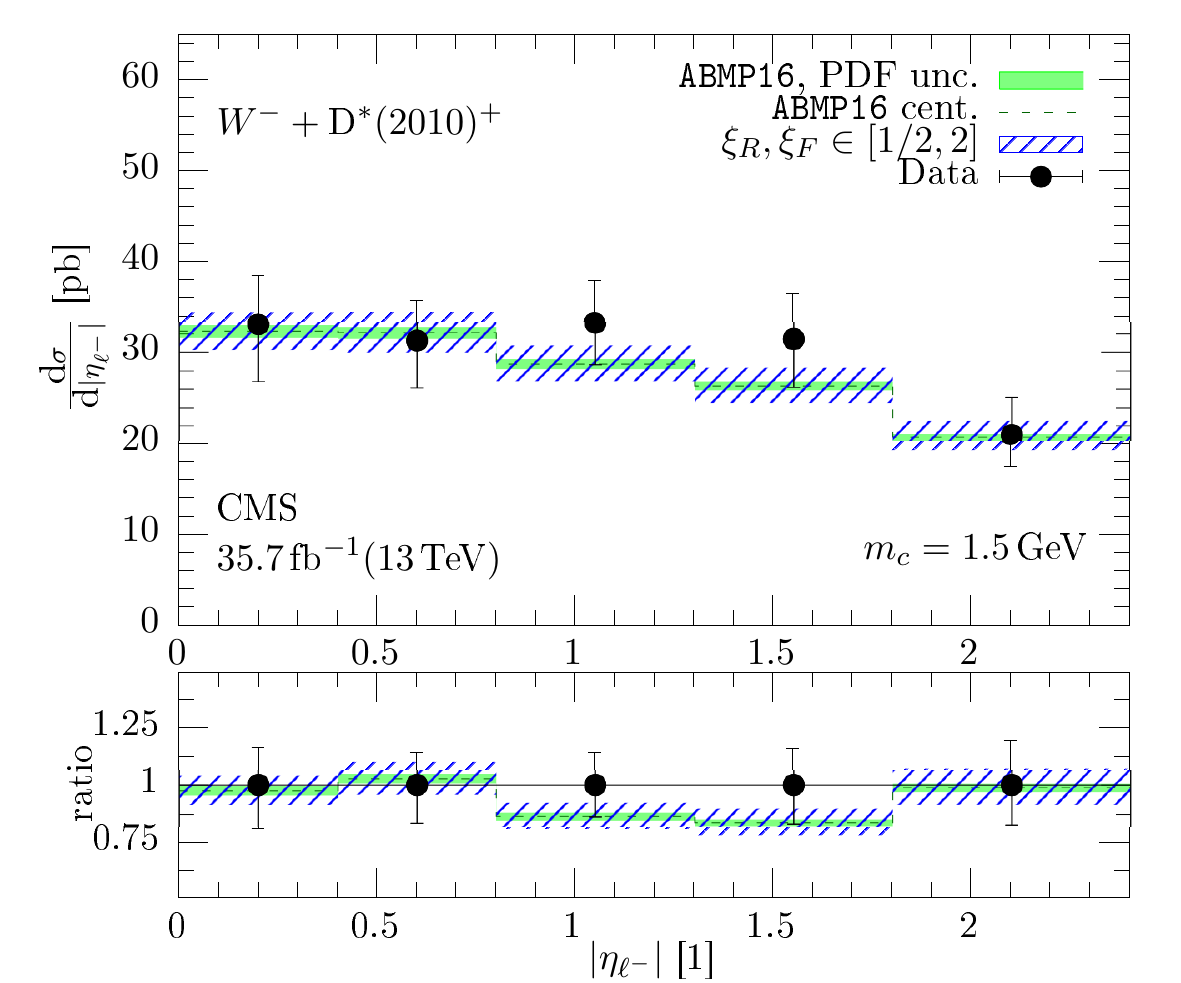}
 \includegraphics[width=0.495\textwidth]{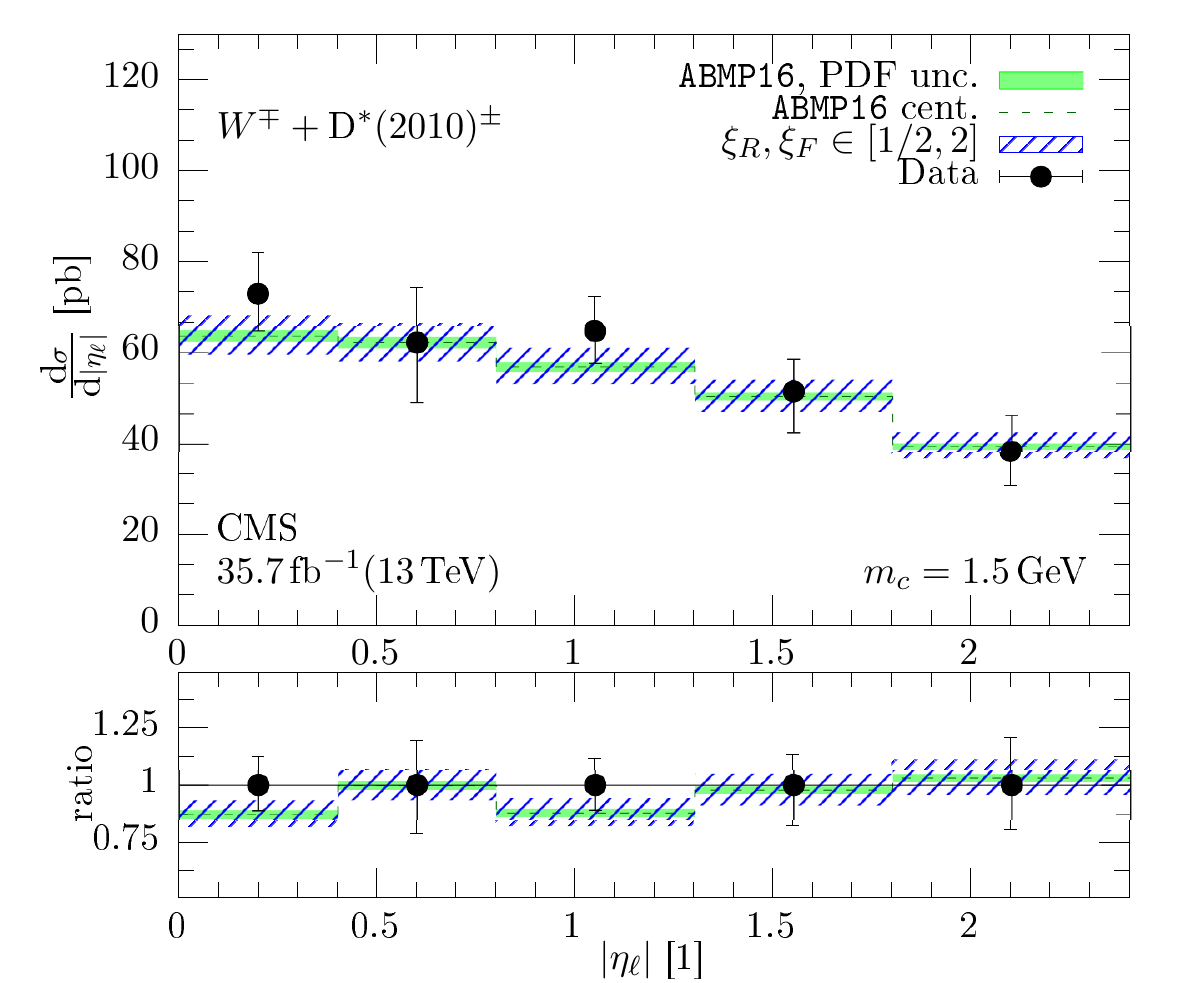}
  \end{center}
  \caption{\label{fig:dmeson-cms} Difference of the differential cross-section for   $W + D^{*}(2010)$-meson production, with $W^\pm \rightarrow \mu^\mp \overset{(-)}{\nu}_\mu$,  where the hardest central $\mu$ and the $D^*(2010)$-meson have charge of opposite sign and 
  that where the $\mu$ and the $D^*(2010)$-meson have charges of the same sign, 
  as a function of the absolute value of the pseudorapidity of the hardest $\mu$.
  The three panels refer respectively to the case of $\mu^+$, $\mu^-$ and their sum. 
  Theoretical predictions at NLO + SMC accuracy, obtained by \texttt{PowHel + PYTHIA8} 
  are compared to experimental data from the CMS collaboration~\cite{Sirunyan:2018hde}. 
  7-point ($\mu_R$, $\mu_F$) scale and PDF uncertainty bands, computed from the 
  30 members
  of the ABMP16\_3\_NLO PDF set, are reported by shaded (blue) and solid (green) bands, 
  respectively. As further input of {\texttt{PowHel}}, we consider 2-loop $\alpha_S$ evolution, 
  $\alpha_S(M_Z) = 0.1066$ consistent with the PDF set, and $m_c = 1.5$~GeV. 
  The predictions include parton shower, hadronization, MPI and beam remnant effects according 
  to the {\texttt{PYTHIA8}} Monash tune. See text for more detail.}
  \end{figure}

\begin{figure}
  \begin{center}
 \includegraphics[width=0.495\textwidth]{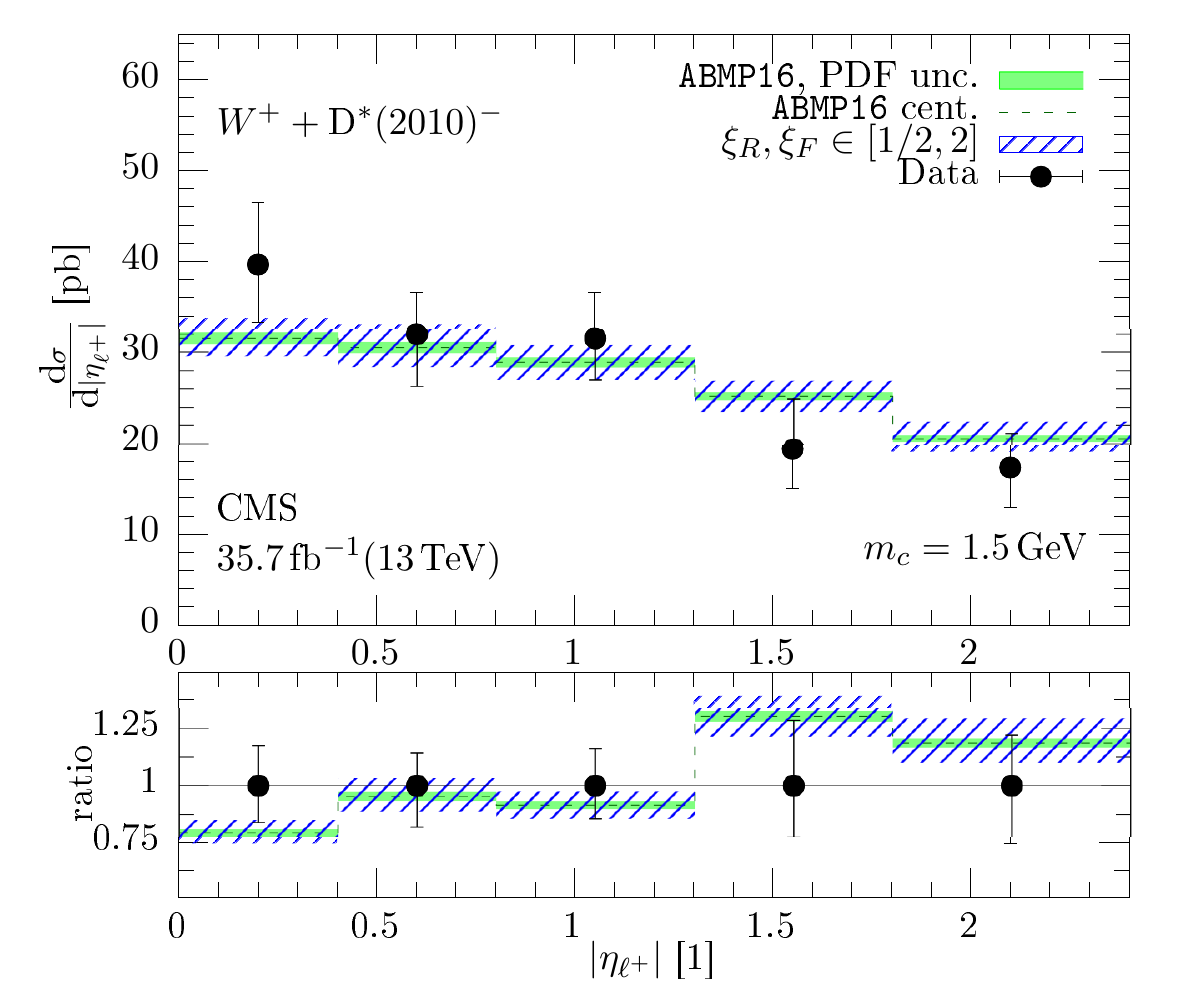}
 \includegraphics[width=0.495\textwidth]{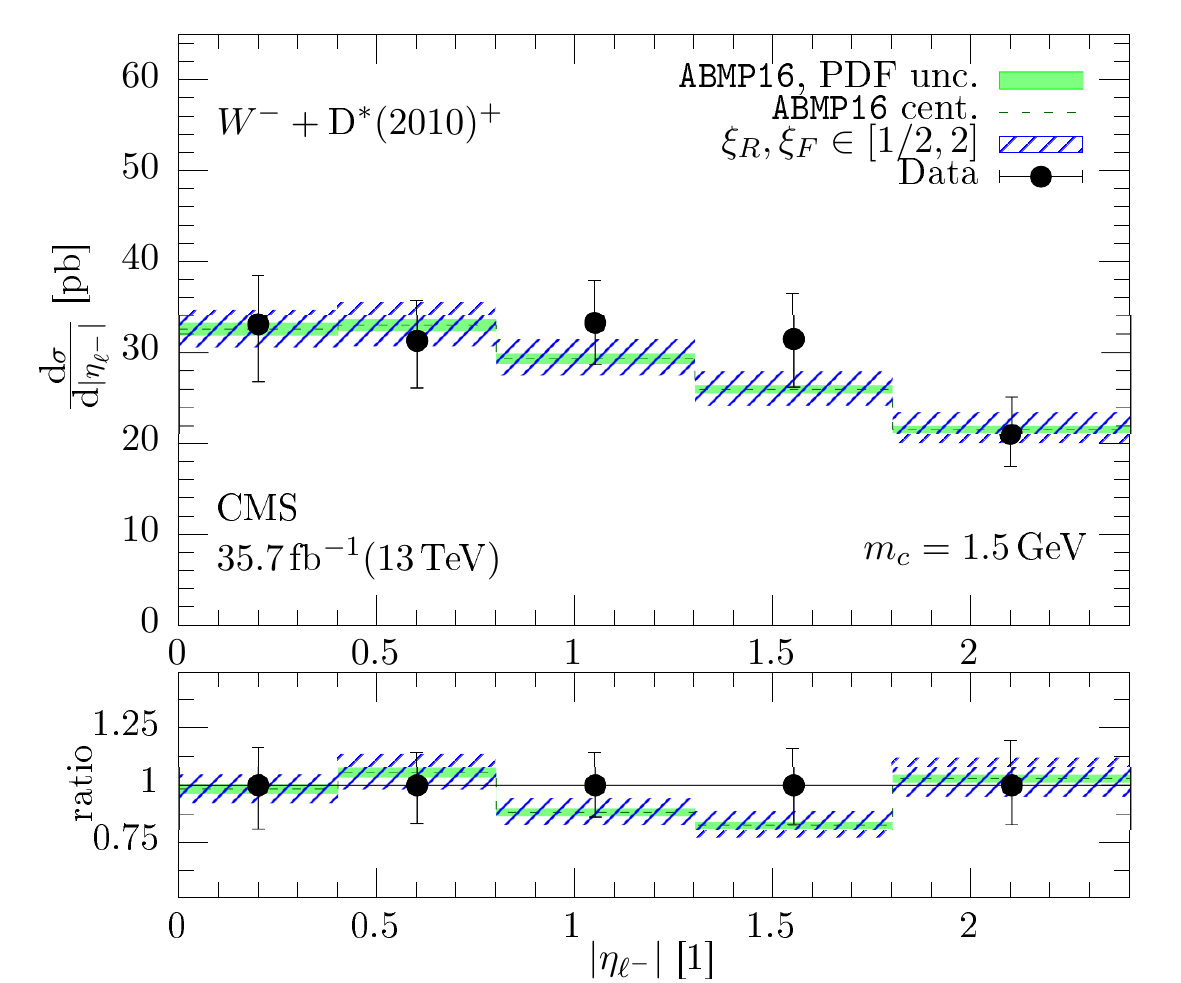}
 \includegraphics[width=0.495\textwidth]{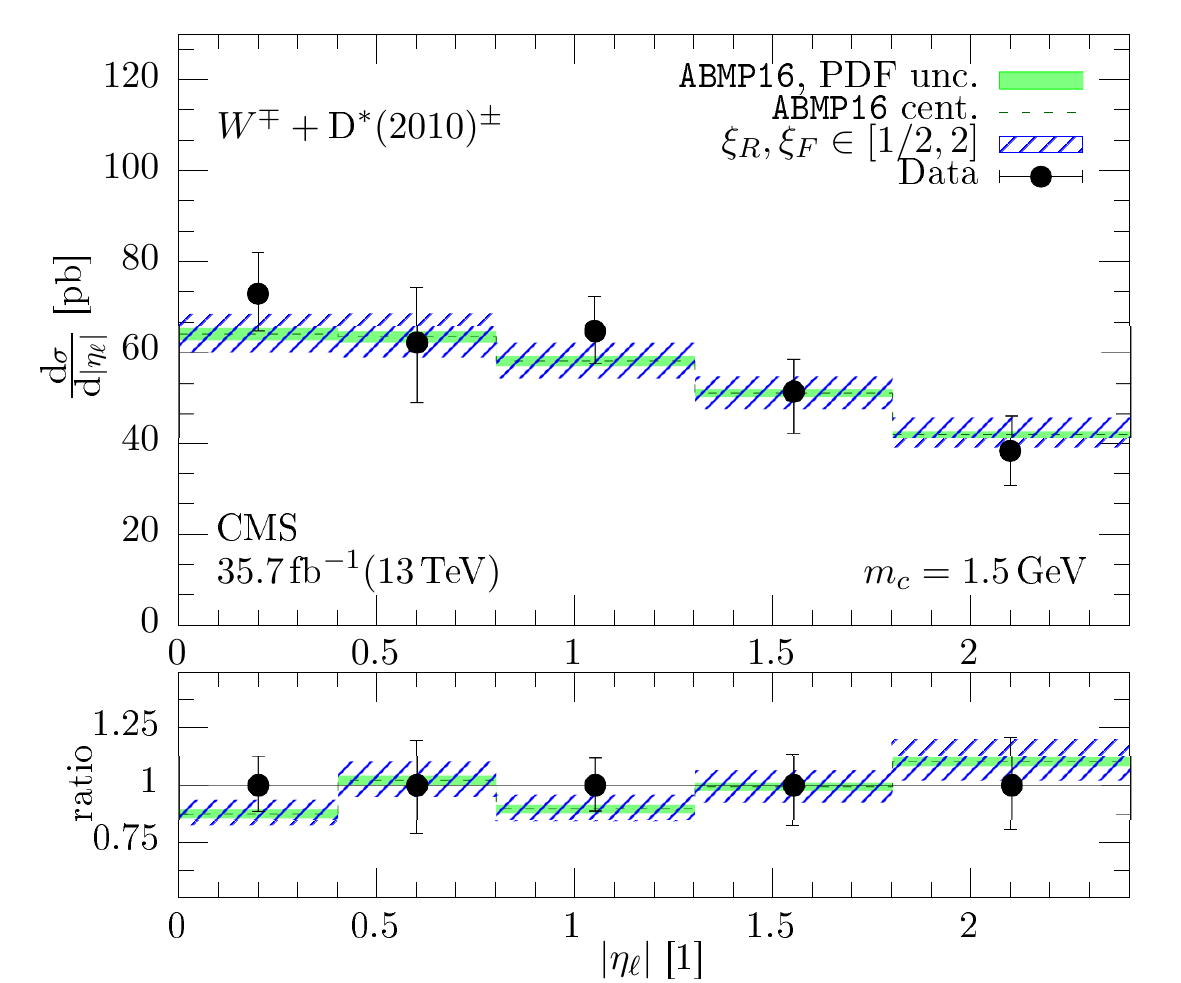}
  \end{center}
  \caption{\label{fig:dmeson-cms-21} Same as in Fig.~\ref{fig:dmeson-cms}, but using the 
    \texttt{PYTHIA8} ATLAS A14 central tune with \texttt{NNPDF2.3LO} PDFs.}
  \end{figure}

\begin{figure}
  \begin{center}
 \includegraphics[width=0.495\textwidth]{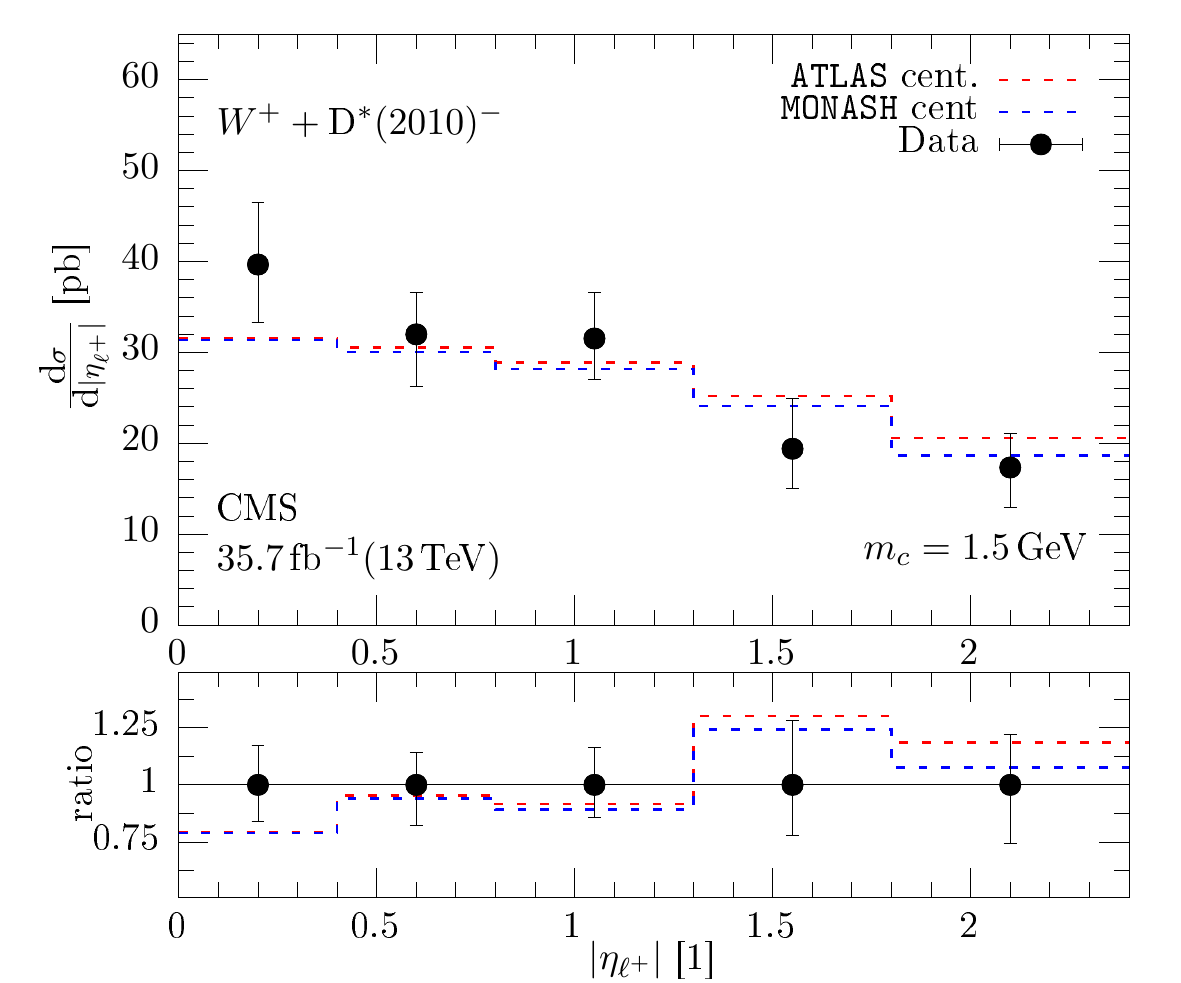}
 \includegraphics[width=0.495\textwidth]{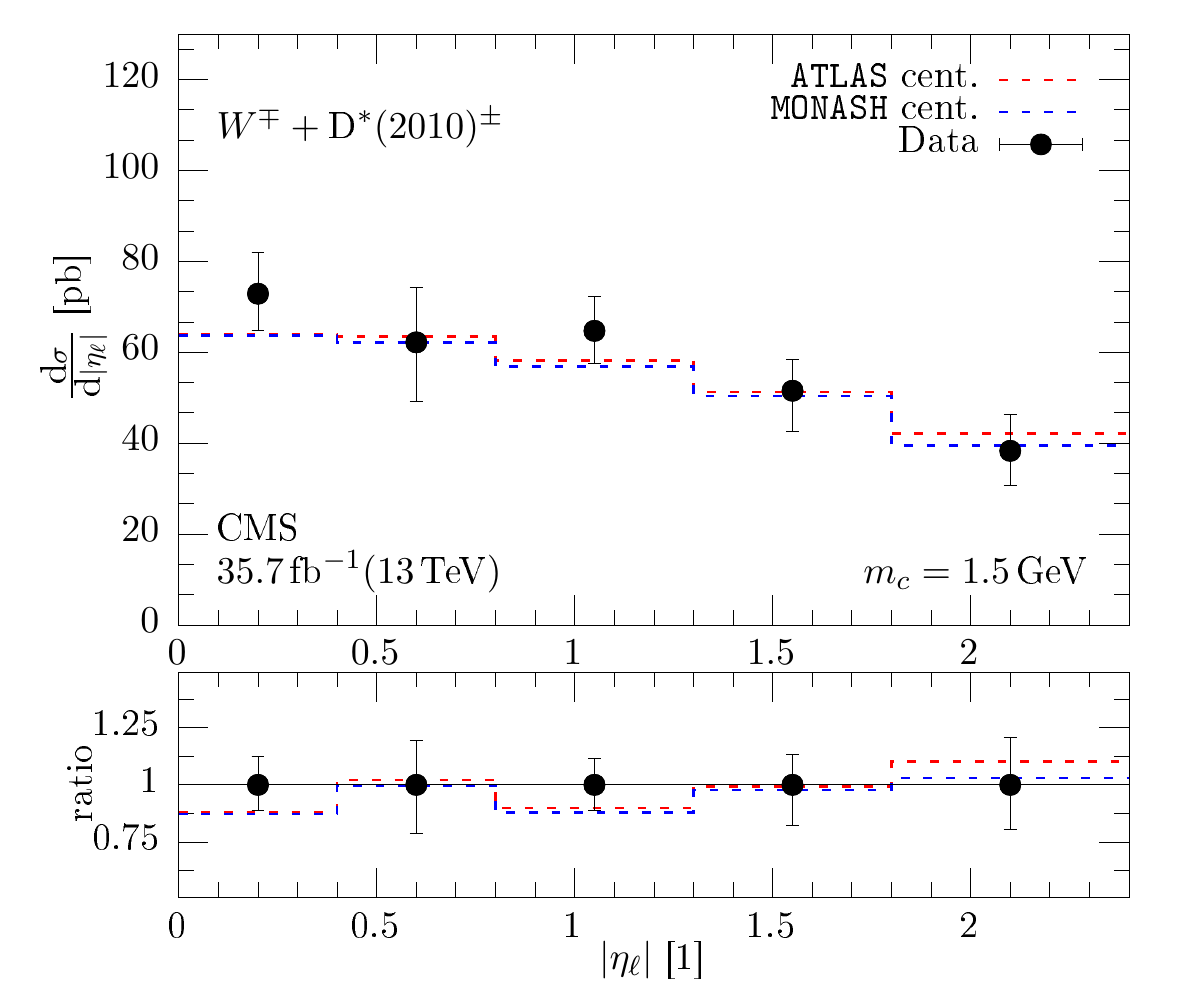}
  \end{center}
  \caption{\label{fig:dmeson-cms-tune}
    Same as in Fig.~\ref{fig:dmeson-cms} and~\ref{fig:dmeson-cms-21}, 
    but limited to central predictions with two different \texttt{PYTHIA8} tunes 
    (Monash and ATLAS A14 central tune with \texttt{NNPDF2.3LO} PDFs). 
    The two panels refer to the $\mu^+$ case (left) and to the 
    $\mu^+ + \mu^-$ case (right), respectively.}
\end{figure}

\begin{figure}
  \begin{center}
 \includegraphics[width=0.495\textwidth]{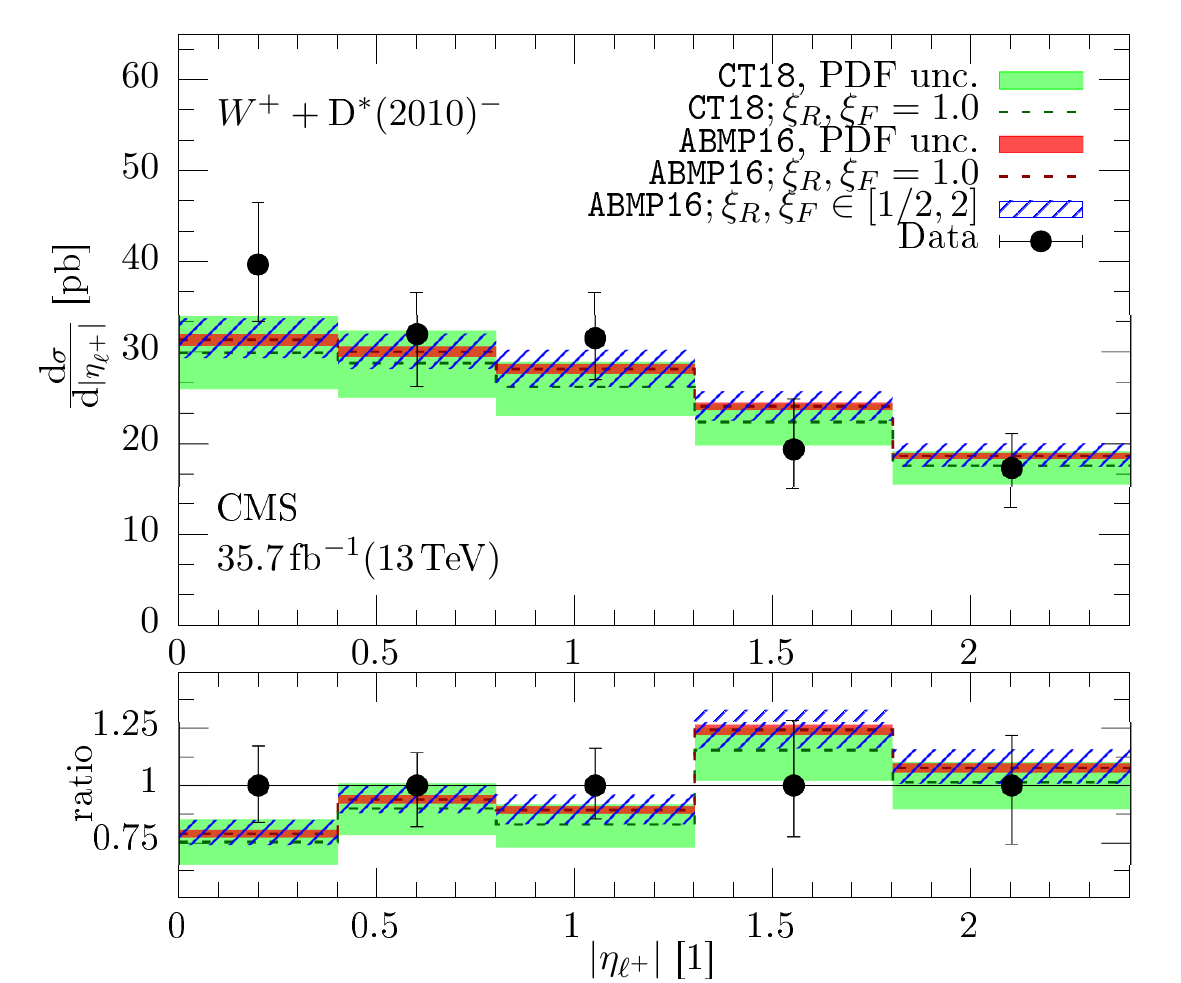}
 \includegraphics[width=0.495\textwidth]{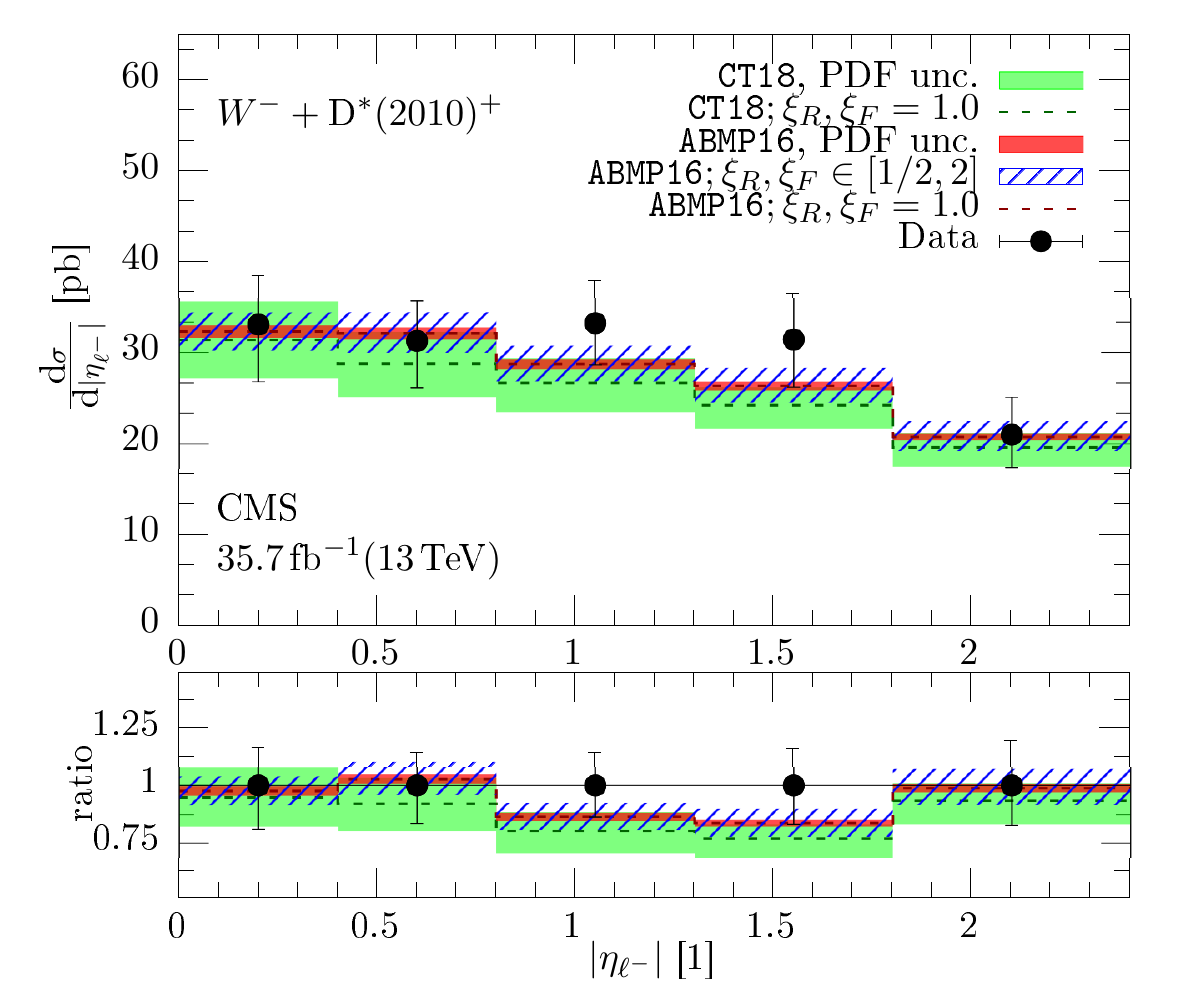}
 \includegraphics[width=0.495\textwidth]{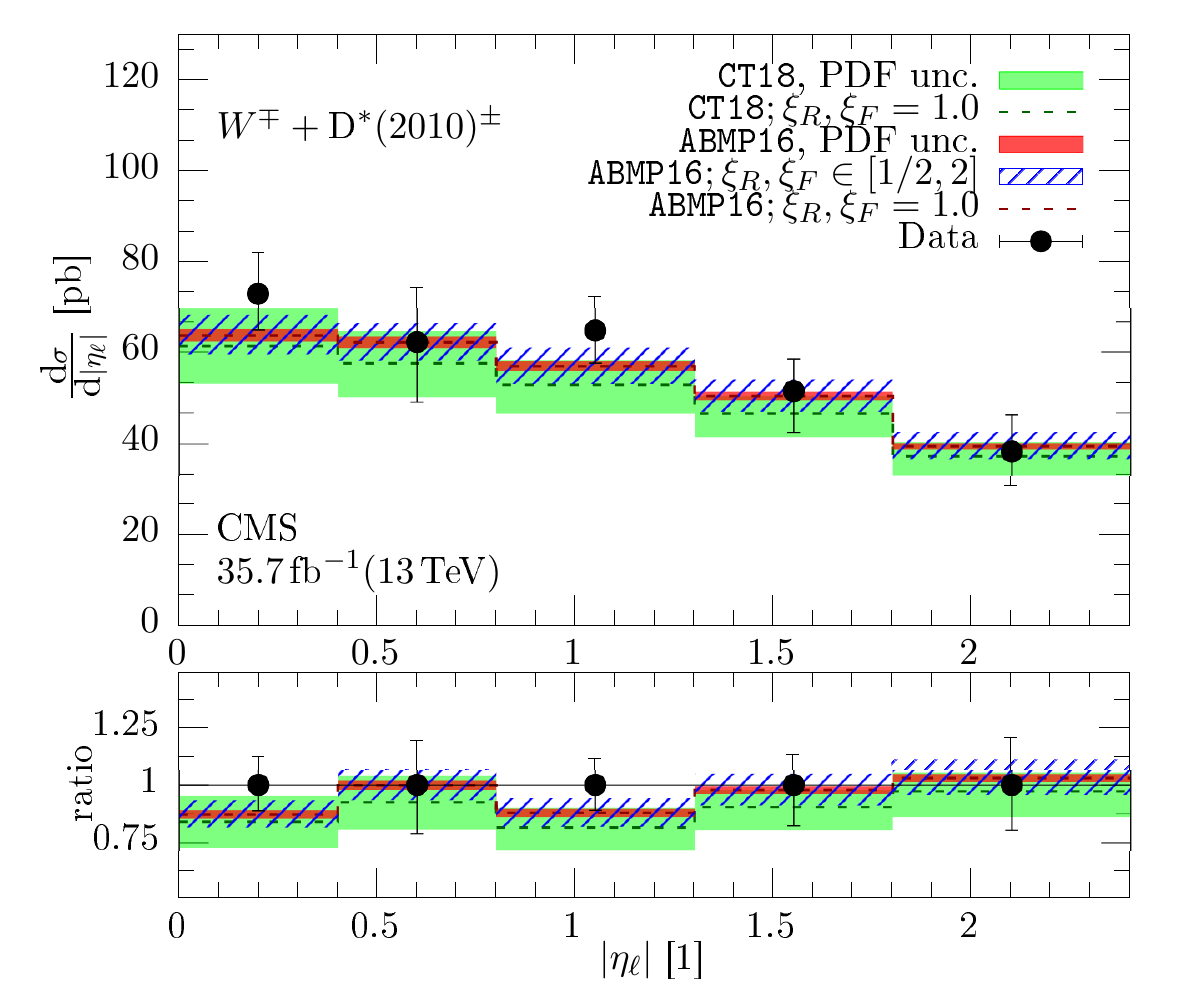}    
  \end{center}
  \caption{\label{fig:dmeson-cms-ct18} Same as Fig.~\ref{fig:dmeson-cms}, 
    but using the CT18NLO PDF set, with 2-loop $\alpha_S$ evolution and 
    $\alpha_S(M_Z)^{n_f=5} = 0.118$ consistent with the PDF set, 
    and with $m_c = 1.4$ GeV. Asymmetric PDF uncertainty bands (green solid), 
    computed at 90\% C.L. corresponding to the tolerance criterion adopted in 
    this PDF fit, refer to this configuration. 
    Predictions from this configuration are compared to those using the 
    ABMP16\_3\_NLO PDF set, with 2-loop $\alpha_S$ evolution and 
    $\alpha_S(M_Z)^{n_f=3} = 0.1066$ ($\alpha_S(M_Z)^{n_f=5} = 0.1191$), 
    consistent with the PDF set, and with $m_c = 1.5$ GeV. 
    For the latter, beside PDF uncertainty bands (red solid) 
    computed at 68\% C.L. as customary for this PDF fit, 
    we also report scale uncertainties (blue hatched bands).} 
  \end{figure}

\begin{figure}
  \begin{center}
 \includegraphics[width=0.495\textwidth]{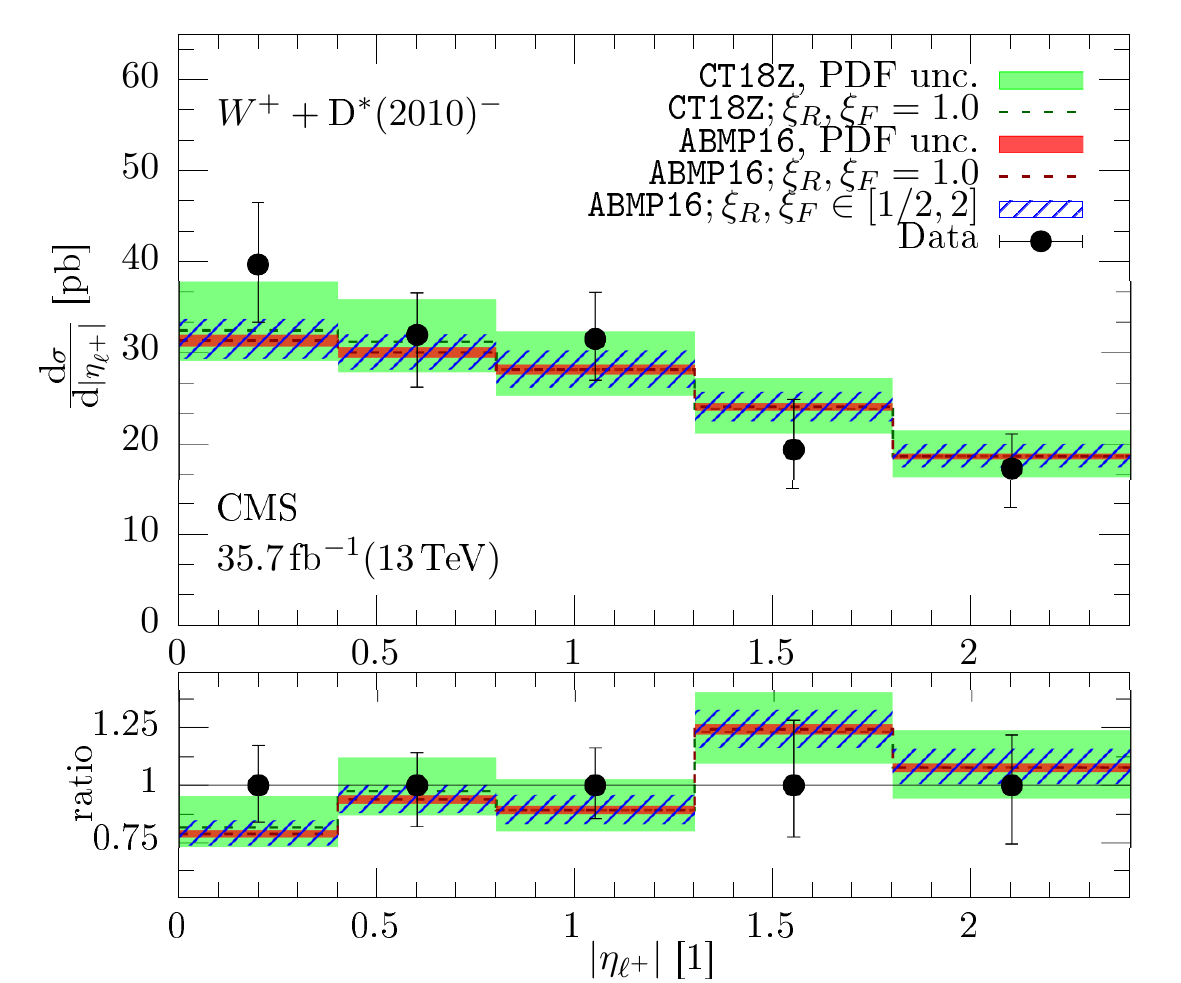}
 \includegraphics[width=0.495\textwidth]{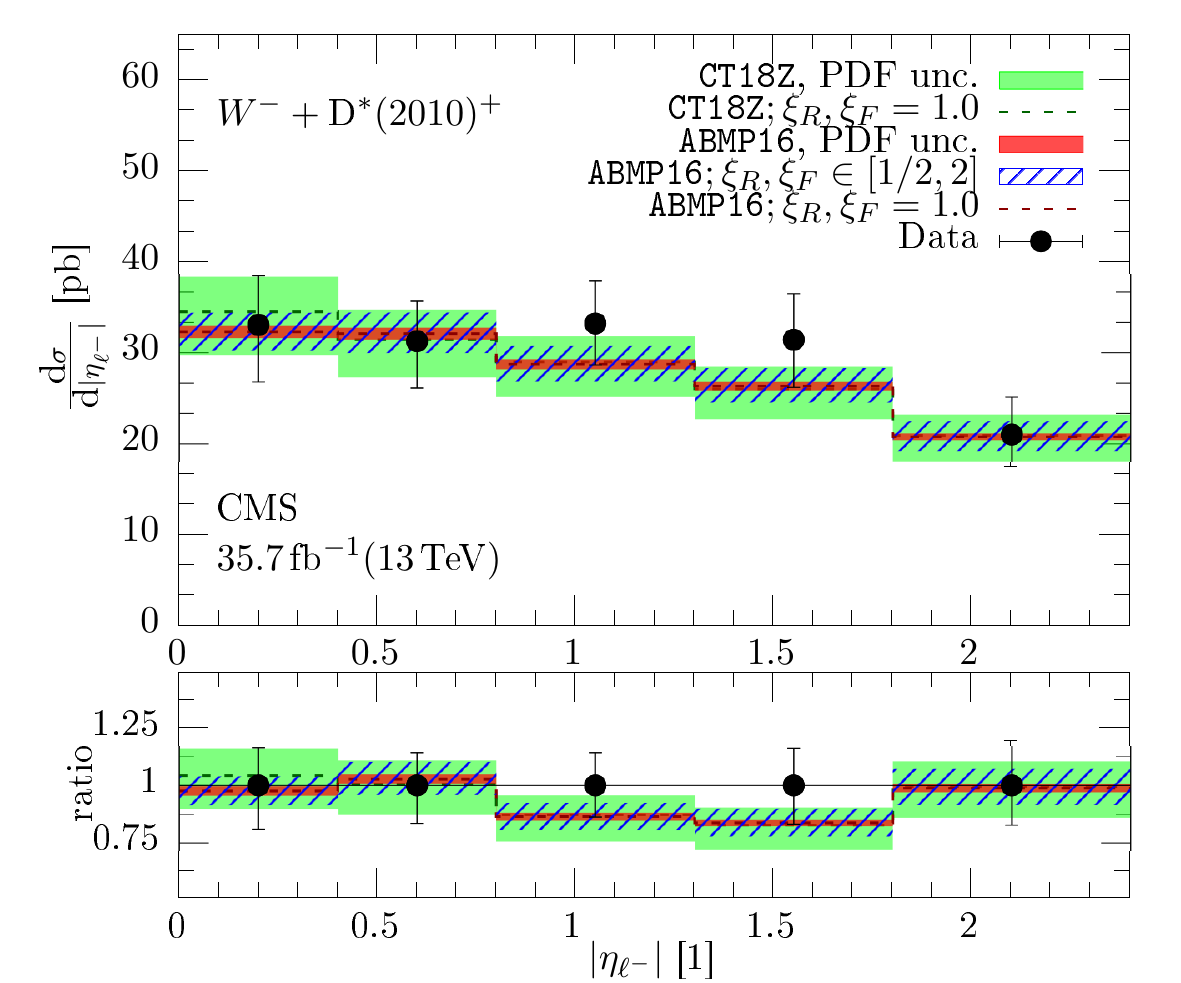}
 \includegraphics[width=0.495\textwidth]{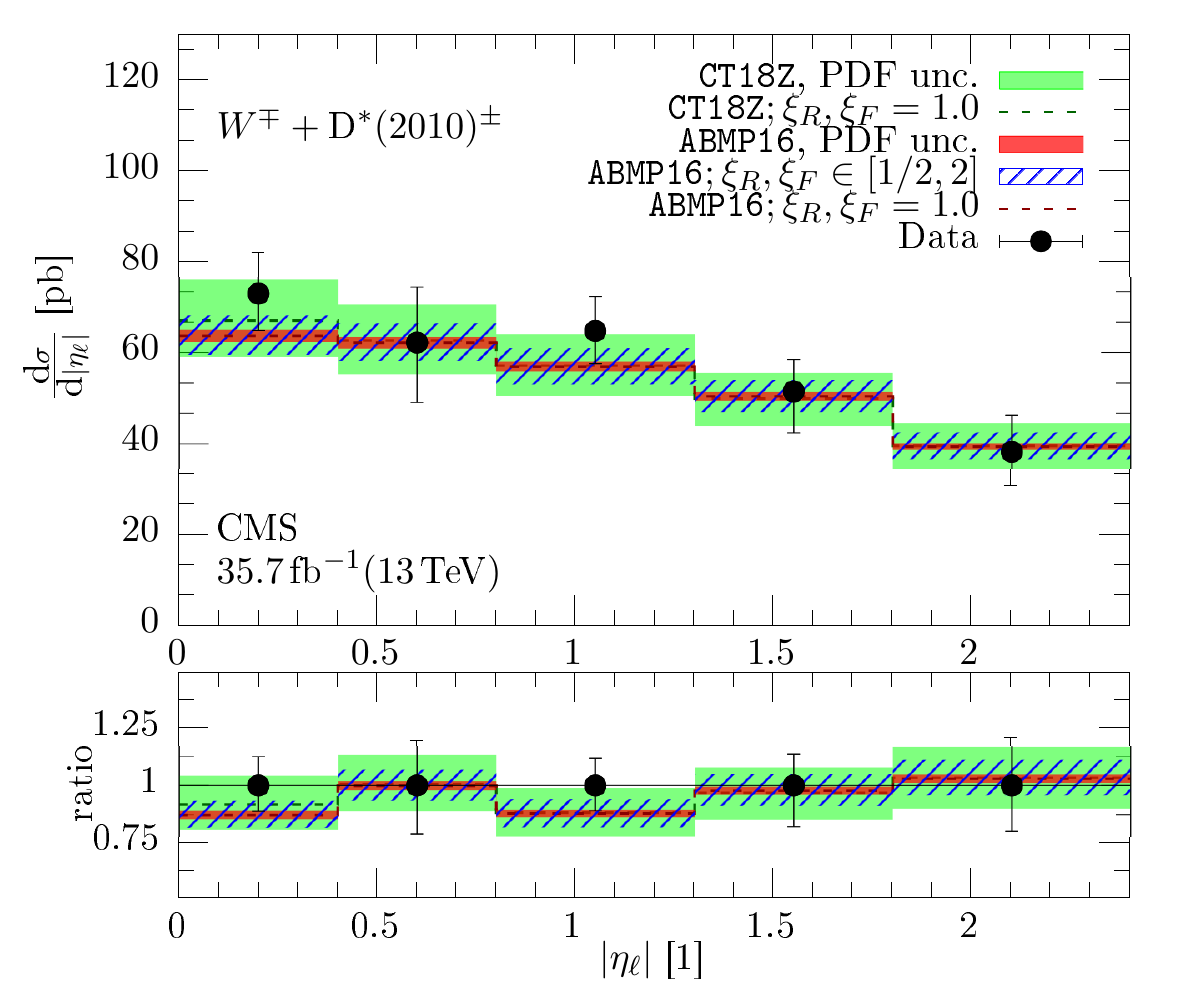}    
  \end{center}
  \caption{\label{fig:dmeson-cms-ct18z}
    Same as Fig.~\ref{fig:dmeson-cms-ct18}, but using the CT18ZNLO PDF set, 
    with 2-loop $\alpha_S$ evolution, $\alpha_S(M_Z)^{n_f=5} = 0.118$ 
    and $m_c = 1.4$ GeV, consistent with the PDF set. 
    Asymmetric PDF uncertainty bands (green solid), 
    computed at 90\% C.L.,
    corresponding to the tolerance criterion adopted in this PDF fit,
    refer to this configuration. 
    Predictions from this configuration are compared to those using the 
    ABMP16\_3\_NLO PDF set, with 2-loop $\alpha_S$ evolution and 
    $\alpha_S(M_Z)^{n_f=3} = 0.1066$ ($\alpha_S(M_Z)^{n_f=5} = 0.1191$), 
    consistent with the PDF set, and with $m_c = 1.5$ GeV. For the latter, 
    beside PDF uncertainty bands (red solid) computed at 68\% C.L. 
    as customary for this PDF fit, we also report scale uncertainties (blue hatched bands).}
  \end{figure}

\begin{figure}
  \begin{center}
 \includegraphics[width=0.495\textwidth]{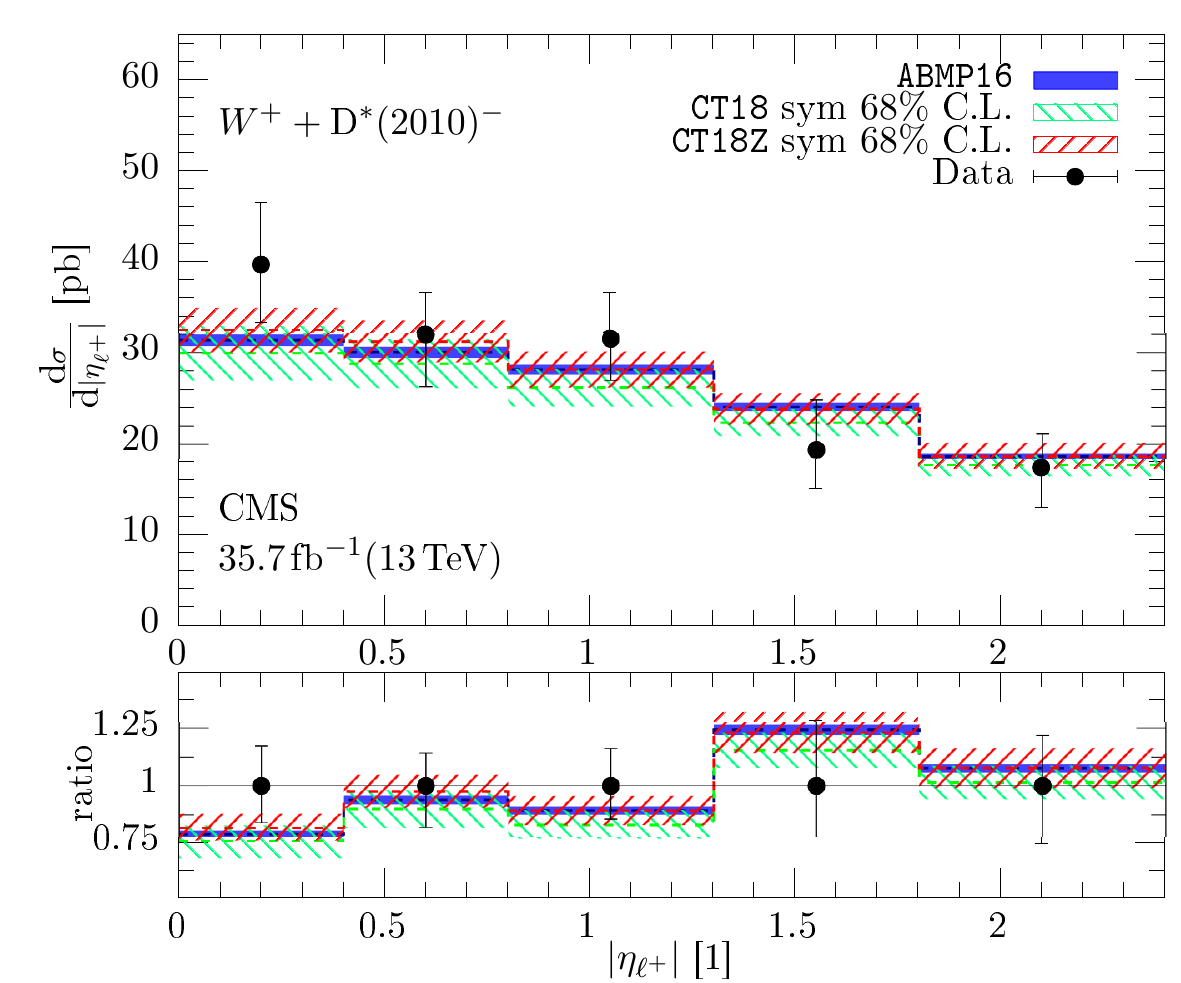}
 \includegraphics[width=0.495\textwidth]{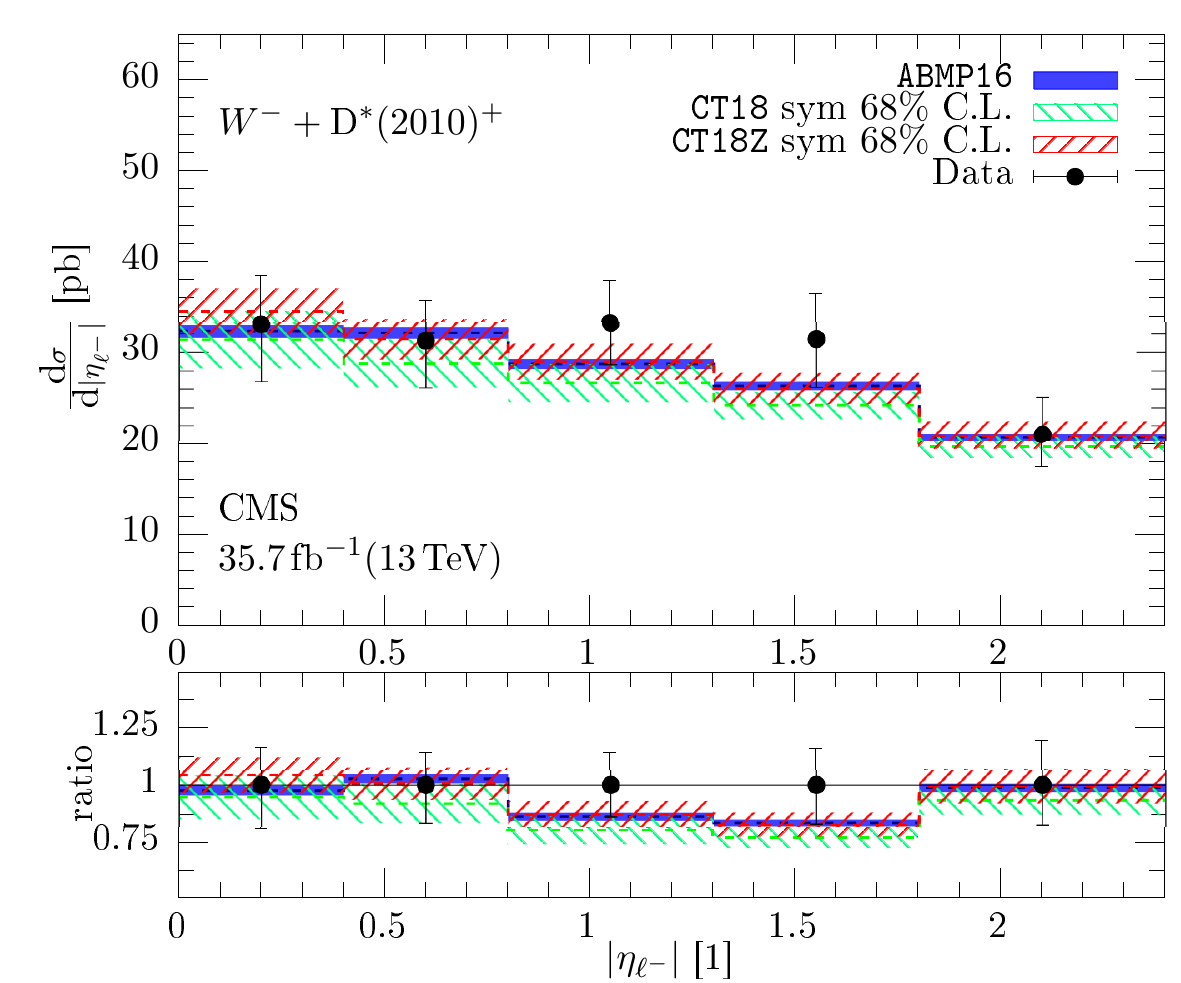}
 \includegraphics[width=0.495\textwidth]{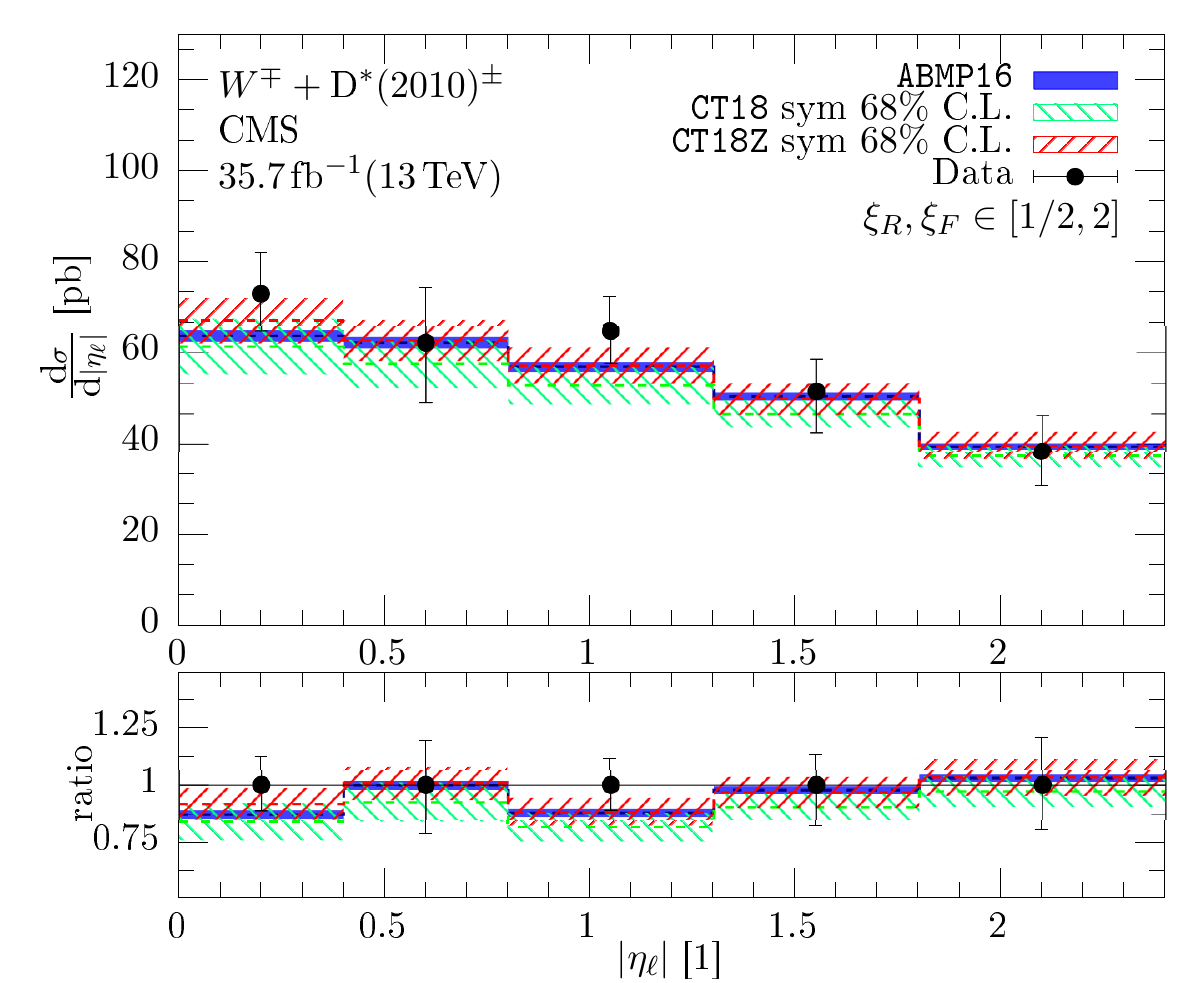}
  \end{center}
  \caption{\label{fig:dmeson-cms-allpdfs} Same as Fig.~\ref{fig:dmeson-cms}, 
    but using the three different NLO PDF + $\alpha_S(M_Z)$ sets ABMP16\_3\_NLO, 
    CT18ZNLO and CT18NLO, each one with its own uncertainties. A value of 
    $m_c = 1.5$~GeV is used in association with the ABMP16\_3\_NLO fit, 
    whereas $m_c = 1.4$~GeV is used in association with the CT18ZNLO and the 
    CT18NLO fits. Differently from Fig.~\ref{fig:dmeson-cms-ct18} 
    and~\ref{fig:dmeson-cms-ct18z}, the CT18NLO and CT18ZNLO uncertainty bands 
    in this figure are symmetrized by computing them with the specific 
    prescription provided by the CT collaboration and scaled to the 68\% C.L. 
    to facilitate a more direct comparison with the ABMP16 uncertainty bands, 
    symmetric by fit construction, customarily provided at this C.L. 
    corresponding to the tolerance criterion adopted in the ABMP16 fit.}
  \end{figure}
  
In Fig.~\ref{fig:dmeson-cms} we compare the predictions obtained by {\texttt{PowHel~+~PY\-THIA8}}, 
using hard-scattering matrix-elements with massive charm, to the CMS experimental data at the particle level. 
The three panels report separately differential distributions of the absolute value of the charged 
muon pseudorapidity for the $\mu^{+}$ and $\mu^{-}$ cases, as well as their sum. 
Theory predictions include parton shower, hadronization, multiple parton interaction (MPI) and beam remnant 
effects, with parameters fixed by the selected \texttt{PYTHIA8} tune. We study the effects of two different 
\texttt{PYTHIA8} tunes, by computing theory predictions with both the default Monash tune~\cite{Skands:2014pea} 
(option 14 in \texttt{PYTHIA8}, current default), used in Fig.~\ref{fig:dmeson-cms}, and one of the recent 
ATLAS A14 central tunes~\cite{TheATLAScollaboration:2014rfk} 
(option 21 in \texttt{PYTHIA8}), used in Fig.~\ref{fig:dmeson-cms-21}. The two selected tune 
options are both based on the same PDF set (NNPDF2.3 LO~\cite{Ball:2012cx}) and both incorporate LHC Run I data 
($\sqrt{s} = 7$ TeV). On the one hand, the Monash tune was designed with the intent of providing an overall 
reasonable description of both minimum-bias and underlying event physics, i.e. of both soft-inclusive physics 
and the processes typically accompanying a hard-scattering (central) collision. On the other hand, the ATLAS A14 
set of tunes focused on events with one or more high-$p_T$ emissions, providing a simultaneous tune of both PS and MPI parameters, 
by making use of ATLAS data for distributions of observables sensitive to initial state interactions, final state interactions and the underlying event. 
Among others, both tunes included Drell-Yan data. However, the Monash tune, differently from the ATLAS one, also includes CMS data.

While we did not make any attempt to vary the LO PDF corresponding to each tune in the SMC not to spoil the 
validity of the tune itself, we consider the effect of the variation of the NLO PDF and scales used as 
input for the generation of the events at the first radiation emission level, by exploiting the reweighting 
opportunities available within the {\texttt{POWHEG-BOX-v2}} framework, that we imported in {\texttt{PowHel}}. 
In particular, the hatched (blue) uncertainty bands in Fig. 3 and 4 refer to the 7-point renormalization and factorization scale variation around the central value $(\mu_R,\mu_F) = \mu_0 = H_T/2$, where $H_T$ is computed according to Eqs.~(\ref{eq:scale})-(\ref{eq:scaleWcc}).
The central predictions refer to the {ABMP16\_3\_NLO} PDF set,  
that we use as our default,   together with its $\alpha_S(M_Z)^{n_f=3} = 0.1066$ value 
(corresponding to $\alpha_S(M_Z)^{n_f=5} = 0.1191$), $\alpha_S$ two-loop evolution and 
$m_c = 1.5$ GeV~\footnote{The ABMP16 QCD analysis provides a simultaneous fit of PDFs, $\alpha_S (M_Z)$ 
and heavy-quark mass values, considering the correlations between these quantities.
In that framework, the charm mass value is fitted in the \msbar\, mass renormalization scheme, 
obtaining $m_c^\msbar = 1.175 \pm 0.033$~GeV at NLO. The conversion of this value to the pole mass renormalization scheme, i.e. the scheme
that we use in our computation, with coefficients presently known up to four-loops~\cite{Marquard:2015qpa}, 
does not lead to a convergent perturbative expansion in $\alpha_S$. For more details on conversion formulas between 
different mass renormalization schemes see e.g. Ref.~\cite{Garzelli:2020fmd} and references therein.}.
In this case, PDF uncertainties, also shown in Fig.~\ref{fig:dmeson-cms} and Fig.~\ref{fig:dmeson-cms-21}, 
amounting to approximately $\pm 2\%$, turn out to be much smaller than scale uncertainties, amounting to approximately to
$\pm (6 - 8)\%$ and slightly asymmetric around the central value.   
Changing tune does not affect in a sizable way the size of the scale and PDF uncertainty bands, that are
approximately the same in Fig.~\ref{fig:dmeson-cms} and~\ref{fig:dmeson-cms-21}. 
It plays a role, instead, in modifying the normalization and the shape of central predictions, 
with those obtained with the ATLAS A14 central tune slightly enhanced (up to a factor $\lesssim$ 10\% 
in the most forward direction) with respect to those with the Monash tune, as shown in Fig.~\ref{fig:dmeson-cms-tune}.

The conclusion that the PDF uncertainty is smaller than scale uncertainty drawn for the ABMP16\_3\_NLO PDF 
case in the discussion of Fig.~\ref{fig:dmeson-cms} and \ref{fig:dmeson-cms-21}, 
does not apply to every PDF set. For example, in case of the {CT18NLO} and {CT18ZNLO} PDF sets, 
PDF uncertainties are larger, as visible in Fig.~\ref{fig:dmeson-cms-ct18}, 
or of comparable size, as visible in Fig.~\ref{fig:dmeson-cms-ct18z},  
with respect to scale uncertainties~\footnote{Scale uncertainty is computed here for the ABMP16\_3\_NLO PDF fit only. 
Repeating the computation for the CT18NLO and CT18ZNLO PDF fits, may lead to modifications of the size of the band. 
However, we expect that for the case study at hand and the considered PDF sets these modifications are small, 
i.e. modifications of the percentual size of the scale uncertainty band amounting to 
$\Delta\mu\% \rightarrow \Delta\mu\% \, \pm\, (1 - 2) \%.$, depending on the tolerance criterion used and, 
moderately, also by the prescription for computing PDF uncertainties.}. 
In particular, the {CT18NLO} and {CT18ZNLO} PDF uncertainty bands shown in Fig.~\ref{fig:dmeson-cms-ct18} and~\ref{fig:dmeson-cms-ct18z} 
refer to the default prescription provided by the CT collaboration, leading to asymmetric uncertainties at the 90\% Confidence Level (C.L.), 
corresponding to the tolerance criterion adopted in their QCD analysis and to the fact that, when performing the PDF fit, 
they did not symmetrize the error bars of the experimental datapoints included. On the other hand, 
the {CT18NLO} and {CT18ZNLO} PDF uncertainty bands shown in Fig.~\ref{fig:dmeson-cms-allpdfs} 
refer to the symmetrized uncertainty prescription by the CT collaboration. Additionally, 
in Fig.~\ref{fig:dmeson-cms-allpdfs} we have rescaled the {CT18NLO} and {CT18ZNLO} PDF uncertainty 
bands to the 68\% C.L. to facilitate the comparison with the {ABMP16\_3\_NLO} uncertainty band, 
which is symmetric by construction, considering that the error bars affecting the experimental datapoints 
were symmetrized when performing this PDF fit. This band size corresponds to the tolerance criterion 
$\Delta \chi^2 = 1$ adopted by default in the ABMP16 PDF fits and in many other PDF fits.
The residual difference in the size of the PDF uncertainty bands visible in Fig.~\ref{fig:dmeson-cms-allpdfs} 
is related to the different theory input, parameterization assumptions and data included in the various PDF analyses which have led to these PDF sets. 
In particular, both the ABMP16 and the CT18ZNLO analyses include recent precise Drell-Yan data at the LHC, 
which turn out to be quite relevant for constraining flavour separation in quark PDFs~\footnote{In \texttt{PowHel} we consider PDF fits with NLO accuracy, 
in full consistency with the perturbative accuracy of the hard-scattering matrix-elements. 
Recent doubts on the use of Drell-Yan data in NNLO PDF fits have been raised e.g. in Ref.~\cite{Alekhin:2021xcu}, 
due to the lack of publicly available NNLO integrators for computing fiducial cross-sections for Drell-Yan production 
capable of accounting for the effects of linear power corrections~\cite{Ebert:2020dfc} which can arise in case of symmetric 
cuts on leptons from $W$-boson decays. As shown in Ref.~\cite{Alekhin:2021xcu}, the size and presence of these power corrections depends 
on the adopted NNLO IR regularization method and of the specific analysis cuts, and it is far from being 
negligible in many experimentally analyzed configurations for various theory setups/tools.}. 
These data are instead absent from the CT18NLO PDF analysis. 
As a consequence, the CT18NLO PDF uncertainty band is larger ($ \sim \pm 10$\%) than the 
CT18ZNLO one ($ \sim \pm 8$\%) for central $|\eta_\mu|$. They approach each other in size at the middle of the $|\eta_\mu|$ spectrum, 
and they end to be roughly of the same size ($ \sim \pm 5$\%) for $|\eta_\mu| \sim 2$.  

On the other hand, the difference in the absolute value of central predictions when using different PDFs 
is related partly to the various ingredients of the PDF fits (see discussion above) and partly to the 
$m_c$ values that we adopted in the computation, i.e. $m_c = 1.4$~GeV for the CT18ZNLO and CT18NLO PDF sets and   
$m_c = 1.5$~GeV for the ABMP16\_3\_NLO PDF set. The CT18NLO PDF fit was
actually performed by the CT collaboration using as input $m_c = 1.3$ GeV, which, for full consistency, we should have also used 
for predictions with it. Decreasing $m_c$ leads to an increase of the cross-sections. Therefore, 
if we would have used the latter value, we would have got central predictions slightly enhanced, i.e. in 
slightly better agreement with the CT18ZNLO and ABMP16\_3\_NLO PDF cases. We used instead 
$m_c = 1.4$ GeV to facilitate the comparison with the CT18ZNLO PDF case, 
considering that the latter were fitted and provided with this $m_c$ value by the CT collaboration.

Overall, when considering their (PDF + scale)  uncertainties, our theory predictions with 
ABMP16\_3\_NLO and 
CT18ZNLO PDFs are compatible in all bins with the experimental data. The best theory/data agreement 
is obtained with the CT18ZNLO predictions. Predictions with the CT18NLO PDFs lie slightly below the 
previous ones, but, if one considers the nominal uncertainty bands at 90\% C.L shown in 
Fig.~\ref{fig:dmeson-cms-ct18}, the conclusion of agreement with experimental data in all bins 
apply also to the latter. In all three cases, we observe that the theoretical predictions in the 
first bin of the $\mu^+$ distribution lie close to the lower limit of the experimental data error bars. 
However, at the present status of uncertainty no significant tension is observed. The corresponding cross sections with their respective uncertainties can be found in Tab. \ref{tab:ProdRunsPowHel_CMS}, while the ratios of the integrated fiducial cross sections with uncertainties determined in the most conservative way, \textit{i.e.} assuming 
no correlations 
of scales and PDFs in numerators and denominators, can be found in Tab. \ref{tab:ProdRunsPowHel_CMS_R}.

Notwithstanding the satisfactory level of agreement between our theory predictions and 
the experimental data, we believe that new NLO PDF analyses including $W + c$ data would be 
valuable, considering the present tensions/inconsistencies among various datasets already included in the PDF analyses 
which have made the strange quark one of the longstanding issues in PDF 
determination.

%%%%%%%%%%%%%%%%%%%
\begin{table}[!th]
\addtolength{\tabcolsep}{-2.8pt}
\centering
\begin{tabular}{|c|c|c|c|c|c|c|c|}
\hline\hline
Process &  PDF & $\sigma^M_\mathrm{MC}$ [pb]& $\sigma^A_\mathrm{MC}$ [pb] & $\delta_\mathrm{scale}$ & $\delta_\mathrm{PDF}$&$\delta^{68\%}_\mathrm{PDF}$ & $\sigma^{{\rm CMS}}$ [pb] \bigstrut\\
\hline\hline
\multirow{3}{*}{$ W^+ + D^{*-} $} &  
{\texttt{ABMP16}} &  $62$ & $64$ & $ \substack{+6.9\%\\-6.4\%}$ & $ \pm \,2\%$& $ \pm \,2\%$ & \multirow{3}{*}{$ 65 \pm 5 \,({\rm stat})\substack{+10 \\ -10} \,({\rm sys}) $}  \bigstrut\\
\cline{2-7}
 & {\texttt{CT18Z}} & $63$ & $64$& $-$  & $\substack{+14.8\% \\ -10.5\%}$& $ \pm \,7.2\%$ & \bigstrut\\
\cline{2-7}
 &{\texttt{CT18}} & $58$  & $59$&  $-$  & $\substack{+11.1\% \\ -11.3\%}$ & $ \pm \,8.4\% $ & \bigstrut\\
\cline{1-8}
\multirow{3}{*}{$W^- + D^{*+}$} &{\texttt{ABMP16}} & $66$ &$67$ & $\substack {+6.8\% \\ -6.5\%}$ & $\pm \,2\% $ & $ \pm \,2\%$ & \multirow{3}{*}{$ 71 \pm 6 \,({\rm stat})\substack{+9 \\ -10} \,({\rm sys}) $} \bigstrut\\ 
\cline{2-7}
 & {\texttt{CT18Z}} & $67$  &$68$& $-$  & $\substack{+10.1\%\\ -12.9\%}$& $ \pm \,6.8\%$ & \bigstrut\\
\cline{2-7}
 &{\texttt{CT18}} & $61$ &$63$ & $-$  & $\substack{+10.8\% \\ -11.7\%}$&  $ \pm \,8.1\%$ & \bigstrut\\
%%%%%%%%%%%%%%%%%%%
\hline\hline
\end{tabular}
\caption{
Our predictions for the integrated fiducial NLO~+~SMC cross sections of $W^+ + D^{*}(2010)^-$ and $W^- + D^{*}(2010)^+$ 
productions according to the CMS analysis at $\sqrt{s} = 13$~TeV of Ref. \cite{Sirunyan:2018hde}. PS, hadronization, 
MPI and beam remnant effects are accounted for using either the PYTHIA8 Monash tune (third column) or the ATLAS A14 central tune (fourth column). 
Also reported, for the simulations with the Monash tune, are the uncertainties stemming from scale variation, 
computed using the AMBP16\_3\_NLO PDF set, and the internal uncertainties related to each PDF set considered. 
The column denoted as $\delta^{68\%}_\mathrm{PDF}$ reports symmetrized PDF uncertainties at 68\% C.L.. 
The last column reports the experimental integrated fiducial cross-sections, obtained by us by 
combining the experimental results in all bins of the $|\eta_{\ell^+}|$ and $|\eta_{\ell^-}|$ distributions, respectively.}
\label{tab:ProdRunsPowHel_CMS}
\end{table}
\begin{table}[!th]
\centering
\begin{tabular}{|c|c|c|c|c|c|}
\hline\hline
PDF & $\mathcal{R}^{M}_{D^{*\pm}}$ &$\mathcal{R}^{A}_{D^{*\pm}}$ &  $\delta_\mathrm{scale}$  & $\delta_\mathrm{PDF}$&$\delta^{68\%}_\mathrm{PDF}$ \bigstrut\\
\hline\hline 
{\texttt{ABMP16}} &  $0.94$  & $0.96$&  $ \substack{ +9.3\%\\-8.6\%}$ &  $ \pm \,3.2\%$& $ \pm \,3.2\%$ \bigstrut\\
\cline{1-6}
{\texttt{CT18Z}} & $0.94$ &$0.95$ & $-$ & $\substack{+18.1\%\\ -17.0\%}$& $ \pm \,9.3\%$ \bigstrut\\
\cline{1-6}
{\texttt{CT18}} & $0.95$  & $0.94$& $-$ & $\substack{+15.8\% \\ -16.8\%}$& $ \pm \,11.1\%$ \bigstrut\\
%%%%%%%%%%%%%%%%%%%
\hline\hline
\end{tabular}
\caption{Our predictons for the ratio of integrated fiducial NLO~+~SMC cross sections 
$\mathcal{R}_{D^{*\pm}} = \sigma(W^+ + D^{*-})/\sigma(W^- + D^{*+})$ according to the CMS analysis at 
$\sqrt{s} = 13$ TeV of Ref. \cite{Sirunyan:2018hde}, using either the Monash (second column) or the 
ATLAS A14 (third column) central tune. Also reported, for the simulations with the Monash tune, are the uncertainties, 
computed assuming uncorrelation between numerator and denominator, stemming from scale variation and from PDF 
variation within each considered PDF set. The last column reports symmetrized PDF uncertainties at 68\% C.L..}
\label{tab:ProdRunsPowHel_CMS_R}
\end{table}

Comparing among each other the panels of each of the previously discussed fi\-gu\-res, 
we observe that a better agreement between central theory predictions and experimental 
data is achieved for the sum of the $\mu^+$ and $\mu^-$ production data, 
than for the data for these two channels considered separately. 
Considering that the $W^-c$ and $W^+\bar{c}$ processes are sensitive to a different mixture of
sea and valence quarks, including separately the $\mu^+$ and $\mu^-$
data in PDF fits is important for quark flavour separation and 
for strange-antistrange asymmetry studies. 
In cross section ratios part of the uncertainties cancel, 
when assuming correlated uncertainties between numerator and denominator.
However, in order to get full advantage from the inclusion of $W^-c$ and $W^+\bar{c}$
absolute cross-section data in the PDF fits, it is crucial that the experimental uncertainties, 
at the moment much larger than the scale uncertainties from missing higher orders affecting theory 
predictions, get reduced.

\subsection{Comparison with ATLAS data: $W^\pm + D^{\mp}$ and $W^\pm + j_c (j_{\bar{c}})$ production at 7 TeV}
\label{sub:atlas}

The ATLAS collaboration published a detailed analysis of $W + c$ production data at $\sqrt{s} = 7$~TeV in 
Ref.~\cite{Aad:2014xca}, which includes both the $W + D$-meson and the $W + j_c$ cases, with $W$ decaying leptonically.
Similarly to the CMS analysis discussed in the previous section, even in the ATLAS analysis the $W$ bosons 
are reconstructed from their decay products, and the distributions of the absolute values of the pseudorapidities 
$|y_{\ell^\pm}|$ of the charged leptons $\ell^\pm$ from $W^\pm$ boson decay are measured. 
Like the CMS analysis, the ATLAS analysis reports separately the differential distributions (rapidities) of 
charged leptons emerging from the decays of $W$ bosons of opposite charge. The ATLAS analysis asks for 
exactly one isolated lepton in order to reduce backgrounds, and, differently from the CMS analysis, 
it provides cross sections for the sum of the muonic and the electronic $W$ boson decay channels, 
with cuts on the lepton $p_{T,\ell}~>~20$~GeV, $|\eta_\ell| < 2.5$, on the missing energy 
$E_{T}^{miss} > 25$ GeV, and on the transverse mass of the reconstructed $W$ boson, $m_T^W > 40$ GeV.   
In the $D$-meson analysis, the presence of at least one $D$-meson is required, with $p_{T,D} > 8$ GeV and 
$|\eta_D| < 2.2$, whereas in the $j_c$ analysis, the presence of exactly one anti-$k_T$ $c$-jet 
($R = 0.4$) with $p_{T,\,{j_c}} > 25$ GeV and $|\eta_{j_c}| < 2.5$, that includes at least a
weakly-decaying prompt charmed hadron with $p_{T,\,{h_c}}~>~5$~GeV
and $\Delta R < 0.3$, is required\footnote{In practice, the charmed hadron is identified on the 
basis of its semileptonic decay products, through soft muon tagging techniques.}.
The binning considered in the ATLAS $D$-meson analysis is slightly coarser than the one used in the 
CMS one (4 bins instead of 5) and covers approximately the same pseudorapidity range.
The bin width considered in the ATLAS $j_c$ analysis is much smaller than the one used in case of 
$D$-mesons and covers the same pseudorapidity range.
As for the CMS analysis, all figures and tables reported in the following of this Section refer to 
(OS - SS) fiducial cross sections. The separate role of the OS and SS contributions to these 
cross-sections are quantified and discussed in Appendix B.

\begin{figure}
  \begin{center}
 \includegraphics[width=0.495\textwidth]{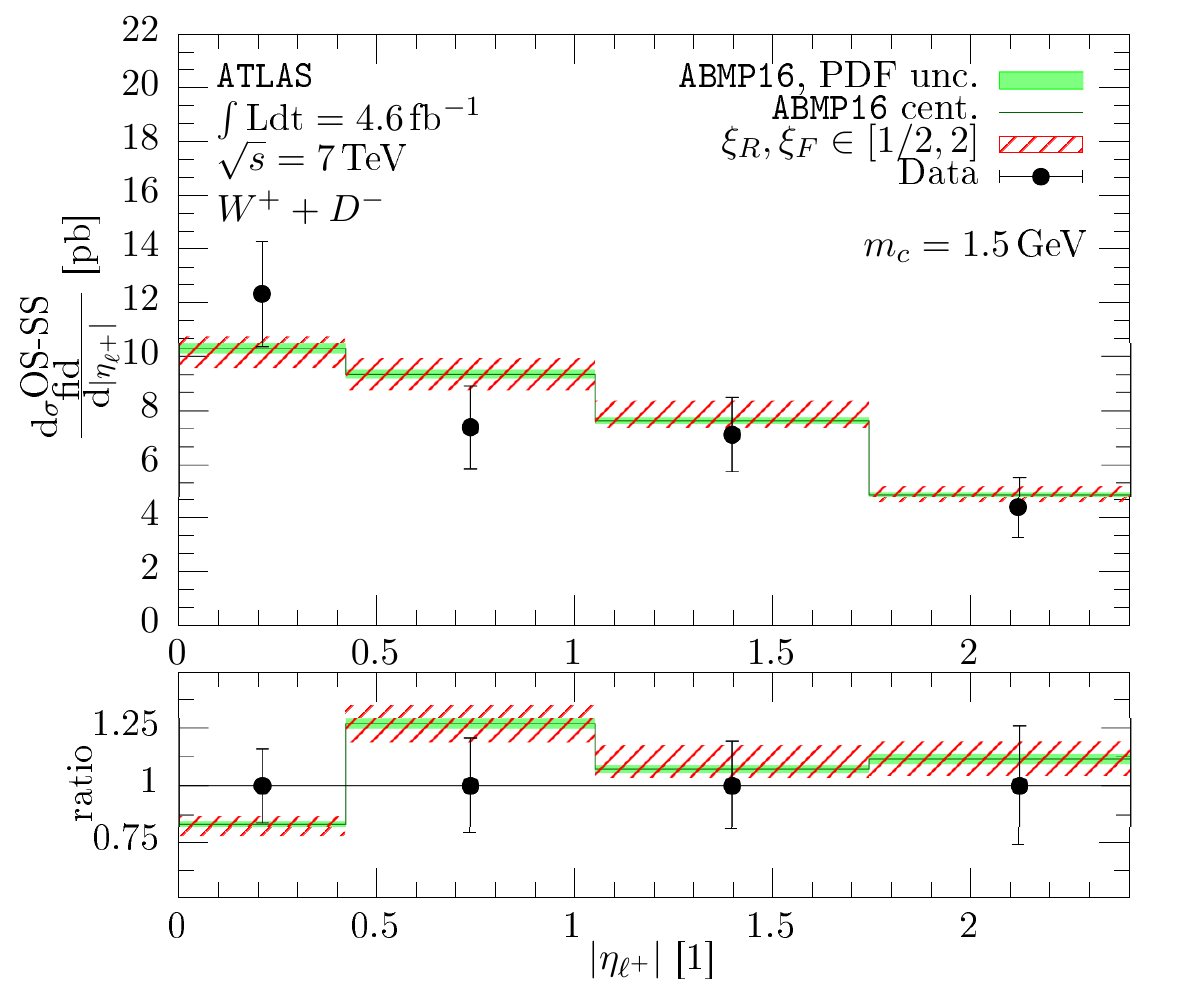}
 \includegraphics[width=0.495\textwidth]{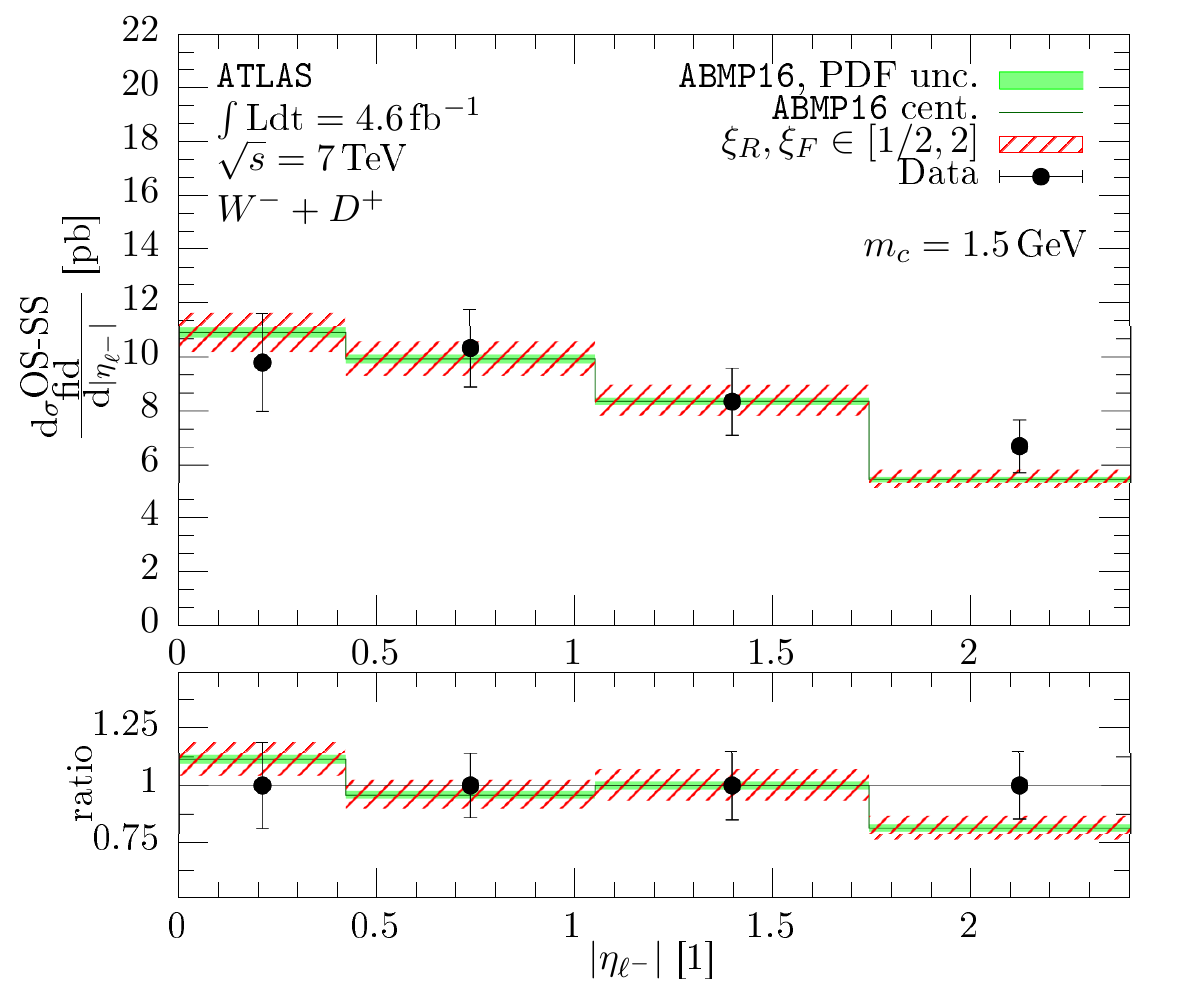}
  \end{center}
  \caption{\label{fig:dmeson-atlas-21} Difference of the differential cross-section for $W + D$-meson production, 
    with $W^\pm \rightarrow \ell^\pm + \overset{(-)}{\nu_\ell}$, where the $\ell^\pm$ and the
    $D^\pm$-meson have charges of opposite sign and that where the $\ell^\pm$ and the $D^\pm$-meson 
    have charges of same sign, as a function of the absolute value of the pseudorapidity of the 
    $\ell^\pm$. Theoretical predictions at NLO + SMC accuracy, obtained by \texttt{PowHel + PYTHIA8} 
    are compared to experimental data from the ATLAS collaboration~\cite{Aad:2014xca}. 
    7-point ($\mu_R$, $\mu_F$) scale and PDF uncertainty bands, computed from the 
    30 members of the ABMP16 PDF set, are reported by shaded (red) and solid (green) bands, 
    respectively. The predictions include parton shower, hadronization, MPI and beam remnant 
    effects according to the {\texttt{PYTHIA8}} ATLAS A14 central tune with NNPDF2.3LO PDFs.}
  \end{figure}

\begin{figure}
  \begin{center}
 \includegraphics[width=0.495\textwidth]{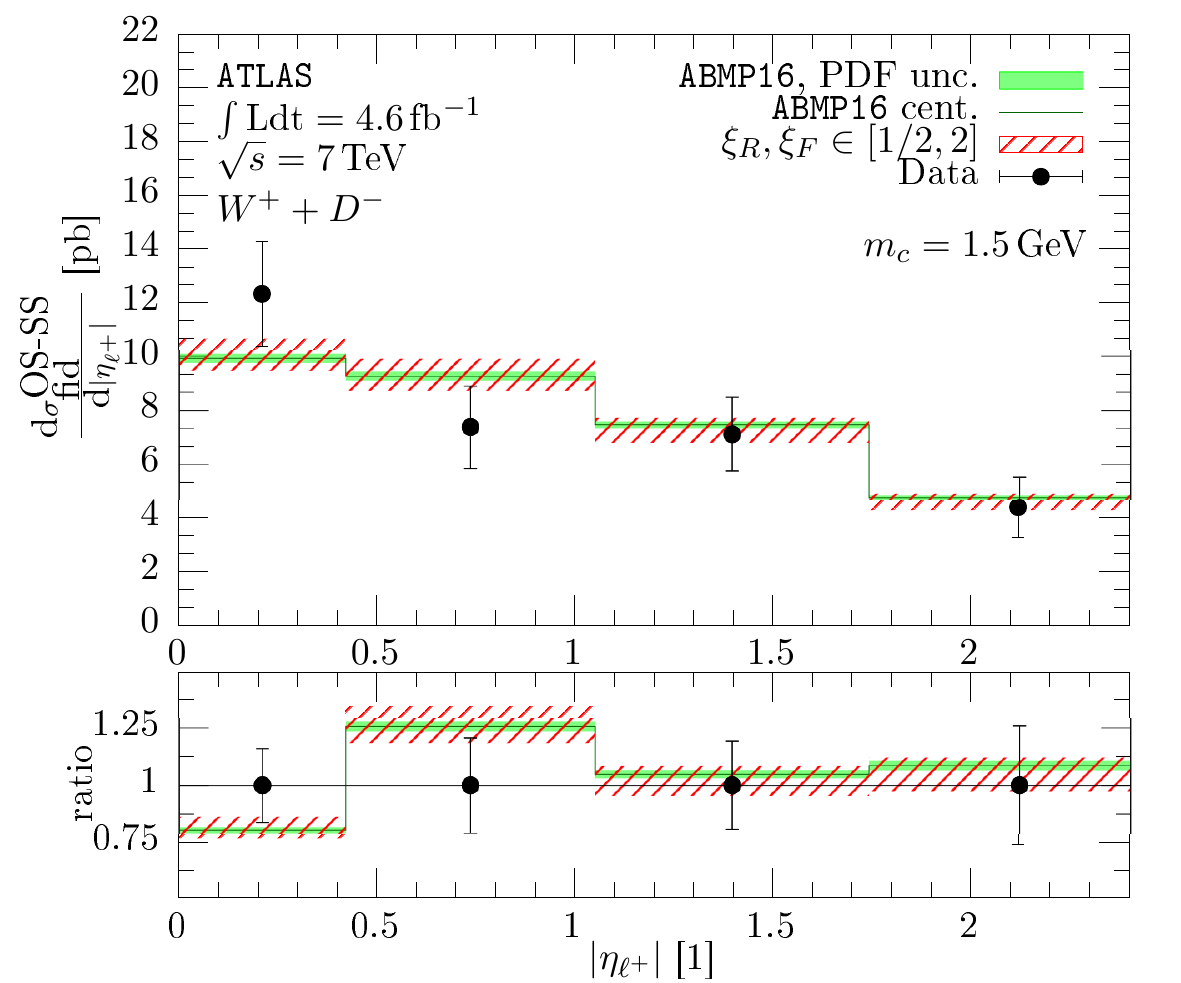}
 \includegraphics[width=0.495\textwidth]{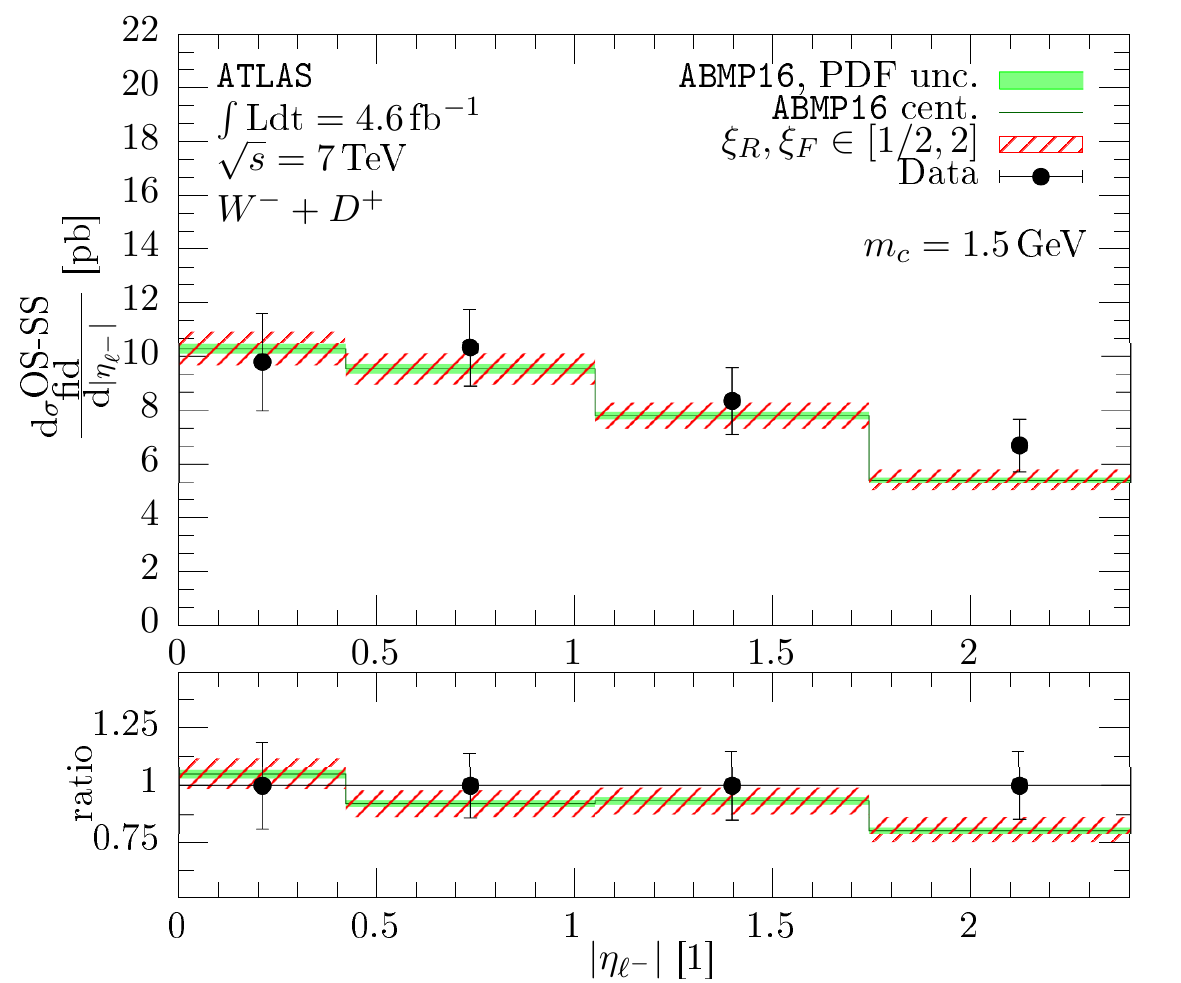}
  \end{center}
  \caption{\label{fig:dmeson-atlas}
Same as in Fig.~\ref{fig:dmeson-atlas-21}, but using the \texttt{PYTHIA8} Monash tune.}
\end{figure}

\begin{figure}
  \begin{center}
 \includegraphics[width=0.495\textwidth]{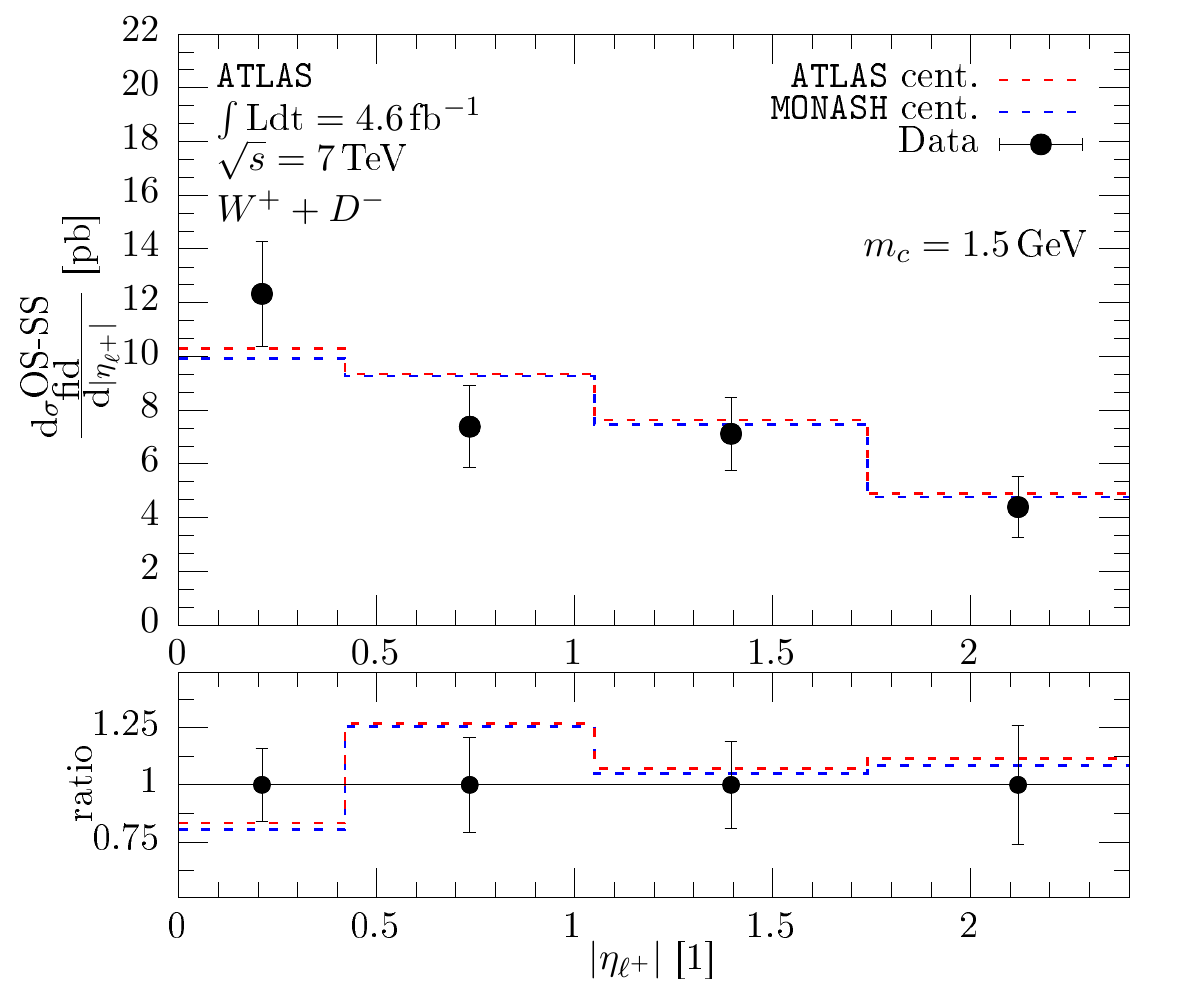}
 \includegraphics[width=0.495\textwidth]{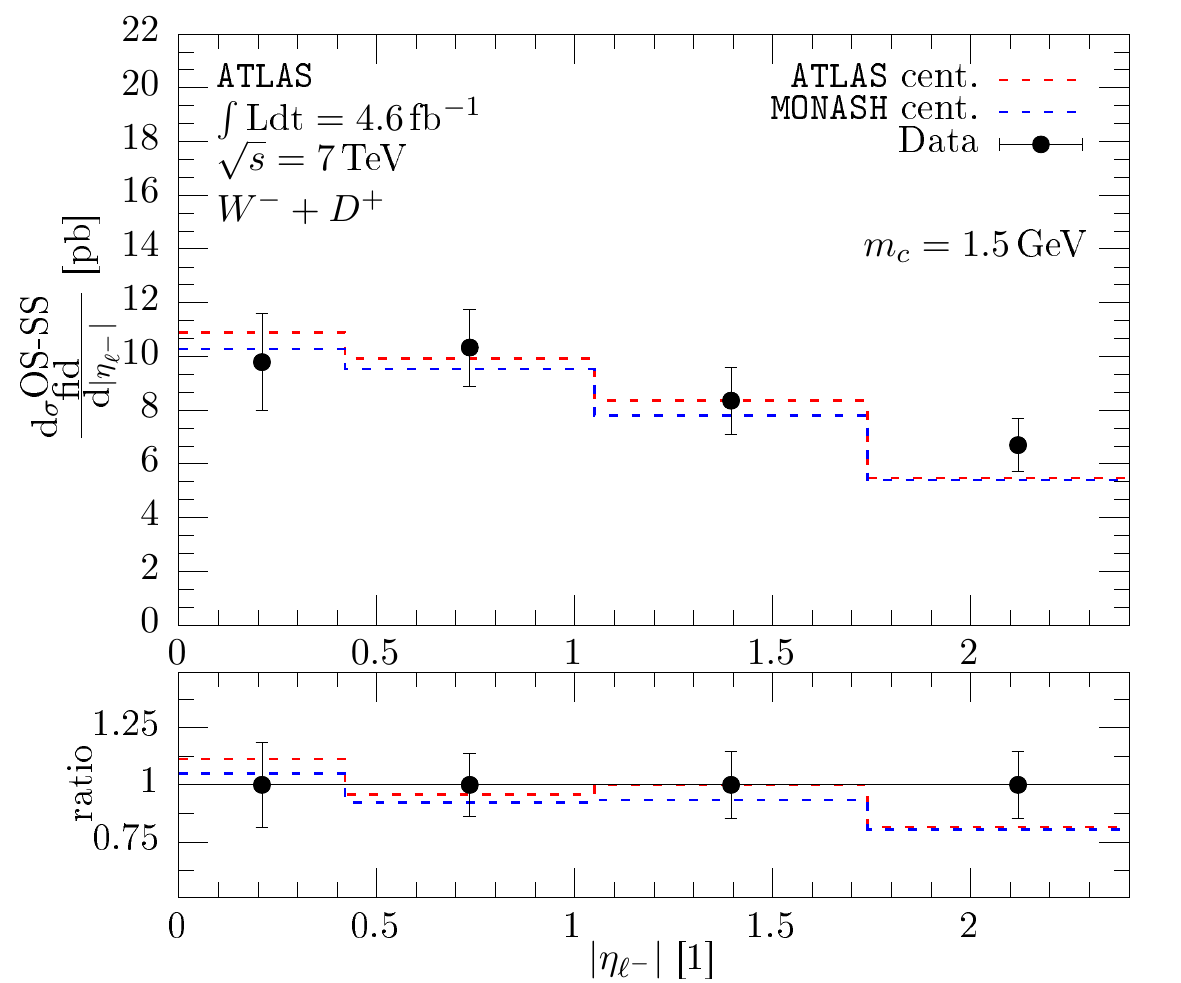}
  \end{center}
  \caption{\label{fig:dmeson-atlas-tune}
    Same as in Fig.~\ref{fig:dmeson-atlas-21} and~\ref{fig:dmeson-atlas}, 
    but limited to central predictions with two different \texttt{PYTHIA8} 
    tunes (Monash and ATLAS A14 central tune with \texttt{NNPDF2.3LO} PDFs).
  }
\end{figure}

In Fig.~\ref{fig:dmeson-atlas-21} we report the comparison of our theory predictions with the 
$W^\pm + D^\mp$ experimental data. We use the same scale and setup as for the CMS analysis, 
except that, in case of the ATLAS analysis, we use as default the 
\texttt{PYTHIA8} ATLAS central A14 tune with NNPDF2.3LO PDFs.  The impact of the choice of a 
different tune is moderate, as can be seen by comparing this figure with 
Fig.~\ref{fig:dmeson-atlas} where the Monash tune is instead used. Central predictions obtained with both 
tunes are compared to each other in Fig.~\ref{fig:dmeson-atlas-tune}, from which one can see that 
they differ by less than 10\% (biggest difference being 7\%), with the ATLAS A14 tune leading to cross sections 
slightly enhanced with respect to the Monash tune in all bins.

We found that the level of agreement between theory predictions and experimental data is similar to the CMS case 
(see discussion in subsection~\ref{sub:cms}). We consider this as a cross-check 
of the compatibility of the ATLAS and CMS data and a proof of the robustness of our conclusions.
In particular, we observe that theory predictions for ($W^\pm + D^\mp$) production are compatible with ATLAS experimental data, 
when considering both the theoretical and experimental uncertainties, in almost all bins.
The only difference worth of notice (though small)  
is observed in the first two bins of the 
$|\eta_{\ell^+}|$ distribution. However, 
considering the present level of experimental uncertainties, this is not sufficient to claim any tension with the data.
For the $|\eta_{\ell^-}|$ distribution, we observe good compatibility of central theory predictions with experimental data in the central pseudorapidity bins, as we also noticed in the CMS case.

Scale uncertainties amount to approximately $\pm (6 - 8)$\% and are larger than PDF uncertainties, 
at least when the latter are computed using the ABMP16\_3\_NLO PDF set. Changing \texttt{PYTHIA8} 
tune does not affect the size of scale and PDF uncertainties, as evident when comparing 
Fig.~\ref{fig:dmeson-atlas} with Fig.~\ref{fig:dmeson-atlas-21}. The sizes of these bands 
closely resemble those obtained in the CMS analysis. For the ATLAS analysis we did not report 
explicitly PDF uncertainties using different PDF fits. In case of CT18NLO and CT18ZNLO PDFs, 
we expect that considerations analogous to those discussed for the CMS analysis apply even in the 
ATLAS analysis. Considering the larger uncertainty bands of these PDF fits 
(see Fig.~\ref{fig:dmeson-cms-allpdfs} for the CMS case), 
we expect a compatibility of theory predictions 
and ATLAS experimental data in all bins.

\begin{figure}
  \begin{center}
   \includegraphics[width=0.495\textwidth]{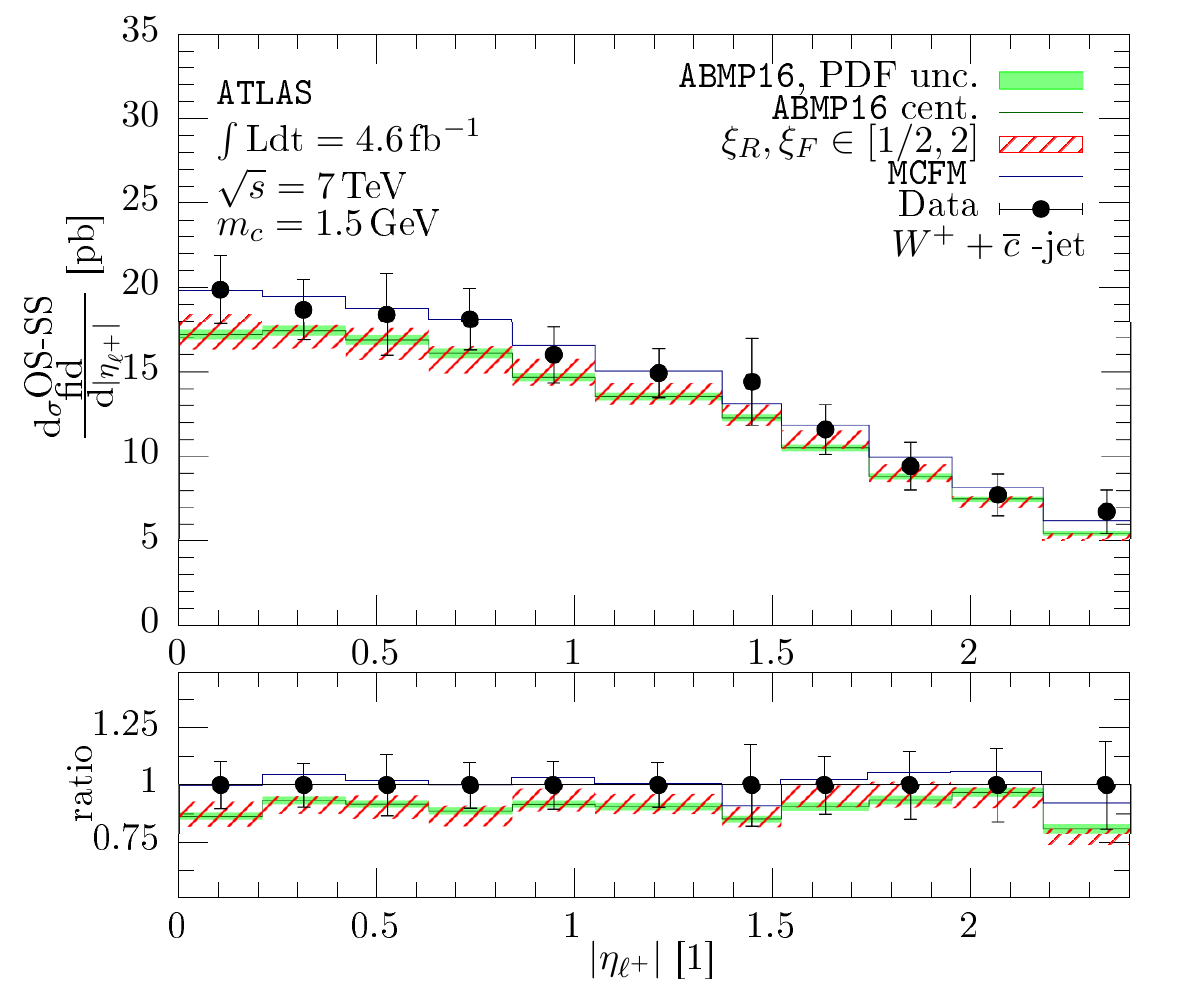}
   \includegraphics[width=0.495\textwidth]{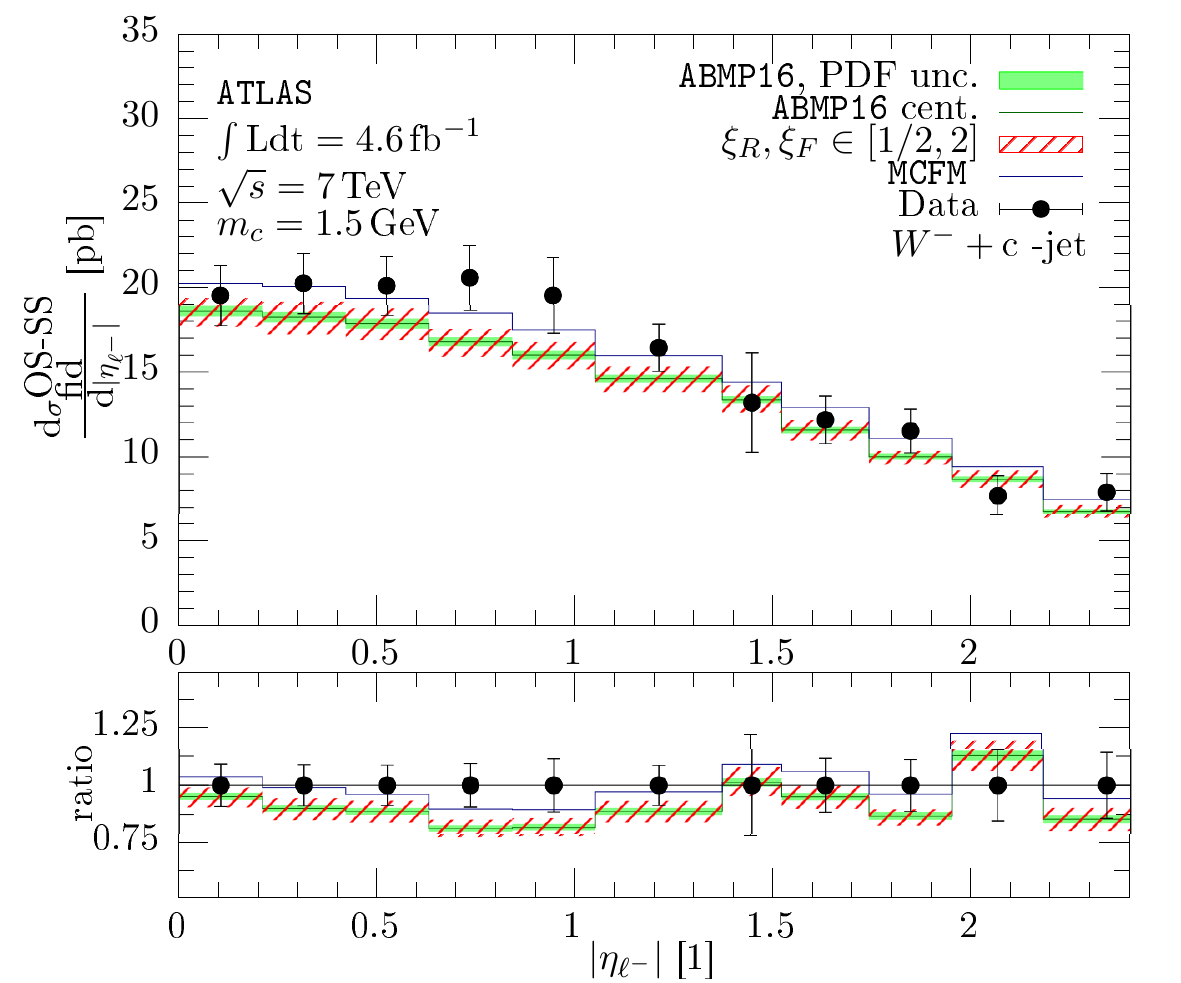}
  \end{center}
  \caption{\label{fig:cjet-atlas-21}
     Same as in Fig.~\ref{fig:dmeson-atlas-21} but for the ATLAS $W + j_c$ analysis cuts of Ref.~\cite{Aad:2014xca}.
     Also reported are the fixed-order NLO predictions. }
  \end{figure}

For completeness we also report the comparison of our theory predictions with experimental data for the $j_c$ ($j_{\bar{c}}$) analysis, 
published in the same ATLAS paper. Considering that, in this case, the transverse momentum cut is bigger 
($p_{T,\,j_c} (p_{T,\,j_{\bar{c}}})>$~25~GeV) than the $p_{T,\,D^\pm}$ cut for the $D$-meson analysis,
accounting for charm finite-mass effects in theory predictions seems to be less relevant. Additionally, 
using fixed-order predictions with hard-scattering matrix-elements, $\alpha_S$ and PDFs in the 3-flavour scheme 
might not be the best solution to compare to $j_c$ ($j_{\bar{c}}$) data at large $p_T$, considering that logarithms of the ratio $(p_T/m_c)$
appearing in the partonic cross sections might become large and then require a resummation. Therefore, 
approaches with heavy-flavour jets defined according to jet flavour algorithms which provide an infrared-safe definition 
of the flavour of a jet and, as a consequence, eliminate the need of resumming the logarithms associated to final state 
$g\rightarrow Q\bar{Q}$ splittings, combined with a massless treatment of heavy-quarks, that allows to effectively resum 
contributions related to initial state $g \rightarrow Q\bar{Q}$ splittings through PDF evolution, have been suggested. 
These approaches are capable of simplifying the calculations and simultaneously reducing scale 
uncertainties~\cite{Banfi:2007gu}, although at the price of loosing all (logarithmic) terms proportional to $m_Q$.
However, the $j_c$ selection criterion applied by the ATLAS collaboration, is based on the identification 
of at least a charmed hadron $h_c$ inside the jet, with a low $p_{T,min}$ cut, i.e. $p_{T,h_c} > 5$ GeV.
Therefore, it might be worth investigating up to which extent charm finite-mass effects have an impact in the $j_c$ selection process.

In practice, the level of agreement between our theory predictions at NLO+SMC accuracy and the experimental data turns out to be 
slightly worse than the one found for the $W + D$-meson analyses, as can be seen by comparing Fig.~\ref{fig:cjet-atlas-21}, 
related to $W + j_c$ production, with Fig.~\ref{fig:dmeson-atlas-21}, related to $W + D$-meson production, 
with theory predictions computed with a consistent input. The uncertainty bands due to 7-point scale and 
ABMP16\_3\_NLO PDF variation in Fig.~\ref{fig:cjet-atlas-21} 
have approximately the same size as those in Fig.~\ref{fig:dmeson-atlas-21}. 
The main difference is the fact that, while, in case of $W~+~D$-meson production, 
central theory predictions in some bins slightly overestimate and in others slighty 
underestimate the central experimental datapoints, in case of $W + j_c$ production the central 
theory predictions systematically underestimate the central experimental datapoints 
by approximately $\sim (5 - 10)$\% and up to $\sim 15\%$ in some bins of the $|\eta_{\ell^-}|$ distribution.
Theory predictions mostly lie close to the lower limit of the experimental uncertainty bands. 
This might point to the limitation of a three-flavour description when working with high-$p_T$ 
$c$-jets and to the need of including higher-order corrections into the theory predictions. 
On the other hand, we usually observe a softening and distortion of transverse momentum distributions for jets due to parton shower and, to a lesser extent, hadronization. Due to the fact that pseudorapidity distributions are integrated in the transverse momentum, if cuts are imposed on the latter (as done in our analysis) this can result in a decrease of the total fiducial yield. We observe this effect when comparing our central NLO+PS predictions against the fixed-order NLO ones, as also shown in Fig.~\ref{fig:cjet-atlas-21}.
The level of agreement between central NLO+PS theory predictions and experimental data is slightly 
higher for forward pseudorapidity bins, than in case of the central pseudorapidity ones.
However, when considering both the experimental and theory uncertainties, 
one can still claim consistency between  predictions and data in almost all bins. 
The larger disagreement is observed in the $0.5 < |\eta_{\ell^-}| < 1$ region, 
which is sensitive to both the $d$- and $s$-quark PDFs. In fact one can notice that the 
experimental differential cross-sections as a function of $|\eta_{\ell^+}|$ have an overall different shape with respect to 
the $|\eta_{\ell^-}|$ ones, for both the $W + D$-meson and $W + j_c$ analyses.  
In the latter case, the pseudorapidity is almost constant (if not increasing) in the region
$0 < |\eta_{\ell^-}| < 1$.  These data refer to LHC Run 1. It will be interesting to cross-check our predictions with 
experimental results with increased statistics when they become available, 
especially considering that our NLO + SMC theory predictions, in contrast to the data, are monotonically decreasing at increasing $|\eta_{\ell^\pm}|$. Unfortunately, 
increasing statistics will not be possible anymore at $\sqrt{s}=7$ TeV, 
but a repetition of data analysis at $\sqrt{s} = 13$ TeV, with the study of additional observables, 
is indeed welcome and in the pipeline of the ATLAS collaboration. 

As mentioned in the Introduction, a study of $W + j_c$ production with massless charm in the 
5-flavour scheme at NNLO has been recently carried out in Ref.~\cite{Czakon:2020coa}. 
The NNLO analysis of Ref.~\cite{Czakon:2020coa}, differently from our one, 
led to central theory predictions which slightly overestimate most of the ATLAS central datapoints, 
although being still in agreement with them when considering the uncertainties.  The comparison with the particle-level 
experimental data~\cite{Aad:2014xca} presented in Ref.~\cite{Czakon:2020coa} is 
complicated (i) by the fact that it is based on fixed-order theory predictions, 
i.e. no parton shower + hadronization + MPI and underlying event effects were included and 
(ii) by the use of the flavoured-$k_T$ jet algorithm that, although being a fully consistent
choice for reconstructing a charm jet in a NNLO calculation preserving its flavour with respect to infrared emissions, 
differs from the anti-$k_T$ jet algorithm used in practice in the LHC experimental analyses published so far.
On the other hand, in our analysis hard-scattering matrix-elements are just accurate to NLO, 
but parton shower, hadronization, MPI and underlying event effects are included using \texttt{PYTHIA8} 
and we still use the anti-$k_T$ jet algorithm.  This jet algorithm at NLO
is still a viable choice not only for light, but even for heavy-flavoured jets if used in association with a 
jet-flavour identification criterion preserving jet flavour in presence of infrared emissions. 
We use the same jet-flavour classification criterion as in the ATLAS analysis, implying that the net average 
contribution to the $W + j_c$ (OS-SS) cross-section from the events with a jet including both a charmed and 
an anticharmed hadron passing the cuts, is null. Further studies and proposals for infrared safe jet 
flavour algorithms for applications in SMC generators and NLO + PS matched computations are indeed welcome.
One more comment is in order concerning the impact of missing higher-order corrections on the stability of our predictions. Taking a look at Fig. 2 of Ref. \cite{Czakon:2020coa}, the NNLO correction results in a moderate shift of normalization with respect to NLO: a uniform $\sim 10\%$ shift for the rapidity of the $\mu^-$ and a  $\sim 10\%$ shift in the central regions for $\mu^+$ which increases up to $\sim 20\%$ as one approaches the forward region. As already discussed, our calculational setup differs from the one of Ref. \cite{Czakon:2020coa} for a number of technical aspects (primarily the fact that we consider massive charm quarks, while Ref.~\cite{Czakon:2020coa} is based on the massless charm approximation, and a different jet algorithm). While extrapolating the findings of Ref.~\cite{Czakon:2020coa} to our case is certainly not a trivial exercise, we consider them as a first indication that the NNLO QCD corrections could lead to uncertainty bands which overlap with our estimates for the NLO uncertainties. For the same purpose, we look forward to new NNLO QCD predictions at $\sqrt{s}=13$ TeV to shed more light also in the case of the CMS analysis discussed in Section \ref{sub:cms}. The obtained integrated fiducial cross sections for the processes at hand and their ratios with uncertainties determined in the most conservative way can be found in Tab. \ref{tab:ProdRunsPowHel_ATLAS} and Tab. \ref{tab:ProdRunsPowHel_ATLAS_R}, respectively.

\begin{figure}
  \begin{center}
 \includegraphics[width=0.495\textwidth]{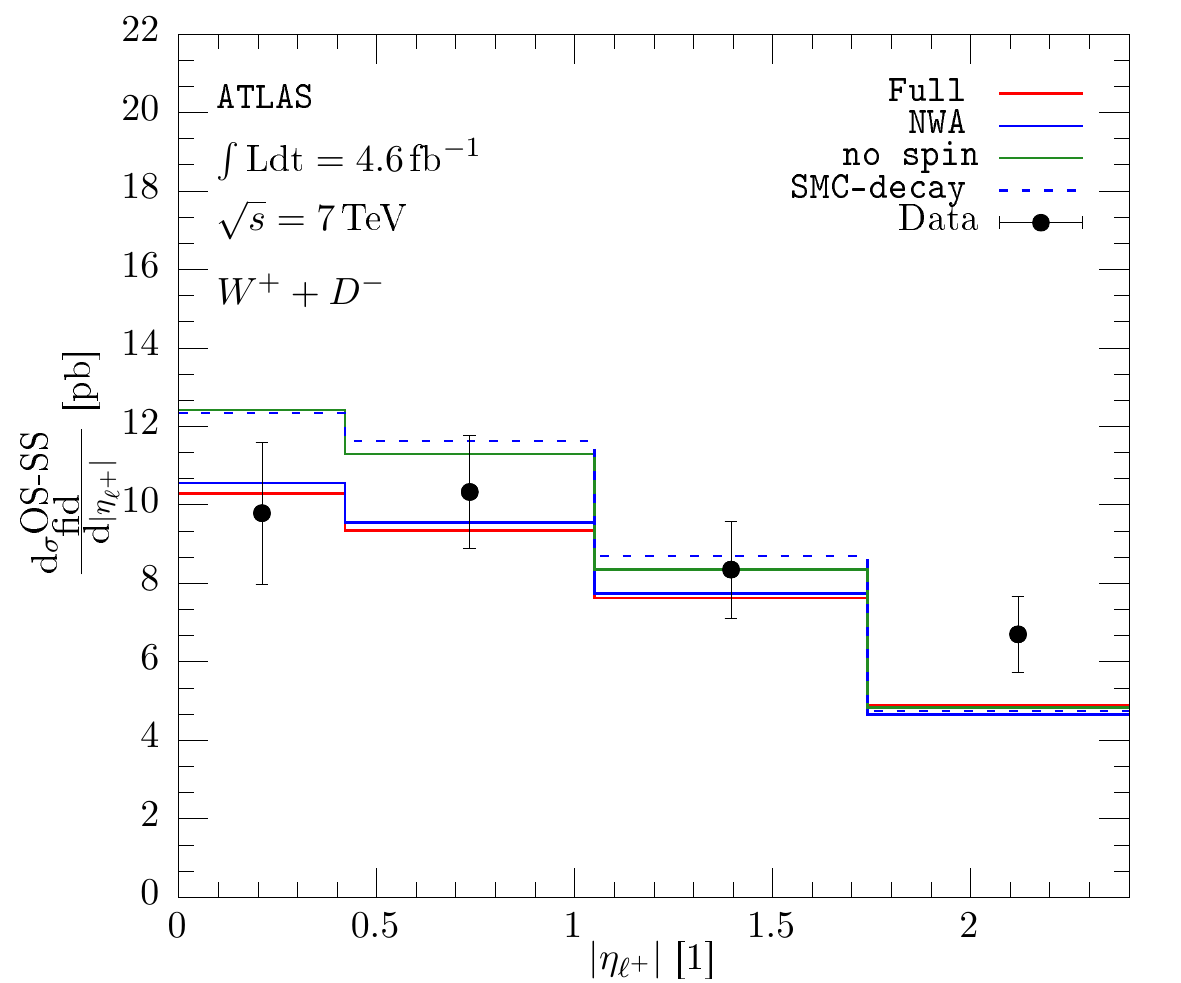}
 \includegraphics[width=0.495\textwidth]{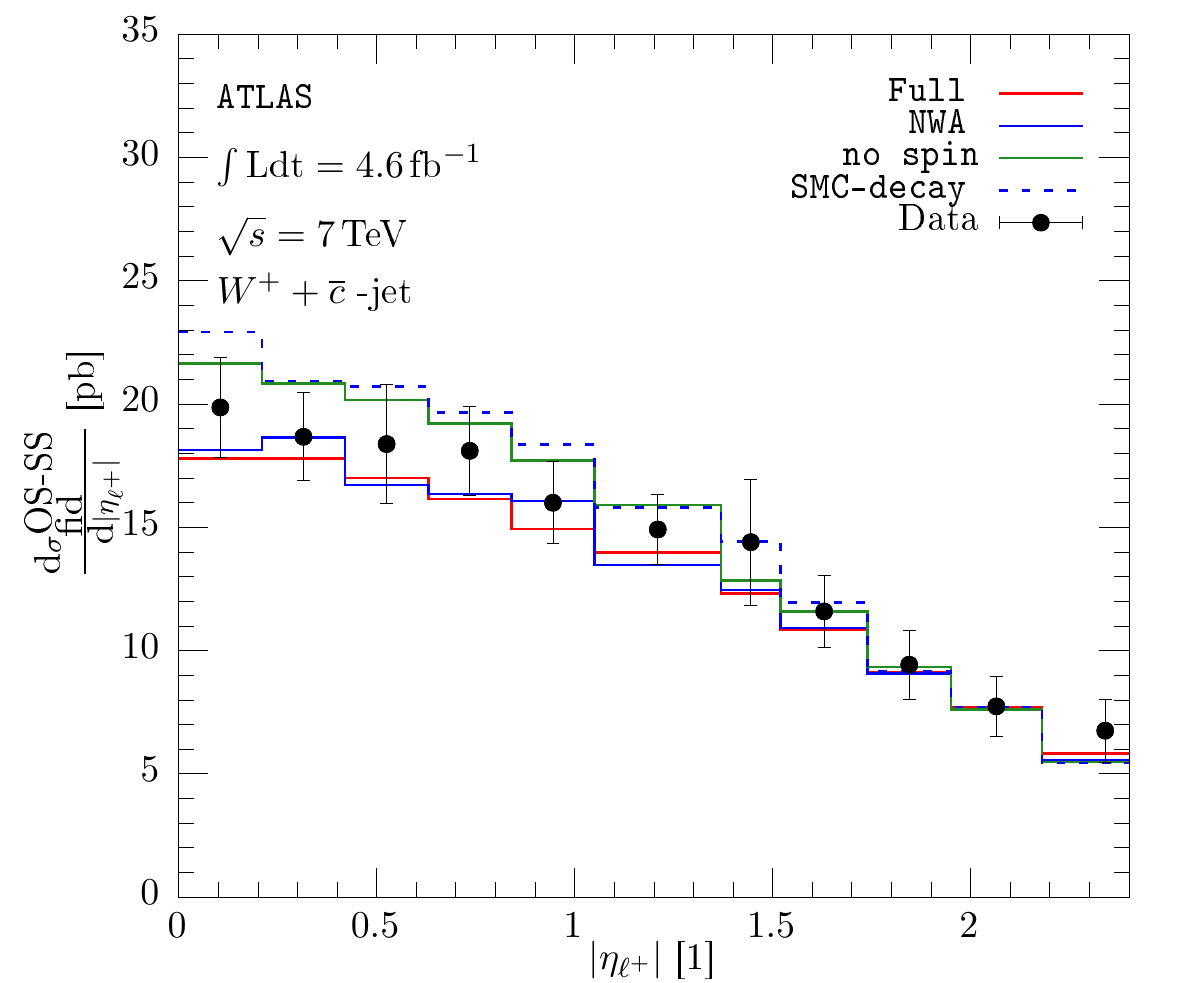}
  \end{center}
  \caption{\label{fig:spincorr-atlas-tune21}
    Left: difference of the differential cross-sections for $W +D$-meson production, 
    with $W$-boson decaying leptonically, from events with an $\ell^+$ and a $D^\pm$-meson with
charge of opposite sign (OS) and that from events with the 
    $\ell^+$ and a $D^\pm$-meson with charge of the same sign (SS), as a function of 
    the absolute value of the  $\ell^+$ pseudorapidity. Theoretical predictions 
    at NLO + SMC accuracy, obtained by \texttt{PowHel + PYTHIA8} using different approximations, 
    are compared to experimental data from the ATLAS collaboration~\cite{Aad:2014xca}. 
    In particular the results of the full simulation, including $W$ off-shellness effects in \texttt{PowHel}, 
    are compared to those with $W$ decayed by \texttt{PowHel} in the NWA approximation including spin-correlation effects, 
    to those in the NWA approximation neglecting spin-correlation effects, and to those where the 
    $W$ boson is kept as stable in \texttt{PowHel} and decayed by \texttt{PYTHIA8}. 
    The {\texttt{PYTHIA8}} ATLAS A14 central tune with NNPDF2.3LO is used in all simulations. 
    Right: same as in the left panel, but for the $W + j_c$    case.}
\end{figure}

\begin{figure}
  \begin{center}
 \includegraphics[width=0.495\textwidth]{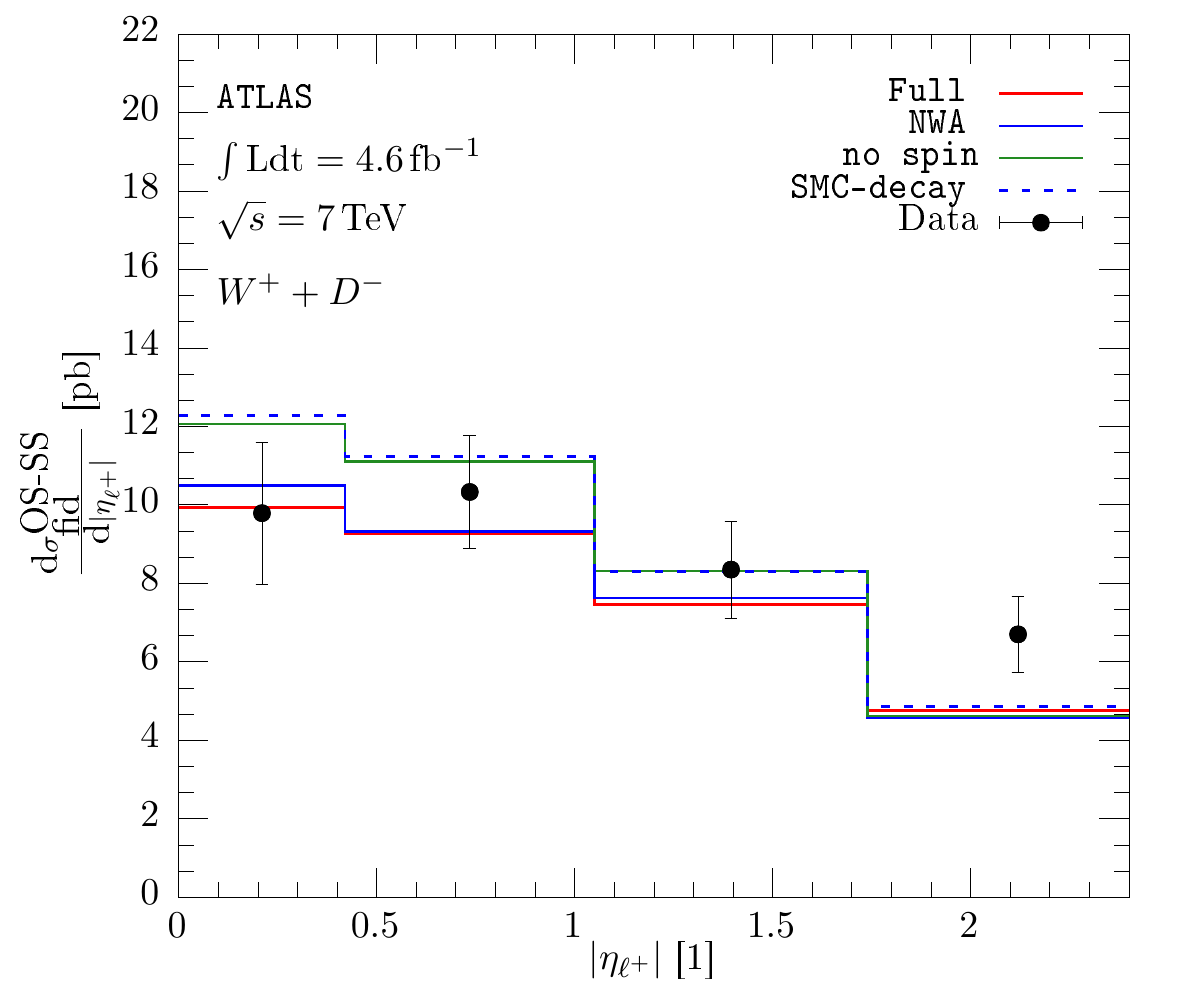}
 \includegraphics[width=0.495\textwidth]{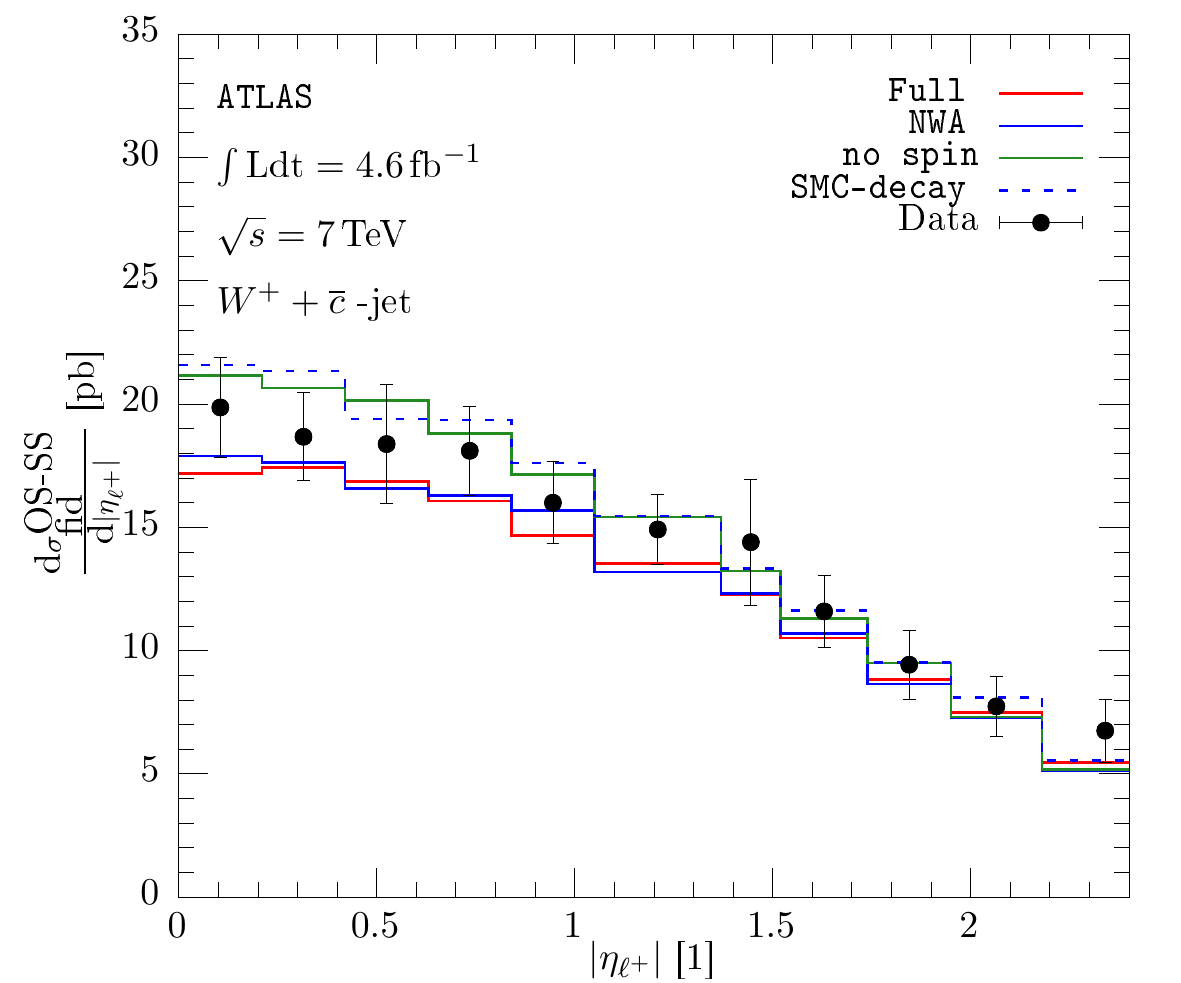}
  \end{center}
  \caption{\label{fig:spincorr-atlas-monash} Same as in Fig.~\ref{fig:spincorr-atlas-tune21}, but for the \texttt{PYTHIA8} Monash tune.}
\end{figure}

Finally, we notice that including or not spin-correlation effects in the description of 
$W^\pm$ production and decay makes sizable changes in the shapes of the $|\eta_{\ell^\pm}|$ distributions, 
as can be seen in Fig.~\ref{fig:spincorr-atlas-tune21} and \ref{fig:spincorr-atlas-monash} for the $W^+$ case. 
This, in turns, can impact the results of PDF fits that intend to make a correct use of these data. 
In order to produce reliable theory predictions to be used in these fits, 
it is thus mandatory to have the most accurate possible description of $W^\pm$ production and decay, 
including full off-shellness and spin-correlation effects.

%%%%%%%%%%%%%%%%%%%
\begin{table}[!th]
  \centering
  \addtolength{\tabcolsep}{-2.8pt}
\begin{tabular}{|c|c|c|c|c|c|c|}
\hline\hline
Process &  PDF & $\sigma^M_\textrm{MC}$ [pb]&$\sigma^A_\textrm{MC}$ [pb] & $\delta_\mathrm{scale}$ & $\delta_\mathrm{PDF}$ & $\sigma^{{\rm ATLAS}}$ [pb] \bigstrut\\
\hline\hline 
$ W^+ + D^- $ & \multirow{4}{*}{\texttt{ABMP16}} & $18.8$ &$19.2$ &  $ \substack{+5.8\%\\-5.3\%}$ & $ \pm 1.5\%$ & $17.8 \pm 1.9 \,({\rm stat}) \pm 0.8 \,({\rm sys})$   \bigstrut\\
\cline{1-1}\cline{3-7}
 $W^- + D^+$&  & $19.8$ & $20.7$ &  $\substack{+5.8\% \\ -5.5\%}$ & $\pm 1.5\%$ & $22.4 \pm 1.8 \,(\rm stat{}) \pm 1.0 \,({\rm sys})$ \bigstrut\\
\cline{1-1}\cline{3-7}
 $W^+ + j_{\bar{c}}$ & & $31.1$  & $ 31.8$&  $\substack{+6.2\%\\-9.1\%}$ & $\pm 2.5\%$ & $33.6 \pm 0.9 \,(\rm stat{}) \pm 1.8 \,({\rm sys})$ \bigstrut\\
\cline{1-1}\cline{3-7}
 $W^- + j_c$ &  & $33.9$& $ 34.6$  &  $\substack{+7.5\%\\ -7.7\%}$& $\pm 2.4\%$ & $37.3 \pm 0.8 \,(\rm stat{}) \pm 1.9 \,({\rm sys})$\bigstrut\\
%%%%%%%%%%%%%%%%%%%
\hline\hline
\end{tabular}
\caption{
    Our predictions for the integrated fiducial NLO~+~SMC cross sections for $W^\pm~+~D^{\mp}$ 
    and $W^\pm + j_c(j_{\bar{c}})$ production according to the $\sqrt{s} = 7$ TeV ATLAS 
    analysis of Ref.~\cite{Aad:2014xca}. PS, hadronization, MPI and beam remnant effects are 
    accounted for using either the Monash tune (third column) or the ATLAS A14 central tune (fourth column). 
    Also reported, for the simulations with the ATLAS A14 central tune, are the uncertainties stemming 
    from scale and PDF variation, both computed using as a basis the ABMP16\_3\_NLO PDF fit, 
    as well as the ATLAS experimental data.
}
\label{tab:ProdRunsPowHel_ATLAS}
\end{table}
\begin{table}[!th]
\centering
\begin{tabular}{|c|c|c|c|c|}
\hline\hline
PDF & $\mathcal{R}^M $ & $\mathcal{R}^A$ & $\delta_\mathrm{scale}$ & $\delta_\mathrm{PDF}$ \bigstrut\\
\hline\hline
\multirow{2}{*}{\texttt{ABMP16}}  &  $\mathcal{R}_{D^\pm}  = 0.95$ &$\mathcal{R}_{D^\pm}  = 0.93$ & $ \substack{ +7.6\% \\ -7.1\%}$  &  $ \pm 2.0\%$  \bigstrut\\
\cline{2-5}
&  $ \mathcal{R}_{j_{\bar{c}/c}}  = 0.92$  &  $ \mathcal{R}_{j_{\bar{c}/c}}  = 0.92$ &   $\substack{+8.9\% \\ -11.0\%}$ & $\pm 3.2\%$\bigstrut\\
%%%%%%%%%%%%%%%%%%%
\hline\hline
\end{tabular}
\caption{
  Our predictions for the ratios of the  integrated fiducial NLO~+~SMC cross sections 
  $\mathcal{R}_{D^\pm} = \sigma(W^+ + D^-)/\sigma(W^- + D^+)$ and  
  $\mathcal{R}_{j_{c}^\pm} = \sigma(W^+ + j_{\bar{c}})/\sigma(W^- + j_c)$ according to the
  $\sqrt{s} = 7$ TeV ATLAS analysis of Ref.~\cite{Aad:2014xca}, using
  either the Monash (second co\-lumn) or the ATLAS A14 (third co\-lumn) central tune.
  Also reported, for the simulations with the ATLAS A14 central tune, are the uncertainties, 
  computed assuming no correlation between numerator and denominator, 
  stemming from scale variation and from PDF variation within the ABMP16\_3\_NLO PDF set.
} 
\label{tab:ProdRunsPowHel_ATLAS_R}
\end{table}

\section{Conclusions}
\label{sec:conclu}

This work presents a NLO QCD + SMC simulation of $W + c$  production, making use of the \texttt{PowHel + PYTHIA8} approach, including charm finite-mass effects in the hard-scattering matrix-elements, treating the $W$-boson off-shell, considering spin-correlation effects and a non-diagonal CKM quark-mixing matrix. This has allowed for consistent comparisons with the already available experimental data at the particle level by the ATLAS and CMS collaborations. Our tool/setup is particularly interesting for simulations in those experimental studies where $D$-mesons with relatively low $p_T$ are reconstructed, cases in which charm finite-mass effects can not be neglected already in the hard-scattering matrix-elements. 
So far, the data at the particle level obtained even in such cases have been analyzed by the experimental collaborations by making use of $W + j$ samples built using as seed hard-scattering matrix-elements with massless partons, 
completely neglecting charm finite-mass effects in the hard-scattering amplitudes. Our work contributes thus to fill a gap, offering to the experimentalists a tool including charm finite-mass and non-diagonal $V_{\mathrm{CKM}}$ effects for their simulations at particle level with NLO QCD + SMC accuracy based on the NLO + PS POWHEG matching formalism. Our tool allows for comparison of the most sophisticated case, where the $W$ boson is produced and decayed off-shell, to the simplified ones where the $W$ boson is produced on-shell and decay in the NWA. Including or not spin-correlations in the latter case turns out to have an important effect on the shape of the distributions for which experimental data are available. 

As was also noticed in Ref.~\cite{Czakon:2020coa}, including non-diagonal $V_{\mathrm{CKM}}$ effects plays an 
important role in the simulations of $W^- + D^+$ meson and $W^- + j_c$ production, due to the fact that in these cases, 
not only the $s$-quark but even the $d$-quark, whose distribution function receives valence contributions differently from the
$s$-quark one, contributes as initial state already at LO. Including non-diagonal $V_{\mathrm{CKM}}$ terms 
(in particular, a non-zero $V_{cd}$ value)  turns out to be indispensable for the correct interpretation of the data, 
in particular for correctly assessing the degree of strange-antistrange sea-quark asymmetry in the proton.

The agreement between our theory predictions and experimental data turns out to be similar for both ATLAS and CMS $W + D$-meson data, which confirms the consistency between the considered datasets and between results at different $\sqrt{s}$.
Also, predictions on leptons of opposite charge show comparable agreement with the data. In our analyses we have considered PDF sets which assume equal distributions for the $s$ and $\bar{s}$ quark. Had we neglected the contribution of $V_{cd}$ in the calculation of matrix elements, the agreement would have dramatically worsened in the case of $W^- + \bar{c}$, where the $d$-quark luminosity plays an important role. This, in turn, would have led to assume a larger $\bar{s}$ distribution function compared to the one of the $s$ quark  in order to reproduce the data.  While this fact alone does not allow us to take firm conclusions on the actual size of the $s-\bar{s}$ asymmetries, it constrains the impact of the latter in PDF fits based on $W+c$ data.

Scale uncertainties amount to approximately $\pm (6 - 8)$\% whereas PDF uncertainties greatly vary when comparing different PDF fits, as a consequence of the different data capable of constraining quark PDFs included and of the PDF extraction methodology employed (including theoretical and parameterization input, as well as statistical aspects). 
We confirm the relevance of including $W + c$ data in NLO PDF fits, already pointed out by other authors, especially for those cases where PDF uncertainties are larger than scale uncertainties (e.g. the CT18NLO PDF case that we discussed). In cases where PDF uncertainties are particularly small (e.g. for the ABMP16\_3\_NLO PDF case that we also considered), including these data in the PDF fit might still be worth, to check consistency / reveal possible tensions with other data, which might lead to a revision of the PDF uncertainties and parameterization.
The potential of including data on the ($W^- + c$)/($W^+ + \bar{c}$) ratio has already been explored by 
the experimental collaborations in their own QCD analyses leading to PDF fits. However, 
to fully exploit the constraining power of absolute cross section data, reduced experimental uncertainties are needed. 

It would be great to systematically include these data even in NNLO PDF fits. This is justified by the much smaller dependence
on unphysical scales. Even greater impact could be achieved if: 1) NNLO predictions would be complemented by parton shower, 
hadronization and MPI effects, in order to be comparable to data at particle level in a consistent way.
This is also important in order to obtain realistic estimates of the SS cross-section contributions (see Appendix B), 
which have to be subtracted from the OS ones, in the comparison with the experimental data available in the form of (OS - SS) cross-sections;
2) in case a 
slicing method is used to regularize NNLO IR divergences, missing power corrections can hamper theoretical precision, 
hence the inclusion of these corrections would be fundamental;  3) heavy-quark finite-mass effects 
would be included in NNLO calculations of $W + c$ production; 4) the uncertainties related to the use of different SMC tunes, 
at present of comparable size with respect to scale uncertainties, would decrease. Also, to make NNLO(+ PS) 
calculations with heavy-flavour jet(s) consistently comparable to experimental data, the jet algorithm used 
in the experimental analyses needs to be changed from the usual anti-$k_T$ to a flavoured jet one. 

While we are confident that many of these developments will take place in the next few years, 
NLO calculations matched to Parton Showers as available in SMC codes can be regarded as 
complementary to fixed-order NNLO predictions in several respects and, for the time being, 
they still represent a competitive tool, although not being the ultimate one, for more
consistent comparisons with the available experimental data on $W~+~c$ production. 
The potential of our generator and of the other a\-vai\-la\-ble tools can be tested/challenged 
by forthcoming analyses, considering $D$-meson distributions with loose $p_T$ cuts and 
observables sensitive to correlations between $D$ mesons and leptons from $W$ decay. 
New analyses in this direction are ongoing within the experimental collaborations and 
we look forward to future de\-ve\-lop\-ments. 

\section*{Acknowledgments}
We are grateful to Lucio Cerrito, Francesco Giuli, Katerina Lipka, Sonja K. Pflitsch, for sharing with us numerous details concerning the LHC experimental analyses.
We are grateful to Sergey Alekhin and Marco Guzzi for useful discussions concerning their PDF fits. 
We thank Sven-Olaf Moch and Pavel Nadolsky for further 
clarifications, and Michael Benzke and Zoltan Trocsanyi for useful suggestions.
Furthermore, we thank the anonimous Referee for attracting our attention towards the SS cross-section contributions.
This research was supported in part by the National Science Foundation under Grant No. NSF PHY-1748958, by grant K 125105 of the National Research, Development and Innovation Fund in Hungary, and by the Bundesministerium f\"ur Bildung und Forschung (contract 05H18GUCC1). 
This publication is based upon work from COST Action CA16201 PARTICLEFACE,
supported by COST (European Cooperation in Science and Technology, 
\texttt{www.cost.eu}).

\cleardoublepage

\appendix

\section{Conversion to the decoupling scheme}

In order to use matrix-elements in the decoupling scheme together with PDFs and $\alpha_S$ in different 
factorization/renormalization schemes, scheme conversions have to be made to get consistent predictions.

In particular we would like to make use of PDFs and $\alpha_S$ 
in the \msbar{} factorization scheme, with a variable number of flavours, depending on the scale. 

The conversion between the two schemes is process dependent and the basis for our derivation can be found in Ref.~\cite{Cacciari:1998it} for a
different process. In our derivation we follow the same steps as in Ref.~\cite{Cacciari:1998it}, 
but we adopt a slightly different notation. To fix notation we assume that the \msbar{} scheme calculation uses $n_f$ massless
quark flavours while in the decoupling scheme we have $\nft = \nf - n$ massless quark flavours, where
$n$ is the number of massive quark flavours we would like to decouple. In commonly used variable flavour number PDF + $\alpha_S$ sets, 
the number of massive quarks which decouple for scales below each threshold scale is just one, 
i.e. massive quarks decouple one at a time, but other solutions would indeed be possible, e.g. 3~+~2 schemes. 
The strong couplings should coincide at the threshold scale made equal to the mass
$m$ of the lightest massive quark we would like to decouple:
\begin{align}
  \alphas{\nf}{(m)} = \alphas{\nft}{(m)} + \mathcal{O}(\alphas{3}{})
  \,,
  \label{eq:MSbarDecouple1}
\end{align}
For both \alphas{}{} definitions the Renormalization Group Equation (RGE) can be used to evolve  them to a different scale:
\begin{align}
  \alphas{\nft}{(\mu_R)} &= \alphas{\nft}{(m)} - b_0^{(\nft)}\alphas{2}{}\log\frac{\mu_R^2}{m^2}
  \,,\nonumber\\
  \alphas{\nf}{(\mu_R)} &= \alphas{\nft}{(m)} - b_0^{(\nf)}\alphas{2}{}\log\frac{\mu_R^2}{m^2}
  \,,
  \label{eq:evolve}
\end{align}
with
\begin{align}
  b_0^{(\nf)} &= \frac{11 C_A - 4\nf T_R}{12\pi}
  \,.
\end{align}
In the equations listed in~(\ref{eq:evolve}) the scale and scheme of the strong coupling in the second terms on 
the right hand sides are arbitrary due to differences being introduced are of higher order. Using these formulae 
we can express the strong coupling in the decoupling scheme in terms of the one in the \msbar{} scheme
first by exploiting Eq.~\eqref{eq:MSbarDecouple1}, then substituting the coupling in the \msbar{} scheme at 
scale $\mu_R$ using the RGE:
\begin{align}
  \alphas{(\nft)}{(\mu_R)} &= \alphas{(\nf)}{(\mu_R)} 
    + \alphas{2}{}\left[b_0^{(\nf)} - b_0^{(\nft)}\right]\log\frac{\mu_R^2}{m^2} =
    \nonumber\\
  &=  \alphas{(\nf)}{(\mu_R)} 
    - \alphas{2}{}\frac{n T_R}{3\pi}\log\frac{\mu_R^2}{m^2} 
  \,,
  \label{eq:MSbarDecouple2}
\end{align}
where in the second term on the right hand side the coupling can be evaluated in any scheme and at any scale because the 
introduced difference is of order $\mathcal{O}(\alphas{3}{})$, thus beyond the perturbative order we are interested in.

If the set of quarks treated as massive in the decoupling scheme is denoted as $\mathcal{H}$ the structure function in the \msbar{} scheme fulfills 
\begin{align}
  F_j^{(\nf)}(x, m) &= F_j^{(\nft)}(x, m)\,,\quad j\not\in\mathcal{H}
  \,,
  \nonumber\\
  F_h^{(\nf)}(x, m_h) &= F_{\overline{h}}^{(\nf)}(x, m_h) = 0\,,\quad h\in \mathcal{H}
  \,,
  \label{eq:StructFunMatch}
\end{align}
where $m$ stands for the mass of the lightest heavy quark in set $\mathcal{H}$. To ease up
notation we assume $\mathcal{H}$ has only one member with identifier $h$ and with mass $m$.
To be still valid in case of multiple massive quarks in $\mathcal{H}$, we keep $n$ to be 
general and assume that $m$ stands for the lightest quark in set $\mathcal{H}$. To find the
compensation terms between the \msbar{} and decoupling schemes the DGLAP equation has to be
considered:
\begin{align}
  \frac{\partial}{\partial\log\mu^2}F_i^{(\nf)}(x, \mu) &= 
    \frac{\alphas{(\nf)}{(\mu)}}{2\pi}\sum_j 
    \int_x^1\frac{\ud z}{z} F_j^{(\nf)}(x/z, \mu) \mathcal{P}_{ij}^{(\nf)}(z)
  \,,
\end{align}
where $\mathcal{P}^{(\nf)}_{ij}(z)$ is the Altarelli-Parisi (AP) kernel for $j\to i$ splitting considering $\nf$ light
quark flavours at momentum fraction $z$. Taking the same assumption as in Ref.~\cite{Cacciari:1998it}, i.e.
the mass of the lightest massive quark is of the same order of the scale, we can
drop $\mathcal{O}(\alphas{2}{})$ terms and the integration of the DGLAP equation between the scale $m$ and
the scale $\mu$ is simply the integration of the coupling prefactor:
\begin{align}
  \int_{\log m^2}^{\log\mu^2}\ud\log\mu^2\alphas{(\nf)}{(\mu)} &=
   \int_{\log m^2}^{\log\mu^2}\ud\log\mu^2\left[\alphas{(\nf)}{(m)} + \alphas{2}{}(\dots)\right] =
   \nonumber\\
   &= \alphas{(\nf)}{(m)}\log\frac{\mu^2}{m^2} + \mathcal{O}(\alphas{2}{})
   \,.
\end{align}
Using this, the integration of the DGLAP equation is trivial:
\begin{align}
  F_i^{(\nf)}(x,\mu) - F_i^{(\nf)}(x,m) &= \frac{\alphas{(\nf)}{(m)}\log\frac{\mu^2}{m^2}}{2\pi}
    \sum_{j}\int_x^1\frac{\ud z}{z} F_j^{(\nf)}(x/z, m)\mathcal{P}_{ij}^{(\nf)}(z)
  \,.
\end{align}
Because the structure function for a massive quark vanishes at scales below the
mass of the quark, the right hand side can be written as:
\begin{align}
  F_i^{(\nf)}(x,\mu) - F_i^{(\nf)}(x,m) &= \frac{\alphas{(\nf)}{(m)}\log\frac{\mu^2}{m^2}}{2\pi}
    \sum_{j\not\in\mathcal{H}}\int_x^1\frac{\ud z}{z} F_j^{(\nf)}(x/z, m)\mathcal{P}_{ij}^{(\nf)}(z)
  \,.
\end{align}
A similar formula can be derived in the decoupling scheme using the corresponding \alphas{}{} and
structure function definitions:
\begin{align}
  F_i^{(\nft)}(x,\mu) - F_i^{(\nft)}(x,m) &= \frac{\alphas{(\nft)}{(m)}\log\frac{\mu^2}{m^2}}{2\pi}
    \sum_{j\not\in\mathcal{H}}\int_x^1\frac{\ud z}{z} F_j^{(\nft)}(x/z, m)\mathcal{P}_{ij}^{(\nft)}(z)
  \,.
  \label{eq:MSbarDecouple3}
\end{align}
In case of the integrated DGLAP equation for the \msbar{} scheme we can exploit 
Eq.\eqref{eq:StructFunMatch} and Eq.\eqref{eq:MSbarDecouple1} to write:
\begin{align}
  F_i^{(\nf)}(x,\mu) - F_i^{(\nft)}(x,m) &= \frac{\alphas{(\nft)}{(m)}\log\frac{\mu^2}{m^2}}{2\pi}
    \sum_{j\not\in\mathcal{H}}\int_x^1\frac{\ud z}{z} F_j^{(\nft)}(x/z, m)\mathcal{P}_{ij}^{(\nf)}(z)
  \,.
  \label{eq:MSbarDecouple4}
\end{align}
Hence the difference between Eq.\eqref{eq:MSbarDecouple4} and Eq.\eqref{eq:MSbarDecouple3} takes the 
form of:
\begin{align}
  F_i^{(\nf)}(x,\mu) - & F_i^{(\nft)}(x,\mu) = \nonumber\\
   = &
  \frac{\alphas{(\nft)}{(m)}\log\frac{\mu^2}{m^2}}{2\pi}
    \sum_{j\not\in\mathcal{H}}\int_x^1\frac{\ud z}{z} F_j^{(\nft)}(x/z, m)
    \left[\mathcal{P}_{ij}^{(\nf)}(z) - \mathcal{P}_{ij}^{(\nft)}(z)\right]
  \,.
\end{align}
Taking a look at the AP kernels, like Eq.(2.33) in Ref.\cite{Campbell:2017hsr}, reveals that only
the $g\to g$ splitting kernel depends on the number of light quark flavours such that the difference
is
\begin{align}
  \mathcal{P}^{(\nf)}_{gg}(z) - \mathcal{P}^{(\nft)}_{gg}(z) &=
    -\frac{2n}{3}T_R\delta(1 - z)
  \,.
\end{align}
Thus, we can conclude that
\begin{align}
  F_j^{(\nf)}(x, \mu) - F_j^{(\nft)}(x, \mu) &= 0\,,\quad j\ne g,\,j\not\in\mathcal{H}
  \,,\nonumber\\
  F_g^{(\nf)}(x, \mu) - F_g^{(\nft)}(x, \mu) &= 
    -\frac{2n}{3}T_R\frac{\alphas{(\nft)}{(m)}\log\frac{\mu^2}{m^2}}{2\pi}
    F_g^{(\nft)}(x, m)
  \,.
\end{align}
Using
\begin{align}
  \alphas{}{} F_g^{(\nft)}(x, m) &= \alphas{}{} F_g^{(\nft)}(x, \mu) + \mathcal{O}(\alphas{2}{})
  \,,
\end{align}
we find
\begin{align}
  F_j^{(\nf)}(x, \mu) &= F_j^{(\nft)}(x, \mu) + \mathcal{O}(\alphas{2}{})\,,
    \quad j\ne g\,j\not\in\mathcal{H}\,,
  \nonumber\\
  F_g^{(\nf)}(x, \mu) &= F_g^{(\nft)}(x, \mu)\left(1 - \alphas{(\nft)}{(m)}\frac{n T_R}{3\pi}
    \log\frac{\mu^2}{m^2}\right) 
    + \mathcal{O}(\alphas{2}{})
  \,.
  \label{eq:MSbarDecouple5}
\end{align}
In case of $W+c$ production the Born level partonic processes can only have initial states 
$\overset{(-)}{q} g$ and $g \overset{(-)}{q}$. Due to symmetry and to ease notation we only
consider the $q g$ channel. The integrand of the convolution integral of the
Collinear Factorization Theorem 
reads in the \msbar{} scheme
\begin{align}
  F_{q}^{(\nf)}(x_1, \mu_F) F_{g}^{(\nf)}(x_1, \mu_F) \sigma_{qg}^{(0)} =
  F_{q}^{(\nf)}(x_1, \mu_F) F_{g}^{(\nf)}(x_1, \mu_F)\,\alphas{(\nf)}{(\mu_R)}\,f_{qg}^{(0)}
  \,,
\end{align}
where $f_{qg}^{(0)}$ is the \msbar{} partonic cross section for the initial state $q\,g$
with strong coupling factorized out. To turn the Factorization Theorem from the \msbar{} scheme to a form valid in the
decoupling scheme, Eqs.\eqref{eq:MSbarDecouple2} and \eqref{eq:MSbarDecouple5} can be used:
\begin{align}
  & F_{q}^{(\nf)}(x_1, \mu_F) F_{g}^{(\nf)}(x_1, \mu_F)\,\alphas{(\nf)}{(\mu_R)}\,f_{qg}^{(0)} =
  F_{q}^{(\nft)}(x_1, \mu_F) F_{g}^{(\nft)}(x_1, \mu_F)\,\alphas{(\nft)}{(\mu_R)}\,f_{qg}^{(0)}
  \cdot 
  \nonumber\\
  &\cdot \left(1 - \alphas{(\nft)}{(m)}\frac{n T_R}{3\pi}\log\frac{\mu_F^2}{m^2}\right)
  \cdot \left(1 + \alphas{(\nft)}{(m)}\frac{n T_R}{3\pi}\log\frac{\mu_R^2}{m^2}\right) =
  \nonumber\\
  &= \left(1 + \alphas{(\nft)}{(m)}\frac{n T_R}{3\pi}\log\frac{\mu_R^2}{\mu_F^2}\right)
  F_{q}^{(\nft)}(x_1, \mu_F) F_{g}^{(\nft)}(x_1, \mu_F)\,\alphas{(\nft)}{(\mu_R)}\,f_{qg}^{(0)}
  \,.
\end{align}
This expression allows us to use PDFs and $\alpha_S$ with $\nft$ light quark flavours in the 
decoupling scheme together with a partonic cross-section in the \msbar{} scheme.  Or, by multiplying 
with the inverse of the prefactor and expanding it perturbatively,  it allows for using a PDF and corresponding
\alphas{}{} defined in the \msbar{} scheme in a calculational framework defined in the decoupling scheme.

Note that as long as the non-physical scales coincide the compensation term is identically zero, so it is only needed for scale
variations if $\mu_R = \mu_F$ and in general when different choices are made for the factorization and 
renormalization scales. Note also that the compensation term is of order $\mathcal{O}(\alphas{2}{})$ hence it is the same order as the NLO correction
to the partonic cross-section. Because of this, the compensation only affects the Born contribution in the NLO calculation, where terms of order 
$\mathcal{O}(\alpha_S^3)$ are dropped.

\section{Contributions of OS and SS events to the fiducial cross-sections}

As already mentioned in Section~\ref{sec:pheno}, in both the ATLAS and CMS experimental ana\-ly\-ses that we considered in this work, 
results for differential cross sections are defined as the difference between the so-called opposite-sign and same-sign contributions 
(denoted as ``OS'' and ``SS'', respectively). This applies to both the experimental data points and to the corresponding theoretical predictions. 
OS contributions are those where the reconstructed $W$ boson and the selected $D$ meson (or $c$-jet, $j_c$) have opposite charges, 
e.g. $W^\pm + D^\mp$. Conversely, SS contributions are those where the reconstructed $W$ boson and the selected $D$ meson (or $j_c$) 
have the same charge, e.g. $W^\pm+D^\pm$. The (OS~-~SS) subtraction has been introduced by the experimental analyses 
with the goal of reducing contaminations from various backgrounds to the $W^+ \bar{c}$ ($W^- c$) signal, 
which gives rise to hadron-level events mostly populating the OS sample/cross section.  
This subtraction procedure comes with a cost, meaning that a fraction of $W^+ \bar{c}$ ($W^- c$) 
signal events can be possibly categorised as SS after shower evolution. As we will see, this effect is far from being negligible.
In fact, 
$W^+ \bar{c}$ ($W^- c$) production at the underlying-Born level, after one or more real emissions, can give rise to
events with multiple charm quarks/hadrons, which in turn have the potential to populate either the OS or the SS sample, 
depending on the final state content and kinematics. In particular,
configurations with multiple charm quarks can appear at two stages:
\begin{itemize}
\item[1] Already at the NLO/LHE level, due to the fact that $W^+ c \bar{c}$ ($W^- c \bar{c}$) 
  Feynman diagrams give rise to a real correction contribution to $W^+ \bar{c}$ ($W^- c$) production. 
\item[2] After parton shower, i.e. even $W^+ \bar{c}$ ($W^- c$) underlying Born configurations for which the 
  first real emission  does not bring in any additional $c$-quark ($\bar{c}$-antiquark) (i.e. {\texttt{PowHel/POWHEGBOX}} 
    Les Houches events with a single $c$ (or $\bar{c}$) quark at the partonic level), 
    can still give rise to final states with more than one $c$ (or $\bar{c}$) quark due to parton shower effects. 
    These translate into hadron-level events with multiple charm hadrons. 
\end{itemize}

In Table \ref{tab:anatomy_CMS} we report our findings for the fiducial cross section of 
$W^\pm + D^*(2010)$-meson production according to the CMS analysis at $\sqrt{s}$ = 13 TeV~\cite{Sirunyan:2018hde}, 
with the (OS~-~SS) central predictions, obtained integrating those reported at the differential level in 
Fig.~\ref{fig:dmeson-cms} of Section~\ref{sub:cms}, splitted into OS and SS contributions. The label ``$W^\pm c + X$'' 
denotes results obtained using the full $W^+ \bar{c}$ and $W^- c$ LHE samples, whereas the ``$W^+ c\,\bar{c}\,$'' and ``$W^- c\,\bar{c}\,$'' categories restrict to the
contribution to the $W^+ \bar{c}$ and $W^- c$  predictions, originated by the subset of 
events including a $c\bar{c}$-pair at the LHE level (i.e. corresponding to case 1 listed above), 
analysed after having gone through parton shower, hadronization, MPI and beam remnant effects, 
accounted for by the \texttt{PowHel/POWHEGBOX} LHE interface to {\texttt{PYTHIA8}} using the Monash tune. 

First of all, we observe that the SS (OS) cross-section contribution corresponding to ``$W^+ c \bar{c}$'' 
events is larger than the SS (OS)  ``$W^- c \bar{c}$'' one, which reflects differences between the valence 
$u$ and $d$ quark distributions in the proton. On the other hand, for each $W^\pm c\,\bar{c}$ category the 
OS and SS cross-sections have equal size at the LHE level. OS and SS cross-sections turn out to be fully 
compatible among each other, within the quoted statistical uncertainties, also after
including the full SMC perturbative and non-perturbative physics effects. 
Hence we can conclude that the (OS~-~SS) subtraction effectively suppresses the contributions induced by the 
$pp \to W^\pm c\,\bar{c}$ LHE events present in our $W^+ c$ and $W^- \bar{c}$ event samples. 

Additionally, it appears clear that SS contributions from $W^\pm c\,\bar{c}$ represent only a fraction of the 
total SS background: the rest comes from final states induced by $W$~+~single charm Les Houches events, 
which turn into SS background after shower evolution. This is a genuine effect of the parton shower that 
cannot be estimated with a fixed-order calculation, where all the ($W$ + single charm + $X$) parton-level final states, 
with $X$ not including any further charm quark, are categorized as OS. 

Overall, the size of the SS contributions from $W^\pm c + X$ (\textit{i.e.} the full sample) amounts to 
$11\%-13\%$ of the (OS~-~SS) fiducial cross section we obtain after applying the CMS analysis cuts. 
As already mentioned, this is far from being negligible, implying that accurate estimates of the SS 
cross section are mandatory for realistic comparisons of theory predictions with the available (OS~-~SS) experimental data.

Our findings for the ATLAS analysis at 7 TeV for the $W^\pm + D$-meson and $W^\pm~+~j_c (j_{\bar{c}})$ 
cases are reported in Tables \ref{tab:anatomy_ATLAS_had} and  \ref{tab:anatomy_ATLAS_jet}, respectively, 
corresponding to the central predictions shown in Fig.~\ref{fig:dmeson-atlas-21} and \ref{fig:cjet-atlas-21} 
of Section~\ref{sub:atlas}, obtained with \texttt{PowHel~+~PYTHIA8} with the ATLAS A14 tune. 
The conclusions that can be drawn from these tables are qualitatively the same as the ones for the 
CMS case discussed above, therefore we will not repeat them here. However, 
it is worth noting that the relative size of the SS contributions is smaller for ATLAS than for the CMS case, 
amounting to $4\%-6\%$ ($2\%-3\%$) of the (OS~-~SS) cross section for the $W^\pm+D^\mp$-meson ($W^\pm + j_c (j_{\bar{c}})$) ATLAS analyses. 
Thus, the relative importance of the SS background appears to increase with the 
center-of-mass energy and the use of looser cuts (indeed, the applied $p_{T,D}$ and $\eta_D$ 
cuts are more inclusive in the CMS analysis, than in the ATLAS ones, see Section~\ref{sec:pheno} 
for more detail) and to decrease when shifting from the hadron-level ($W^\pm + D^\mp$) to the jet-level 
($W^\pm+j_c (j_{\bar{c}})$) viewpoint. The first observation is understood as a consequence of the fact that the 
LHE partons enter the shower evolution with larger energies on the average, and the probability of OS~$\rightarrow$~SS 
conversions increases as a consequence of the larger available phase space for shower activity. 
The second fact looks also reasonable considering that charmed jets are more inclusive observables compared to the 
individual hadrons and thus genuine shower effects like the ones that lead to OS~$\rightarrow$~SS event conversions should have a smaller impact.

To conclude this Section, we emphasize that fixed-order calculations, at least at the level of 
perturbative accuracy that we are considering in this work, 
do not allow for a correct SS cross-section evaluation for the considered experimental configurations.
In fact, a substantial fraction of the total SS contribution is instead generated by 
PS effects on top of LHE events including one and only one $c$ (or $\bar{c}$) quark, 
as follows from comparing the ``$W c$'' and ``$W c \bar{c}\,$'' SS categories for each experimental configuration. 
This is particularly important in case of the $W^\pm + D^\mp$-meson analyses considered here, 
whereas the effect is smaller, but still non negligible, in case of the $W^\pm + j_c (j_{\bar{c}})$ study. 
Therefore, PDF collaborations aiming at including $W + c$ production data in their NLO or NNLO fits, 
should incorporate or, at least, account for these PS effects in their QCD analyses. 
Fixed-order predictions have the tendency to underestimate the SS background. This, ultimately, 
might impact the extraction of the strange quark PDFs, leading to a possible underestimate of the latter.

\begin{table}[h!]
\begin{center}
  \begin{tabular}{c|c|c|c}
    \hline \hline\\[-0.4cm]
  LHE Partonic Process  & $\sigma^{\rm OS-SS}_{\rm MC}$ [pb] & $\sigma^{\rm OS}_{\rm MC}$ [pb] & $\sigma^{\rm SS}_{\rm MC}$ [pb]  \\[0.2cm]
    \hline\hline\\[-0.4cm]
     $W^{+} \, c \, + X$  & 62(1) & 70(1) & 7.8(4) \\[0.2cm]
     $W^{-}  \, c \, + X$  & 66(1) &  73(1) & 7.3(4) \\[0.2cm]
     $W^{\pm}  \, c \, + X$ &  128(2)  & 143(2) & 15.1(6) \\[0.2cm]
     \hline
     $W^{+} \, c\,\bar{c}$  & -0.1(1) &  1.4(1) & 1.5(1) \\[0.2cm]
     $W^{-}  \, c\,\bar{c}$  & 0.0(1) &  0.9(1) & 0.9(1) \\[0.2cm]
     $W^{\pm}  \, c\,\bar{c}$  & -0.1(2) & 2.3(2) & 2.4(2) \\[0.2cm]
   \hline \hline
\end{tabular}
\end{center}
\caption{ \texttt{PowHel + PYTHIA8} predictions for the OS~-~SS, OS and SS  inclusive fiducial cross-sections for $W^\pm + D^*(2010)$-meson
production and their sum, with cuts according to the  $\sqrt{s}=13$ TeV CMS analysis of Ref.~\cite{Sirunyan:2018hde}, already considered in Section~\ref{sub:cms}. Parton shower, hadronization, MPI and beam remnant effects are accounted for using the Monash {\texttt{PYTHIA8}} tune. The first three lines are obtained from the use of our full $W^+  \bar{c}$ and $W^- c$ event samples, whereas in the last three lines the hadron-level contributions due to $W^+ \bar{c}$ and $W^- c$  LHE events including a $c\bar{c}$ pair already at the first radiation emission level (case~1 discussed in the text) are singled out.
  Predictions in the first column, first three lines, correspond to the integration over lepton rapidity of the central predictions shown in the three panels of Fig.~\ref{fig:dmeson-cms}, obtained using the ABMP16 PDF set. The quoted uncertainties account for the degree of statistical accuracy of the LHE files used for this study, containing a few million events.}
\label{tab:anatomy_CMS}
\end{table}

\begin{table}[h!]
\begin{center}
  \begin{tabular}{c|c|c|c}
    \hline \hline\\[-0.4cm]
   Process  & $\sigma^{\rm OS-SS}_{\rm MC}$ [pb] & $\sigma^{\rm OS}_{\rm MC}$ [pb] & $\sigma^{\rm SS}_{\rm MC}$ [pb]  \\[0.2cm]
    \hline\hline\\[-0.4cm]
     $W^{+} \, c \, + X$  & 19.2(3) & 20.4(3) & 1.19(6) \\[0.2cm]
     $W^{-}  \, c \, + X$  & 20.7(3) & 21.7(3) & 0.93(6) \\[0.2cm]
     \hline
     $W^{+} \, c\,\bar{c}$  & 0.05(6)  &  0.68(4) & 0.63(4) \\[0.2cm]
     $W^{-}  \, c\,\bar{c}$  &  -0.03(4)  &  0.36(3) &  0.39(3) \\[0.2cm]
   \hline \hline
\end{tabular}
\end{center}
\caption{
  Same as Table~\ref{tab:anatomy_CMS}, but for $W^\pm + D$-meson
  production according to the $\sqrt{s}=7$~TeV ATLAS analysis of Ref.~\cite{Aad:2014xca}, with \texttt{PYTHIA8} configured according to the ATLAS A14 central tune.}
\label{tab:anatomy_ATLAS_had}
\end{table}

\begin{table}[h!]
\begin{center}
  \begin{tabular}{c|c|c|c}
    \hline \hline\\[-0.4cm]
   Process  & $\sigma^{\rm OS-SS}_{\rm MC}$ [pb] & $\sigma^{\rm OS}_{\rm MC}$ [pb] & $\sigma^{\rm SS}_{\rm MC}$ [pb]  \\[0.2cm]
    \hline\hline\\[-0.4cm]
     $W^{+} \, c \, + X$  & 31.8(4) & 32.9(4) & 1.06(6) \\[0.2cm]
     $W^{-}  \, c \, + X$  & 34.6(4) &  35.3(4) & 0.67(4) \\[0.2cm]
     \hline
     $W^{+} \, c\,\bar{c}$  & 0.03(6)  &  0.78(4)  &  0.75(4)  \\[0.2cm]
     $W^{-}  \, c\,\bar{c}$  &  0.03(4)  &  0.41(3) &  0.38(3)  \\[0.2cm]
   \hline \hline
\end{tabular}
\end{center}
\caption{Same as Table~\ref{tab:anatomy_CMS}, but for $W^\pm + j_c (j_{\bar{c}})$  production according to the $\sqrt{s}=7$~TeV ATLAS analysis of Ref.~\cite{Aad:2014xca}, with \texttt{PYTHIA8} configured according to the ATLAS A14 central tune.}
\label{tab:anatomy_ATLAS_jet}
\end{table}

\cleardoublepage

\bibliographystyle{JHEP} 
\bibliography{paperwc}

\providecommand{\href}[2]{#2}\begingroup\raggedright\begin{thebibliography}{10}

\bibitem{Accardi:2016ndt}
A.~Accardi {\em et.~al.}, {\it {A Critical Appraisal and Evaluation of Modern
  PDFs}},  {\em Eur. Phys. J.} {\bf C76} (2016), no.~8 471,
  [\href{http://xxx.lanl.gov/abs/1603.08906}{{\tt arXiv:1603.08906}}].

\bibitem{Zenaiev:2015rfa}
{\bf PROSA} Collaboration, O.~Zenaiev {\em et.~al.}, {\it {Impact of
  heavy-flavour production cross sections measured by the LHCb experiment on
  parton distribution functions at low x}},  {\em Eur. Phys. J.} {\bf C75}
  (2015), no.~8 396, [\href{http://xxx.lanl.gov/abs/1503.04581}{{\tt
  arXiv:1503.04581}}].

\bibitem{Bertone:2018dse}
V.~Bertone, R.~Gauld, and J.~Rojo, {\it {Neutrino Telescopes as QCD
  Microscopes}},  {\em JHEP} {\bf 01} (2019) 217,
  [\href{http://xxx.lanl.gov/abs/1808.02034}{{\tt arXiv:1808.02034}}].

\bibitem{Zenaiev:2019ktw}
{\bf PROSA} Collaboration, O.~Zenaiev, M.~V. Garzelli, K.~Lipka, S.~O. Moch,
  A.~Cooper-Sarkar, F.~Olness, A.~Geiser, and G.~Sigl, {\it {Improved
  constraints on parton distributions using LHCb, ALICE and HERA heavy-flavour
  measurements and implications for the predictions for prompt
  atmospheric-neutrino fluxes}},  {\em JHEP} {\bf 04} (2020) 118,
  [\href{http://xxx.lanl.gov/abs/1911.13164}{{\tt arXiv:1911.13164}}].

\bibitem{Alekhin:2014sya}
S.~Alekhin, J.~Blumlein, L.~Caminadac, K.~Lipka, K.~Lohwasser, S.~Moch,
  R.~Petti, and R.~Placakyte, {\it {Determination of Strange Sea Quark
  Distributions from Fixed-target and Collider Data}},  {\em Phys. Rev.} {\bf
  D91} (2015), no.~9 094002, [\href{http://xxx.lanl.gov/abs/1404.6469}{{\tt
  arXiv:1404.6469}}].

\bibitem{Alekhin:2017olj}
S.~Alekhin, J.~Blumlein, and S.~Moch, {\it {Strange sea determination from
  collider data}},  {\em Phys. Lett.} {\bf B777} (2018) 134--140,
  [\href{http://xxx.lanl.gov/abs/1708.01067}{{\tt arXiv:1708.01067}}].

\bibitem{Faura:2020oom}
F.~Faura, S.~Iranipour, E.~R. Nocera, J.~Rojo, and M.~Ubiali, {\it {The
  Strangest Proton?}},  {\em Eur. Phys. J.} {\bf C80} (2020), no.~12 1168,
  [\href{http://xxx.lanl.gov/abs/2009.00014}{{\tt arXiv:2009.00014}}].

\bibitem{Abdolmaleki:2019acd}
{\bf xFitter Developers' Team} Collaboration, H.~Abdolmaleki {\em et.~al.},
  {\it {Probing the strange content of the proton with charm production in
  charged current at LHeC}},  {\em Eur. Phys. J.} {\bf C79} (2019), no.~10 864,
  [\href{http://xxx.lanl.gov/abs/1907.01014}{{\tt arXiv:1907.01014}}].

\bibitem{FPF:2021}
M.~V. Garzelli, ``{\it{QCD opportunities with forward neutrinos during the
  HL-LHC phase}}.'' Poster presentation at the APS 2021 April Meeting Quarks to
  Cosmos, 2021.

\bibitem{Sirunyan:2019qia}
{\bf CMS} Collaboration, A.~M. Sirunyan {\em et.~al.}, {\it {A search for the
  standard model Higgs boson decaying to charm quarks}},  {\em JHEP} {\bf 03}
  (2020) 131, [\href{http://xxx.lanl.gov/abs/1912.01662}{{\tt
  arXiv:1912.01662}}].

\bibitem{ATLAS:2021zwx}
{\bf ATLAS} Collaboration, {\it {Direct constraint on the Higgs-charm coupling
  from a search for Higgs boson decays to charm quarks with the ATLAS
  detector}}, .

\bibitem{Iwamoto:2017ytj}
S.~Iwamoto, G.~Lee, Y.~Shadmi, and Y.~Weiss, {\it {Tagging new physics with
  charm}},  {\em JHEP} {\bf 09} (2017) 114,
  [\href{http://xxx.lanl.gov/abs/1703.05748}{{\tt arXiv:1703.05748}}].

\bibitem{Aaboud:2017phn}
{\bf ATLAS} Collaboration, M.~Aaboud {\em et.~al.}, {\it {Search for dark
  matter and other new phenomena in events with an energetic jet and large
  missing transverse momentum using the ATLAS detector}},  {\em JHEP} {\bf 01}
  (2018) 126, [\href{http://xxx.lanl.gov/abs/1711.03301}{{\tt
  arXiv:1711.03301}}].

\bibitem{Sirunyan:2017xgm}
{\bf CMS} Collaboration, A.~M. Sirunyan {\em et.~al.}, {\it {Search for dark
  matter produced in association with heavy-flavor quark pairs in proton-proton
  collisions at $\sqrt{s}=13$ TeV}},  {\em Eur. Phys. J. C} {\bf 77} (2017),
  no.~12 845, [\href{http://xxx.lanl.gov/abs/1706.02581}{{\tt
  arXiv:1706.02581}}].

\bibitem{Branco:2011iw}
G.~C. Branco, P.~M. Ferreira, L.~Lavoura, M.~N. Rebelo, M.~Sher, and J.~P.
  Silva, {\it {Theory and phenomenology of two-Higgs-doublet models}},  {\em
  Phys. Rept.} {\bf 516} (2012) 1--102,
  [\href{http://xxx.lanl.gov/abs/1106.0034}{{\tt arXiv:1106.0034}}].

\bibitem{Aaltonen:2007dm}
{\bf CDF} Collaboration, T.~Aaltonen {\em et.~al.}, {\it {First measurement of
  the production of a $W$ boson in association with a single charm quark in $p
  \bar{p}$ collisions at $\sqrt{s}$ = 1.96-TeV}},  {\em Phys. Rev. Lett.} {\bf
  100} (2008) 091803, [\href{http://xxx.lanl.gov/abs/0711.2901}{{\tt
  arXiv:0711.2901}}].

\bibitem{Abazov:2008qz}
{\bf D0} Collaboration, V.~M. Abazov {\em et.~al.}, {\it {Measurement of the
  ratio of the $p \bar{p} \to W^+ c^-$ jet cross section to the inclusive $p
  \bar{p} \to W +$ jets cross section}},  {\em Phys. Lett. B} {\bf 666} (2008)
  23--30, [\href{http://xxx.lanl.gov/abs/0803.2259}{{\tt arXiv:0803.2259}}].

\bibitem{Aaltonen:2012wn}
{\bf CDF} Collaboration, T.~Aaltonen {\em et.~al.}, {\it {Observation of the
  Production of a W Boson in Association with a Single Charm Quark}},  {\em
  Phys. Rev. Lett.} {\bf 110} (2013), no.~7 071801,
  [\href{http://xxx.lanl.gov/abs/1209.1921}{{\tt arXiv:1209.1921}}].

\bibitem{CMS:2021oxn}
{\bf CMS} Collaboration, A.~Tumasyan {\em et.~al.}, {\it {Measurements of the
  associated production of a W boson and a charm quark in proton-proton
  collisions at $\sqrt{s}$ = 8 TeV}},
  \href{http://xxx.lanl.gov/abs/2112.00895}{{\tt arXiv:2112.00895}}.

\bibitem{Chatrchyan:2013uja}
{\bf CMS} Collaboration, S.~Chatrchyan {\em et.~al.}, {\it {Measurement of
  Associated W + Charm Production in pp Collisions at $\sqrt{s}$ = 7 TeV}},
  {\em JHEP} {\bf 02} (2014) 013,
  [\href{http://xxx.lanl.gov/abs/1310.1138}{{\tt arXiv:1310.1138}}].

\bibitem{Aad:2014xca}
{\bf ATLAS} Collaboration, G.~Aad {\em et.~al.}, {\it {Measurement of the
  production of a $W$ boson in association with a charm quark in $pp$
  collisions at $\sqrt{s} =$ 7 TeV with the ATLAS detector}},  {\em JHEP} {\bf
  05} (2014) 068, [\href{http://xxx.lanl.gov/abs/1402.6263}{{\tt
  arXiv:1402.6263}}].

\bibitem{Sirunyan:2018hde}
{\bf CMS} Collaboration, A.~M. Sirunyan {\em et.~al.}, {\it {Measurement of
  associated production of a W boson and a charm quark in proton-proton
  collisions at $\sqrt{s} =$ 13 TeV}},  {\em Eur. Phys. J.} {\bf C79} (2019),
  no.~3 269, [\href{http://xxx.lanl.gov/abs/1811.10021}{{\tt
  arXiv:1811.10021}}].

\bibitem{Aaij:2015cha}
{\bf LHCb} Collaboration, R.~Aaij {\em et.~al.}, {\it {Study of $W$ boson
  production in association with beauty and charm}},  {\em Phys. Rev. D} {\bf
  92} (2015), no.~5 052001, [\href{http://xxx.lanl.gov/abs/1505.04051}{{\tt
  arXiv:1505.04051}}].

\bibitem{Giele:1995kr}
W.~T. Giele, S.~Keller, and E.~Laenen, {\it {QCD corrections to $W$ boson plus
  heavy quark production at the Tevatron}},  {\em Phys. Lett. B} {\bf 372}
  (1996) 141--149, [\href{http://xxx.lanl.gov/abs/hep-ph/9511449}{{\tt
  hep-ph/9511449}}].

\bibitem{Campbell:2015qma}
J.~M. Campbell, R.~K. Ellis, and W.~T. Giele, {\it {A Multi-Threaded Version of
  MCFM}},  {\em Eur. Phys. J. C} {\bf 75} (2015), no.~6 246,
  [\href{http://xxx.lanl.gov/abs/1503.06182}{{\tt arXiv:1503.06182}}].

\bibitem{Stirling:2012vh}
W.~J. Stirling and E.~Vryonidou, {\it {Charm production in association with an
  electroweak gauge boson at the LHC}},  {\em Phys. Rev. Lett.} {\bf 109}
  (2012) 082002, [\href{http://xxx.lanl.gov/abs/1203.6781}{{\tt
  arXiv:1203.6781}}].

\bibitem{Denner:2009gj}
A.~Denner, S.~Dittmaier, T.~Kasprzik, and A.~Muck, {\it {Electroweak
  corrections to W + jet hadroproduction including leptonic W-boson decays}},
  {\em JHEP} {\bf 08} (2009) 075,
  [\href{http://xxx.lanl.gov/abs/0906.1656}{{\tt arXiv:0906.1656}}].

\bibitem{Kallweit:2014xda}
S.~Kallweit, J.~M. Lindert, P.~Maierh\"ofer, S.~Pozzorini, and M.~Sch\"onherr,
  {\it {NLO electroweak automation and precise predictions for W+multijet
  production at the LHC}},  {\em JHEP} {\bf 04} (2015) 012,
  [\href{http://xxx.lanl.gov/abs/1412.5157}{{\tt arXiv:1412.5157}}].

\bibitem{Kallweit:2015dum}
S.~Kallweit, J.~M. Lindert, P.~Maierhofer, S.~Pozzorini, and M.~Sch\"onherr,
  {\it {NLO QCD+EW predictions for V + jets including off-shell vector-boson
  decays and multijet merging}},  {\em JHEP} {\bf 04} (2016) 021,
  [\href{http://xxx.lanl.gov/abs/1511.08692}{{\tt arXiv:1511.08692}}].

\bibitem{Biedermann:2017yoi}
B.~Biedermann, S.~Br\"auer, A.~Denner, M.~Pellen, S.~Schumann, and J.~M.
  Thompson, {\it {Automation of NLO QCD and EW corrections with Sherpa and
  Recola}},  {\em Eur. Phys. J. C} {\bf 77} (2017) 492,
  [\href{http://xxx.lanl.gov/abs/1704.05783}{{\tt arXiv:1704.05783}}].

\bibitem{Frixione:2002ik}
S.~Frixione and B.~R. Webber, {\it {Matching NLO QCD computations and parton
  shower simulations}},  {\em JHEP} {\bf 06} (2002) 029,
  [\href{http://xxx.lanl.gov/abs/hep-ph/0204244}{{\tt hep-ph/0204244}}].

\bibitem{Alwall:2014hca}
J.~Alwall, R.~Frederix, S.~Frixione, V.~Hirschi, F.~Maltoni, O.~Mattelaer,
  H.~S. Shao, T.~Stelzer, P.~Torrielli, and M.~Zaro, {\it {The automated
  computation of tree-level and next-to-leading order differential cross
  sections, and their matching to parton shower simulations}},  {\em JHEP} {\bf
  07} (2014) 079, [\href{http://xxx.lanl.gov/abs/1405.0301}{{\tt
  arXiv:1405.0301}}].

\bibitem{Nason:2004rx}
P.~Nason, {\it {A New method for combining NLO QCD with shower Monte Carlo
  algorithms}},  {\em JHEP} {\bf 11} (2004) 040,
  [\href{http://xxx.lanl.gov/abs/hep-ph/0409146}{{\tt hep-ph/0409146}}].

\bibitem{Frixione:2007vw}
S.~Frixione, P.~Nason, and C.~Oleari, {\it {Matching NLO QCD computations with
  Parton Shower simulations: the POWHEG method}},  {\em JHEP} {\bf 11} (2007)
  070, [\href{http://xxx.lanl.gov/abs/0709.2092}{{\tt arXiv:0709.2092}}].

\bibitem{Czakon:2020coa}
M.~Czakon, A.~Mitov, M.~Pellen, and R.~Poncelet, {\it {NNLO QCD predictions for
  W+c-jet production at the LHC}},
  \href{http://xxx.lanl.gov/abs/2011.01011}{{\tt arXiv:2011.01011}}.

\bibitem{Collins:1978wz}
J.~C. Collins, F.~Wilczek, and A.~Zee, {\it {Low-Energy Manifestations of Heavy
  Particles: Application to the Neutral Current}},  {\em Phys. Rev.} {\bf D18}
  (1978) 242.

\bibitem{Cacciari:2008gp}
M.~Cacciari, G.~P. Salam, and G.~Soyez, {\it {The anti-$k_t$ jet clustering
  algorithm}},  {\em JHEP} {\bf 04} (2008) 063,
  [\href{http://xxx.lanl.gov/abs/0802.1189}{{\tt arXiv:0802.1189}}].

\bibitem{Banfi:2006hf}
A.~Banfi, G.~P. Salam, and G.~Zanderighi, {\it {Infrared safe definition of jet
  flavor}},  {\em Eur. Phys. J.} {\bf C47} (2006) 113--124,
  [\href{http://xxx.lanl.gov/abs/hep-ph/0601139}{{\tt hep-ph/0601139}}].

\bibitem{Rojo:2021gdq}
J.~Rojo, {\it {Progress in the NNPDF global analyses of proton structure}},  in
  {\em {55th Rencontres de Moriond on QCD and High Energy Interactions}}, 4,
  2021.
\newblock \href{http://xxx.lanl.gov/abs/2104.09174}{{\tt arXiv:2104.09174}}.

\bibitem{Ball:2021leu}
R.~D. Ball {\em et.~al.}, {\it {The Path to Proton Structure at One-Percent
  Accuracy}},  \href{http://xxx.lanl.gov/abs/2109.02653}{{\tt
  arXiv:2109.02653}}.

\bibitem{Ball:2017nwa}
{\bf NNPDF} Collaboration, R.~D. Ball {\em et.~al.}, {\it {Parton distributions
  from high-precision collider data}},  {\em Eur. Phys. J.} {\bf C77} (2017),
  no.~10 663, [\href{http://xxx.lanl.gov/abs/1706.00428}{{\tt
  arXiv:1706.00428}}].

\bibitem{Bailey:2020ooq}
S.~Bailey, T.~Cridge, L.~A. Harland-Lang, A.~D. Martin, and R.~S. Thorne, {\it
  {Parton distributions from LHC, HERA, Tevatron and fixed target data: MSHT20
  PDFs}},  {\em Eur. Phys. J.} {\bf C81} (2021), no.~4 341,
  [\href{http://xxx.lanl.gov/abs/2012.04684}{{\tt arXiv:2012.04684}}].

\bibitem{Hou:2019efy}
T.-J. Hou {\em et.~al.}, {\it {New CTEQ global analysis of quantum
  chromodynamics with high-precision data from the LHC}},  {\em Phys. Rev.}
  {\bf D103} (2021), no.~1 014013,
  [\href{http://xxx.lanl.gov/abs/1912.10053}{{\tt arXiv:1912.10053}}].

\bibitem{Alioli:2010xd}
S.~Alioli, P.~Nason, C.~Oleari, and E.~Re, {\it {A general framework for
  implementing NLO calculations in shower Monte Carlo programs: the POWHEG
  BOX}},  {\em JHEP} {\bf 06} (2010) 043,
  [\href{http://xxx.lanl.gov/abs/1002.2581}{{\tt arXiv:1002.2581}}].

\bibitem{Bevilacqua:2011xh}
G.~Bevilacqua, M.~Czakon, M.~V. Garzelli, A.~van Hameren, A.~Kardos, C.~G.
  Papadopoulos, R.~Pittau, and M.~Worek, {\it {HELAC-NLO}},  {\em Comput. Phys.
  Commun.} {\bf 184} (2013) 986--997,
  [\href{http://xxx.lanl.gov/abs/1110.1499}{{\tt arXiv:1110.1499}}].

\bibitem{Frixione:1995ms}
S.~Frixione, Z.~Kunszt, and A.~Signer, {\it {Three jet cross-sections to
  next-to-leading order}},  {\em Nucl. Phys. B} {\bf 467} (1996) 399--442,
  [\href{http://xxx.lanl.gov/abs/hep-ph/9512328}{{\tt hep-ph/9512328}}].

\bibitem{Frixione:1997np}
S.~Frixione, {\it {A General approach to jet cross-sections in QCD}},  {\em
  Nucl. Phys. B} {\bf 507} (1997) 295--314,
  [\href{http://xxx.lanl.gov/abs/hep-ph/9706545}{{\tt hep-ph/9706545}}].

\bibitem{Sjostrand:2019zhc}
T.~Sjöstrand, {\it {The PYTHIA Event Generator: Past, Present and Future}},
  {\em Comput. Phys. Commun.} {\bf 246} (2020) 106910,
  [\href{http://xxx.lanl.gov/abs/1907.09874}{{\tt arXiv:1907.09874}}].

\bibitem{Sj_strand_2015}
T.~Sj\"ostrand, S.~Ask, J.~R. Christiansen, R.~Corke, N.~Desai, P.~Ilten,
  S.~Mrenna, S.~Prestel, C.~O. Rasmussen, and P.~Z. Skands, {\it An
  introduction to pythia 8.2},  {\em Computer Physics Communications} {\bf 191}
  (Jun, 2015) 159--177.

\bibitem{Sjostrand:2006za}
T.~Sjostrand, S.~Mrenna, and P.~Z. Skands, {\it {PYTHIA 6.4 Physics and
  Manual}},  {\em JHEP} {\bf 05} (2006) 026,
  [\href{http://xxx.lanl.gov/abs/hep-ph/0603175}{{\tt hep-ph/0603175}}].

\bibitem{Bellm:2019zci}
J.~Bellm {\em et.~al.}, {\it {Herwig 7.2 release note}},  {\em Eur. Phys. J. C}
  {\bf 80} (2020), no.~5 452, [\href{http://xxx.lanl.gov/abs/1912.06509}{{\tt
  arXiv:1912.06509}}].

\bibitem{Corcella:2000bw}
G.~Corcella, I.~G. Knowles, G.~Marchesini, S.~Moretti, K.~Odagiri,
  P.~Richardson, M.~H. Seymour, and B.~R. Webber, {\it {HERWIG 6: An Event
  generator for hadron emission reactions with interfering gluons (including
  supersymmetric processes)}},  {\em JHEP} {\bf 01} (2001) 010,
  [\href{http://xxx.lanl.gov/abs/hep-ph/0011363}{{\tt hep-ph/0011363}}].

\bibitem{Corcella:2002jc}
G.~Corcella, I.~G. Knowles, G.~Marchesini, S.~Moretti, K.~Odagiri,
  P.~Richardson, M.~H. Seymour, and B.~R. Webber, {\it {HERWIG 6.5 release
  note}},  \href{http://xxx.lanl.gov/abs/hep-ph/0210213}{{\tt hep-ph/0210213}}.

\bibitem{Bevilacqua:2017cru}
G.~Bevilacqua, M.~V. Garzelli, and A.~Kardos, {\it {$t\bar{t}b\bar{b}$
  hadroproduction with massive bottom quarks with PowHel}},
  \href{http://xxx.lanl.gov/abs/1709.06915}{{\tt arXiv:1709.06915}}.

\bibitem{Garzelli:2014aba}
M.~V. Garzelli, A.~Kardos, and Z.~Tr\'ocs\'anyi, {\it {Hadroproduction of
  $t\bar{t}b\bar{b}$ final states at LHC: predictions at NLO accuracy matched
  with Parton Shower}},  {\em JHEP} {\bf 03} (2015) 083,
  [\href{http://xxx.lanl.gov/abs/1408.0266}{{\tt arXiv:1408.0266}}].

\bibitem{Buckley:2014ana}
A.~Buckley, J.~Ferrando, S.~Lloyd, K.~Nordstrom, B.~Page, M.~Rofenacht,
  M.~Schonherr, and G.~Watt, {\it {LHAPDF6: parton density access in the LHC
  precision era}},  {\em Eur. Phys. J.} {\bf C75} (2015) 132,
  [\href{http://xxx.lanl.gov/abs/1412.7420}{{\tt arXiv:1412.7420}}].

\bibitem{Cacciari:1998it}
M.~Cacciari, M.~Greco, and P.~Nason, {\it {The P(T) spectrum in heavy flavor
  hadroproduction}},  {\em JHEP} {\bf 05} (1998) 007,
  [\href{http://xxx.lanl.gov/abs/hep-ph/9803400}{{\tt hep-ph/9803400}}].

\bibitem{Campbell:2019dru}
J.~Campbell and T.~Neumann, {\it {Precision Phenomenology with MCFM}},  {\em
  JHEP} {\bf 12} (2019) 034, [\href{http://xxx.lanl.gov/abs/1909.09117}{{\tt
  arXiv:1909.09117}}].

\bibitem{Campbell:2005bb}
J.~M. Campbell and F.~Tramontano, {\it {Next-to-leading order corrections to Wt
  production and decay}},  {\em Nucl. Phys. B} {\bf 726} (2005) 109--130,
  [\href{http://xxx.lanl.gov/abs/hep-ph/0506289}{{\tt hep-ph/0506289}}].

\bibitem{PhysRevD.98.030001}
{\bf Particle Data Group} Collaboration, M.~e.~a. Tanabashi, {\it Review of
  particle physics},  {\em Phys. Rev. D} {\bf 98} (Aug, 2018) 030001.

\bibitem{Alekhin:2018pai}
S.~Alekhin, J.~Bl\"umlein, and S.~Moch, {\it {NLO PDFs from the ABMP16 fit}},
  {\em Eur. Phys. J. C} {\bf 78} (2018), no.~6 477,
  [\href{http://xxx.lanl.gov/abs/1803.07537}{{\tt arXiv:1803.07537}}].

\bibitem{Hou:2019qau}
T.-J. Hou {\em et.~al.}, {\it {Progress in the CTEQ-TEA NNLO global QCD
  analysis}},  \href{http://xxx.lanl.gov/abs/1908.11394}{{\tt
  arXiv:1908.11394}}.

\bibitem{Catani:2004nc}
S.~Catani, D.~de~Florian, G.~Rodrigo, and W.~Vogelsang, {\it {Perturbative
  generation of a strange-quark asymmetry in the nucleon}},  {\em Phys. Rev.
  Lett.} {\bf 93} (2004) 152003,
  [\href{http://xxx.lanl.gov/abs/hep-ph/0404240}{{\tt hep-ph/0404240}}].

\bibitem{CT18private}
{\bf CT18} Collaboration. Private communication, 2021.

\bibitem{Alioli:2008tz}
S.~Alioli, P.~Nason, C.~Oleari, and E.~Re, {\it {NLO Higgs boson production via
  gluon fusion matched with shower in POWHEG}},  {\em JHEP} {\bf 04} (2009)
  002, [\href{http://xxx.lanl.gov/abs/0812.0578}{{\tt arXiv:0812.0578}}].

\bibitem{Alioli:2008gx}
S.~Alioli, P.~Nason, C.~Oleari, and E.~Re, {\it {NLO vector-boson production
  matched with shower in POWHEG}},  {\em JHEP} {\bf 07} (2008) 060,
  [\href{http://xxx.lanl.gov/abs/0805.4802}{{\tt arXiv:0805.4802}}].

\bibitem{Skands:2014pea}
P.~Skands, S.~Carrazza, and J.~Rojo, {\it {Tuning PYTHIA 8.1: the Monash 2013
  Tune}},  {\em Eur. Phys. J. C} {\bf 74} (2014), no.~8 3024,
  [\href{http://xxx.lanl.gov/abs/1404.5630}{{\tt arXiv:1404.5630}}].

\bibitem{TheATLAScollaboration:2014rfk}
{\bf ATLAS} Collaboration, {\it {ATLAS Pythia 8 tunes to 7 TeV data}},  {\em
  \href{http://cds.cern.ch/record/1974411}{ATL-PHYS-PROC-2014-273}} (11, 2014).

\bibitem{Ball:2012cx}
R.~D. Ball {\em et.~al.}, {\it {Parton distributions with LHC data}},  {\em
  Nucl. Phys. B} {\bf 867} (2013) 244--289,
  [\href{http://xxx.lanl.gov/abs/1207.1303}{{\tt arXiv:1207.1303}}].

\bibitem{Marquard:2015qpa}
P.~Marquard, A.~V. Smirnov, V.~A. Smirnov, and M.~Steinhauser, {\it {Quark Mass
  Relations to Four-Loop Order in Perturbative QCD}},  {\em Phys. Rev. Lett.}
  {\bf 114} (2015), no.~14 142002,
  [\href{http://xxx.lanl.gov/abs/1502.01030}{{\tt arXiv:1502.01030}}].

\bibitem{Garzelli:2020fmd}
M.~V. Garzelli, L.~Kemmler, S.~Moch, and O.~Zenaiev, {\it {Heavy-flavor
  hadro-production with heavy-quark masses renormalized in the ${\overline{\rm
  MS}}$, MSR and on-shell schemes}},  {\em JHEP} {\bf 04} (2021) 043,
  [\href{http://xxx.lanl.gov/abs/2009.07763}{{\tt arXiv:2009.07763}}].

\bibitem{Alekhin:2021xcu}
S.~Alekhin, A.~Kardos, S.~Moch, and Z.~Trocsanyi, {\it {Precision studies for
  Drell-Yan processes at NNLO}},
  \href{http://xxx.lanl.gov/abs/2104.02400}{{\tt arXiv:2104.02400}}.

\bibitem{Ebert:2020dfc}
M.~A. Ebert, J.~K.~L. Michel, I.~W. Stewart, and F.~J. Tackmann, {\it
  {Drell-Yan $q_{T}$ resummation of fiducial power corrections at N$^{3}$LL}},
  {\em JHEP} {\bf 04} (2021) 102,
  [\href{http://xxx.lanl.gov/abs/2006.11382}{{\tt arXiv:2006.11382}}].

\bibitem{Banfi:2007gu}
A.~Banfi, G.~P. Salam, and G.~Zanderighi, {\it {Accurate QCD predictions for
  heavy-quark jets at the Tevatron and LHC}},  {\em JHEP} {\bf 07} (2007) 026,
  [\href{http://xxx.lanl.gov/abs/0704.2999}{{\tt arXiv:0704.2999}}].

\bibitem{Campbell:2017hsr}
J.~Campbell, J.~Huston, and F.~Krauss, {\em {The Black Book of Quantum
  Chromodynamics}: {A Primer for the LHC Era}}.
\newblock Oxford University Press, 12, 2017.

\end{thebibliography}\endgroup

\end{document}